\newcommand{\cev}[1]{\reflectbox{\ensuremath{\vec{\reflectbox{\ensuremath{#1}}}}}}

\documentclass[12pt,a4]{article}
\usepackage{soul}
\newcommand{\rf}[1]{(\ref{#1})}
\usepackage{amsthm,amsmath,latexsym,amssymb,amsfonts,amscd}
\usepackage{graphicx,lscape,fancyhdr,array,euscript,wrapfig} 
\usepackage{ifthen,empheq}
\usepackage{sidecap}
\usepackage{rotating,framed}
\usepackage{ifpdf}
\usepackage{verbatim}
\usepackage{color}

\numberwithin{equation}{section}
\usepackage{tikz}
\usepackage{hyperref}
\usepackage[sorting=none,	maxnames=99,	firstinits=true,	doi=false,	citestyle=numeric-comp,	backend=bibtex	]{biblatex}

\AtEveryBibitem{\clearfield{title}}
\renewbibmacro{in:}{}
\newcolumntype{x}[1]{%
>{\centering\hspace{0pt}}m{#1}}%
\newcolumntype{w}[1]{%
>{\raggedright\hspace{0pt}}m{#1}}%
\newcolumntype{z}[1]{%
>{\raggedleft\hspace{0pt}}m{#1}}%

\newcommand{\DO}{{D_{\Omega}}}

\newcommand{\pl}{\partial}
\newcommand{\be}{\begin{equation}}
\newcommand{\ee}{\end{equation}}

\renewcommand{\d}{\textrm{d}}

\newcommand{\mm}{{\ensuremath{\underline{m}}}}
\newcommand{\nn}{{\ensuremath{\underline{n}}}}
\newcommand{\rr}{{\ensuremath{\underline{r}}}}
\newcommand{\kk}{{\ensuremath{\underline{k}}}}

\newcommand{\aA}{{\ensuremath{\mathcal{A}}}}

\newcommand{\ga}{\alpha}
\newcommand{\gb}{\beta}
\newcommand{\gc}{\gamma}
\newcommand{\gd}{\delta}

\newcommand{\MW}{\mathcal{W}}
\newcommand{\MB}{\mathcal{B}}
\newcommand{\MS}{\mathcal{S}}
\newcommand{\MA}{\mathcal{\aA}}
\newcommand{\Mxi}{\mathcal{\xi}}

\newcommand{\fud}[2]{{}^{#1}{}_{#2}\,}
\newcommand{\fdu}[2]{{}_{#1}{}^{#2}\,}

\newcommand{\homo}[2]{{\Gamma_{#1}\left\langle#2\right\rangle}}

\newcommand{\B}{\boldsymbol{B}} 
\newcommand{\W}{\boldsymbol{W}} 
\newcommand{\flC}{{\mathbf{C}}} 
\newcommand{\flomega}{{\boldsymbol{\omega}}} 
\newcommand{\flxi}{{\boldsymbol{\xi}}} 
\newcommand{\pC}{\hat{C}} 
\newcommand{\pomega}{{\hat{\omega}}} 
\newcommand{\pxi}{{\xi}} 
\newcommand{\sC}{{\tilde{C}}} 
\newcommand{\somega}{{\tilde{\omega}}} 
\newcommand{\sxi}{{\tilde{\xi}}} 
\newcommand{\adD}{{D}} 
\newcommand{\tadD}{{\tilde{D}}} 

\newcommand{\PVW}{{\mathbb{W}}}
\newcommand{\PVB}{{\mathbb{B}}}
\newcommand{\PVS}{{\mathbb{S}}}

\newcommand{\formJ}{{\boldsymbol{\mathsf{J}}}}
\newcommand{\formL}{{\boldsymbol{\mathsf{L}}}}
\newcommand{\formK}{{\boldsymbol{\mathsf{K}}}}
\newcommand{\formU}{{\boldsymbol{\mathsf{U}}}}

\newcommand{\formj}{{\boldsymbol{\mathsf{j}}}}
\newcommand{\compA}{{\mathsf{A}}}
\newcommand{\compB}{{\mathsf{B}}}
\newcommand{\compC}{{\mathsf{C}}}
\newcommand{\compJ}{{\mathsf{J}}}
\newcommand{\compj}{{\mathsf{j}}}
\newcommand{\curj}{{\mathsf{j}}}

\newcommand{\besubeqs}{\begin{subequations}}
\newcommand{\esubeqs}{\end{subequations}}

\newcommand{\rhs}{right-hand side \,}
\newcommand{\eomsWS}{equations of motion}

\begin{document}

\begin{center}

{\Large\bfseries Higher Spins and Matter Interacting in Dimension Three} \\

\vskip 0.1\textheight

Pan \textsc{Kessel},${}^1$ Gustavo \textsc{Lucena G\'{o}mez},${}^{1,2}$ Evgeny \textsc{Skvortsov},${}^{1,3}$ Massimo \textsc{Taronna},${}^{1}$

\vspace{2cm}

{\em ${}^{1}$ Albert Einstein Institute, Am M\"{u}hlenberg 1, Golm, D-14476 Germany}\\
\vspace*{5pt}
{\em ${}^{2}$ Institute of Physics AS CR, Na Slovance 2, Prague 8, Czech Republic} \\
\vspace*{5pt}
{\em ${}^{3}$ Lebedev Institute of Physics, Leninsky ave. 53, 119991 Moscow, Russia}

\vskip 0.05\textheight

{\footnotesize \texttt{pan.kessel, gustavo.lucena-gomez, evgeny.skvortsov, massimo.taronna@aei.mpg.de}}

\vskip 0.05\textheight

	{\bf Abstract }  

\end{center} 
\begin{quotation}
\noindent  
The spectrum of Prokushkin--Vasiliev Theory is puzzling in light of the Gaberdiel--Gopakumar conjecture because it generically contains an additional sector besides higher-spin gauge and scalar fields. We find the unique truncation of the theory avoiding this problem to order 2 in perturbations around AdS$_3$. The second-order backreaction on the physical gauge sector induced by the scalars is computed explicitly. The cubic action for the physical fields is determined completely. The subtle issue of the allowed class of pseudo-local field redefinitions is discussed.

\end{quotation}

\newpage

\tableofcontents

\section{Introduction}

Higher-spin theories have been studied in significant detail in the last years.\footnote{See e.g. \cite{Giombi:2009wh,Henneaux:2010xg,Campoleoni:2010zq}. For a non-exhaustive list of reviews we refer to \cite{Vasiliev:1999ba,Bekaert:2005vh,Sagnotti:2011qp,Gaberdiel:2012uj,Taronna:2012gb,Didenko:2014dwa,Gomez:2014dwa,Bekaert:2010hw}.} In particular the three-dimensional case has received significant attention due to a conjecture by Gaberdiel and Gopakumar\cite{Gaberdiel:2012uj} about a duality between $\mathcal{W}_N$-minimal models and three-dimensional higher-spin theories containing one complex scalar field and with gauge algebra $\textrm{hs}(\lambda)$.
More precisely, $\mathcal{W}_N$-minimal models are two-dimensional conformal field theories which are given by Wess--Zumino--Witten coset models of the form
\begin{equation}
\frac{\textrm{SU}(N)_k \otimes \textrm{SU}(N)_1}{\textrm{SU}(N)_{k+1}} \,.
\end{equation} 
This conjecture was put forward for the t'Hooft limit thereof in which $N,k \rightarrow \infty$ at fixed
\begin{equation}
0 \leq \lambda=\frac{N}{N+k} \leq 1 \,.
\end{equation}
The above t'Hooft coupling is to be identified with the $\lambda$ parameter of the $\textrm{hs}(\lambda)$ higher-spin theory. This duality is interesting because two-dimensional conformal field theories are among the best understood interacting quantum field theories. Furthermore higher-spin theories are much simpler than full string theories \cite{Chang:2012kt} particularly in three dimensions \cite{Gaberdiel:2014cha,Gaberdiel:2015mra} because higher-spin gauge fields are non-propagating in this case. 
The Gaberdiel--Gopakumar conjecture therefore provides a relatively simple example of an AdS/CFT duality from which one might hope to understand the general nature of these dualities in more detail. 

An explicit construction of an $\textrm{hs}(\lambda)$ higher-spin theory in three dimensions is Prokushkin--Vasiliev Theory\cite{Prokushkin:1998bq}. Its physical field content is given by a complex scalar\footnote{More precisely we will restrict ourselves to a truncation of Prokushkin--Vasiliev Theory with this physical field content. Off the start Prokushkin--Vasiliev Theory contains two complex scalars \cite{Prokushkin:1998bq}.} with $m^2=-1+\lambda^2$ and a tower of higher-spin gauge fields with spin $s=1,2,3,\dots,\infty$ obeying Fronsdal equations\cite{Fronsdal:1978rb} at order 1 in perturbations around AdS$_3$. This is precisely the spectrum required by the Gaberdiel--Gopakumar conjecture.
However, a priori Prokushkin--Vasiliev Theory contains an additional sector consisting of Killing tensors and a further set of gauge fields which are not related to Fronsdal fields.\footnote{Note that this issue is unrelated to the problem of light states (see e.g. \cite{Gaberdiel:2012uj}).} The field-theoretical interpretation of these fields and their role within the Gaberdiel--Gopakumar duality is to the best of our knowledge unclear. At order 2 in perturbation theory these fields will generically interact with the physical sector. We will refer to them as twisted fields in the following for reasons that will become clear in Section~\ref{sec:general}. In this respect four-dimensional Vasiliev Theory\cite{Vasiliev:1990en} is simpler than the three dimensional one as it is possible to formulate the theory without the need for introducing a twisted sector.

It was known since the work of Vasiliev\cite{Vasiliev:1996xk} that to order 1 in perturbation theory all twisted fields can be set to zero consistently after an appropriate field redefinition. As we will establish in Section~\ref{sec:presentation} this field redefinition is not unique and will lead to free parameters in the second-order equations of motion. We will show that \emph{there exists a unique point in parameter space which allows for trivial solutions of the second-order twisted fields} and one can therefore truncate Prokushkin--Vasiliev Theory to its physical sector at this point. For any other choice of the parameters the twisted fields cannot be set to zero consistently at order 2. Interestingly, there exists another higher-spin theory in three dimensions that is free of twisted fields by construction, which is the $D$-dimensional Vasiliev theory \cite{Vasiliev:2003ev} at $D=3$ which corresponds to $\lambda=1$. We will comment on this further in the conclusions of this paper (see Appendix~\ref{app:Ddimensionaltheory} for technical details).

For vanishing scalar field the physical sector of Prokushkin--Vasliev Theory can be described by $\textrm{hs}(\lambda)\oplus\textrm{hs}(\lambda)$ Chern--Simons Theory\cite{Achucarro:1987vz,Blencowe:1988gj} (see \cite{Campoleoni:2011tn,Gomez:2014kua} for an introduction). For a non-vanishing scalar field much less about the dynamics of the theory is known, and in particular a full action is not yet available. This is because Prokushkin--Vasiliev Theory contains auxiliary coordinates and fields which have to be solved for in order to obtain equations for the physical and twisted fields. To the second-order this becomes a task of considerable technical difficulty and so far this has been studied mostly at the linear level (with \cite{Sezgin:2003pt,Kristiansson:2003xx,Sezgin:2005pv,Giombi:2012ms,Didenko:2009td,Chang:2011mz,Iazeolla:2011cb} among the exceptions). At the linear level the system obeys free equations of motion and therefore the higher-spin fields and the scalar do not interact. 

\emph{In this paper we will systematically extract and analyze the second-order equations of motion of Prokushkin--Vasiliev Theory for both physical and twisted fields}. In particular we will compute the backreaction on the higher-spin gauge fields $\varphi_{\mm(s)}$ due to the scalar $\Phi$ directly from Prokushkin--Vasliev Theory to order 2 in perturbation theory. We do so for $\lambda = \frac12$. For this analysis we reformulate perturbation theory in a manifestly Lorentz-covariant form. The theory at $\lambda=\frac12$ is technically simpler to deal with but we expect its features to be generic.

From the metric-like perspective it is expected that this backreaction has a compact form:
\begin{align}
\square \varphi_{\mm(s)}+...&=\frac{g_s}{s} J_{\mm(s)}\,, \label{standbackr}
\end{align}
with a priori undetermined coefficients $g_s$. Up to terms proportional to the cosmological constant $\Lambda$ and the scalar's equations of motion the canonical currents $J_{\mm(s)}$ read
\begin{equation}
\label{standardcurrent}
J_{\mm(s)}= (-i)^s \, \Phi^* (\cev{\nabla}_\mm - \vec{\nabla}_\mm)^s\Phi + \mathcal{O}(\Lambda)\,.
\end{equation}
In Section~\ref{sec:cubic} \emph{we fix the coefficients $g_s$ by requiring closure of the scalar's gauge transformations at $\lambda=\frac12$ and therefore determine the cubic action of the physical sector}.\footnote{The general case will be presented elsewhere \cite{Charlotte}.}

In order to relate the backreaction obtained directly from Prokushkin--Vasiliev Theory to \eqref{standardcurrent} a field redefinition quadratic in the scalar field $\Phi$ is needed containing terms of the form
\begin{align}\label{eq:quasiloc}
\sum_l \nabla_\mm...\nabla_\mm \overbrace{\nabla_\nn...\nabla_\nn}^{l}\Phi^* \nabla^\nn...\nabla^\nn\nabla_\mm...\nabla_\mm\Phi\,.
\end{align}
Field redefinitions of this type are also necessary to formulate Prokushkin--Vasiliev Theory in a manifestly Lorentz-covariant manner. These redefinitions contain generically an infinite number of derivatives and are therefore potentially non-local. In particular they allow for a complete removal of the backreaction in \eqref{standbackr}, as was first shown in \cite{Prokushkin:1999xq}.
Our analysis highlights the urgent need for a better understanding of the class of allowed field redefinitions in Prokushkin--Vasiliev Theory (see e.g. \cite{Vasiliev:2015wma}). \\

\noindent Summarizing, our results are the following:
\begin{itemize}
\item The second-order Prokushkin--Vasiliev Theory at $\lambda = \frac12$ possesses free parameters specifying the truncation to the physical sector. Only at one point in parameter space can one consistently set all second-order twisted fields to zero. 
\item The backreaction on the second-order physical fields is computed explicitly in a manifestly Lorentz-covariant manner, in particular at the point in parameter space mentioned hereabove.
\item We determine the cubic action describing the physical sector of the theory by enforcing closure of the gauge transformations for the scalar. The coupling constants $g_s$ are thus fixed and read
\begin{equation}
g_s = \frac{1}{(2s-2)!}\,.
\end{equation}
\end{itemize} 
\noindent
Along the way, we reformulate perturbation theory in a manifestly Lorentz-covariant form and we also systematically compute all cohomologies relevant for our second-order analysis. \\

We have structured this paper in such a way that the reader should be able to follow the presentation of our results without any detailed understanding of Prokushkin--Vasiliev Theory. The equations of motion for twisted and physical fields are extracted from Prokushkin--Vasiliev Theory but to second-order this is a technically involved task. After reviewing necessary ingredients for our analysis in Section~\ref{sec:general} we will only quote the extracted equations of motion and discuss their implications in Section~\ref{sec:presentation}. In Section~\ref{sec:cubic} we will discuss the cubic action in a self-contained way. In Section~\ref{sec:Veq} we will then outline how we extract the equations of motion from Prokushkin--Vasiliev Theory leaving the more technical details to the appendices. The reader not interested in the way the results are obtained may simply skip Section~\ref{sec:Veq}, whereas the reader interested in the procedure should read the latter section and then move to Sections~\ref{sec:presentation} and~\ref{sec:cubic}. In Section~\ref{sec:conclusions} we discuss our results and give an outlook.

\section{Ingredients of Higher-Spin Theories}
\label{sec:general}
In the following we summarize all the necessary ingredients for presenting our main results in Section~\ref{sec:presentation}. This section is structured as follows:

\begin{description}
\item[Section~\ref{sec:framemetric}] will review some basic facts of the metric-like and frame-like formulation of higher-spin theories. 
\item[Section~\ref{sec:HSalgebra}] will present the higher-spin algebra $\textrm{hs}(\frac12)$. We will briefly discuss how this algebra is constructed and outline a particularly useful oscillator realization.
\item[Section~\ref{sec:precursors}] details the free equations of motion for the various fields of Prokushkin--Vasiliev Theory. These involve not only scalar fields and higher-spin gauge fields, but also additional fields whose interpretation is not obvious as we will discuss. 
\item[Section~\ref{sec:unfolded}] explains the structure of the non-linear equations of motion for these fields. We will discuss their form at both first and second order in perturbations around an $\textrm{AdS}_3$-background. This section will furthermore introduce the basic quantities we calculated for this work.
\item[Section~\ref{sec:cohomology}] discusses whether some of the interactions terms in the equations of motion of Section~\ref{sec:unfolded} can be removed by field redefinitions. This question will be related to studying the cohomologies of the adjoint and twisted-adjoint covariant derivative.
\item[Section~\ref{sec:conservation}] details how these covariant derivatives can be expressed in Fourier space and uses this fact to derive conservation identities for currents in Prokushkin--Vasiliev Theory.
\end{description}
Some readers might want to skip some of the following subsections as they mostly review well-established material \cite{Vasiliev:1992ix, Prokushkin:1998bq} for the discussion of our results in Section~\ref{sec:presentation}. We summarize all conventions used in this paper in Appendix~\ref{app:notation}.

\subsection{Frame-like and Metric-like Formulation of Higher-Spin Theories}
\label{sec:framemetric}
Historically massless spin-$s$ fields were first described by introducing a totally-symmetric tensor field $\varphi_{\mm_1 \dots \mm_s}$ with vanishing double trace \cite{Fronsdal:1978rb},
\begin{equation}
\varphi^{\rr \kk}{}_{ \rr \kk \mm_1 \dots \mm_{s-4}}=0 \,.
\end{equation}
For non-vanishing cosmological constant $\Lambda$ the free equations of motion then take the form
\begin{align}\label{FronsdalEq}
F_{\mm(s)}&=\Box \varphi_{\mm(s)}-\nabla_{\mm}\nabla^\nn\varphi_{\nn\mm(s-1)}+\frac12\nabla_{\mm}\nabla_{\mm}\varphi^{\nn}{}_{\nn\mm(s-2)}-m^2\varphi_{\mm(s)}+2\Lambda g_{\mm\mm}\varphi\fdu{\mm(s-2)\nn}{\nn}=0\,,
\end{align}
with $m^2=\Lambda(s-(D+s-3)(s-2))$. Furthermore $g_{\mm \nn}$ and $\nabla_\mm$ denote $\textrm{(A)dS}_D$ metric and the $\textrm{(A)dS}_D$ covariant derivative respectively.  The equations of motion are invariant under the following gauge transformations:
\begin{equation}
\delta \varphi_{\mm(s)}=\nabla_\mm\epsilon_{\mm(s-1)}\,,
\end{equation}
for a traceless gauge parameter
\begin{equation}
\epsilon^{\kk}{}_{\kk \mm(s-3)} = 0 \,.
\end{equation}
An alternative approach, pioneered in \cite{Vasiliev:1980as}, is to describe the higher-spin theory in terms of higher-spin generalizations of the spin-2 vielbein and spin-connection. So far a fully non-linear theory can only be formulated using this formalism. In the following we will restrict ourselves to the three-dimensional case, for which the number of spin-connections does not grow with the spin as opposed to what happens in higher dimensions \cite{Lopatin:1987hz}. We denote the spin-$s$ vielbein and spin-connection by
\begin{align}
&e^{a(s-1)}_\mm\,, \qquad \omega^{a(s-1),b}_\mm\,,
\end{align}
where both fields are traceless in their fiber indices. In the three-dimensional case we can furthermore dualize the spin-connection to a symmetric tensor of the same type as the vielbein
\begin{align}
&\omega^{a(s-1)}_\mm \equiv \epsilon\fud{a}{bc}\omega^{a(s-2)b,c}_\mm\,.
\end{align} 
In the following we will denote the $\textrm{(A)dS}_3$ background fields by $h^a_\mm$ and $\varpi^a_\mm$. At the lowest order in perturbations around the $\textrm{(A)dS}_3$ background the Fronsdal field can be identified with the totally-symmetric part of the higher-spin vielbein
\begin{align}
\label{eq:metricframeid}
\varphi_{\mm(s)}= e^{a(s-1)}_{\mm} h^{\phantom{a(}}_{a\,\mm}...h^{\phantom{a(}}_{a\,\mm}\,.
\end{align} 
The free equations of motion are generalizations of the vanishing torsion and Riemann tensor equations in gravity and are given by
\besubeqs \label{frameequations}
\begin{align}
T^{a(s-1)}&\equiv\nabla e^{a(s-1)}-\epsilon\fud{a}{bc}h^b\wedge \omega^{a(s-2)c}=0  \label{eq:zeroTorsionConstr} \,,\\
R^{a(s-1)}&\equiv\nabla \omega^{a(s-1)}+\epsilon\fud{a}{bc}h^b\wedge e^{ a(s-2)c}=0  \label{eq:vanishingRiemann} \,.
\end{align}
\esubeqs
The Fronsdal equations are then found in exactly the same way as for the spin-two case, that is, by solving the zero-torsion constraint \eqref{eq:zeroTorsionConstr} for $\omega=\omega(\nabla e)$ and then plugging the solution into the second equation \eqref{eq:vanishingRiemann}. In three dimensions all higher-spin fields including the spin-2 field are topological, i.e. the Fronsdal equations \eqref{FronsdalEq} or equivalently the frame-like equations \eqref{frameequations} do not describe any local degrees of freedom. 
Also in three dimensions there is the following isomorphism:	
\begin{equation}
\textrm{sp}_2\simeq \textrm{sl}_2\simeq \textrm{so}(1,2)\,,
\end{equation} 
which allows to convert every vector index into two spinorial two-component indices with the help of the matrices $\sigma^{\ga\gb}_\mm \in \{ \mathbb{I},\sigma_1, \sigma_3 \}$. Every symmetric and traceless rank-$k$ $\textrm{so}(1,2)$ tensor is therefore isomorphic to a spinorial symmetric tensor of rank $2k$ 
\begin{align}
\label{eq:spinref}
V^{a(k)}\qquad \longleftrightarrow\qquad V^{\ga(2k)}\,.
\end{align}
This dictionary will be used extensively in the following. Furthermore we will consider only the case of negative cosmological constant in the rest of our discussion.

\subsection{Higher-Spin Symmetry} 
\label{sec:HSalgebra}
The higher-spin algebra is a key ingredient of higher-spin theories. It links together a number of higher-spin fields into a single connection, or more generally into a single module of the algebra \cite{Fradkin:1986ka, Vasiliev:1986qx, Vasiliev:1989re,Vasiliev:2004cm}. The higher-spin algebra is constructed from a certain quotient of the universal enveloping algebra of the $\textrm{AdS}_3$-isometry algebra, which can be also equipped with some discrete elements or further tensored with matrix algebras.

The $\textrm{AdS}_3$-isometry algebra is semi-simple, $\textrm{so}(2,2)\simeq \textrm{sp}(2)\oplus \textrm{sp}(2)$, which leads one to consider quotients of the universal enveloping algebra $\textrm{U}(\textrm{sp}(2))$. One then considers the associative algebra \cite{Vasiliev:1989re, Feigin}, 
\begin{equation}
\textrm{Aq}(2,\nu)=\textrm{U}(\textrm{sp}(2))/ \langle C_2+\frac14(3-2\nu-\nu^2) \rangle \,,
\end{equation}
where the denominator denotes the two-sided ideal generated by the quadratic Casimir $C_2$ subtracted by some number parametrized by $\nu\in\mathbb{R}$. With respect to its commutator the associative algebra forms a Lie algebra which decomposes into (as a Lie algebra) 
\begin{equation}
\textrm{Aq}(2,\nu)= \mathbb{C} \oplus \textrm{hs}(\lambda) \,.
\end{equation}
Here we defined the higher spin algebra $\textrm{hs}(\lambda)$ with $\lambda=\frac12(\nu + 1)$ while $\mathbb{C}$ is the identity component of the universal enveloping algebra.

In this work we will focus on the case $\nu=0$ (i.e. $\lambda=\frac12$) for which a particularly simple oscillator realization of this algebra can be given, which we will briefly review in the following. Let $\{\hat{y}_\ga\}$ be a set of two canonically commuting oscillators, obeying
\begin{align}
\label{eq:osszillators}
[\hat{y}_\ga,\hat{y}_\gb]&=2i\epsilon_{\ga\gb}\,.
\end{align} 
Using this definition we can realize the $\textrm{sp}(2)$ algebra by considering the combinations $T_{\ga \gb} \equiv -\frac{i}4\{\hat{y}_\ga,\hat{y}_\gb\}$, which satisfy
\begin{align}
[T_{\ga\ga},T_{\gb\gb}]&=\epsilon_{\ga\gb}T_{\ga\gb} \,.
\end{align}
The associative algebra $\textrm{Aq}(2,0)$ can then be constructed by considering even functions of these oscillators, i.e. $f(\hat{y})=f(-\hat{y})$.
Using \eqref{eq:osszillators} one can easily check that
\begin{align}
\label{eq:quadrCasimir}
C_2 = -\frac12 T^{\ga \gb} T_{\ga \gb} = - \frac34 \,,
\end{align}
which indeed corresponds to the case $\nu=0$. 
The $\textrm{AdS}_3$-isometry algebra contains two copies of $\textrm{sp}(2)$. It is convenient to introduce a Clifford pair\footnote{We do not collect the Clifford elements into a doublet $\{\psi_i,\psi_j\}=\delta_{ij}$ because, as will be discussed below, the vacuum solution for Prokushkin--Vasiliev theory breaks this symmetry.} $\phi$ and $\psi$ 
\begin{equation}
\label{eq:cliffordElements}
\phi^2=1\,, \qquad \psi^2=1\,, \qquad \textrm{such that}\quad\{\phi,\psi\}=0\,,
\end{equation}
which we further assume to commute with all $\hat{y}_\ga$ oscillators. The Clifford element $\phi$ ensures the doubling of $\textrm{sp}(2)$. There is not yet any particular reason for introducing $\psi$ but as we will see this element is important for the theory to describe non-trivial dynamics. Using these definitions we can realize the $\textrm{AdS}_3$ algebra as follows
\begin{align}
\label{eq:AdSGenerators}
L_{\ga\gb}&\equiv -\frac{i}4\{\hat{y}_\ga,\hat{y}_\gb\}\,, & P_{\ga\gb}&\equiv \phi L_{\ga\gb}\,.
\end{align}
Here and in the following we set the cosmological constant $\Lambda=1$. Using \eqref{eq:osszillators} and \eqref{eq:cliffordElements} one can easily check that the definitions \eqref{eq:AdSGenerators} indeed obey the expected commutation relations
\begin{align}
\label{adsalgebraosc}
[L_{\ga\ga},L_{\gb\gb}]&=\epsilon_{\ga\gb}L_{\ga\gb}\,, &[L_{\ga\ga},P_{\gb\gb}]&=\epsilon_{\ga\gb}P_{\ga\gb}\,, &[P_{\ga\ga},P_{\gb\gb}]&=\epsilon_{\ga\gb}L_{\ga\gb}\,.
\end{align}
The algebra can be effectively dealt with by replacing functions of operators $\hat{y}_\ga$ with functions of ordinary commuting variables $y_\ga$ that are multiplied with the help of the Moyal star-product
\begin{align}
\label{eq:moyalProduct}
(f\star g)(y)= f(y)\exp{i\left(\frac{\cev{\pl}}{\pl y^\ga}\epsilon^{\ga\gb} \frac{\vec{\pl}}{\pl y^\gb}\right)} g(y)=\frac{1}{(2\pi)^2} \int \d^2u \, \d^2v \, e^{i v^\ga u_\ga} \, f(y+u) \, g(y+v) \,,
\end{align}
where any boundary terms are to be dropped when using the integral form.

For $\nu \neq 0$ a deformed oscillator realization can be given but the corresponding star product is not a Moyal product \cite{Korybut:2014jza}. This makes the case $\nu=0$ technically simpler, although we expect it to possess features similar to that of the more general Prokushkin--Vasiliev Theory. 

Note that in the following we will refer somewhat loosely to functions of not only $y_\ga$ but also $\phi$ and $\psi$ as taking value in the higher-spin algebra.

\subsection{Free Equations of Motions}
\label{sec:precursors}
In this section we will explain how free equations of motion for matter and higher-spin gauge fields can be constructed from higher-spin symmetry.
The relevant objects to describe higher-spin fields and matter fields are a connection one-form $\flomega$ and a zero-form $\flC$ respectively which are functions\footnote{Note that we allow for a $y_\ga$-independent components of the gauge connection, which results in an additional spin-1 field component of the connection.} of $y_\ga$, $\psi$ and $\phi$. Unless stated otherwise we consider bosonic fields only which corresponds to restricting the fields to even functions of $y_\alpha$.
As we will discuss $\flC$ and $\flomega$ additionally encode twisted fields which are necessarily present if we want to describe non-trivial dynamics in this language. 

Empty anti-de Sitter space, which is a vacuum solution of the higher-spin theory, can be described by a flat connection $\Omega$
\begin{align}
\d\Omega&=\Omega \, \wedge\star \, \Omega\,,
\end{align}
that can be written in terms of the generators of the $\textrm{AdS}_3$-isometry algebra \eqref{eq:AdSGenerators} as
\begin{align}
\Omega&=\frac12\varpi^{\ga\ga} L_{\ga\ga} +\frac12 h^{\ga\ga} P_{\ga\ga}\,,\label{AdSConnection}
\end{align} 
where again $\varpi^{\ga\ga}$ and $h^{\ga\ga}$ denote the spin-connection and vielbein of $AdS$ space\footnote{Allowing the vacuum connection to have non-zero values for higher-spin fields one can describe matter fields on a more general background, e.g. a higher-spin black hole \cite{Ammon:2011ua,Kraus:2012uf,Cabo-Bizet:2014wpa}, but in a linearized approximation and therefore neglecting the backreaction of matter fields. In this work we will however only consider a pure $\textrm{AdS}_3$-background.}. The free equations are then given by
\begin{equation}
\DO \flomega  = 0\,, \qquad
\DO \flC = 0\,, \label{omegaCeq}
\end{equation}
where we have introduced the $\textrm{AdS}_3$ covariant derivative $\DO$:
\begin{align}
\label{covDerNonLLTDef}
\DO \mathbf{F}&=\d \mathbf{F}-\Omega \wedge \star \mathbf{F} + (-1)^{|\mathbf{F}|} \mathbf{F} \wedge \star \Omega\,,
\end{align}
where $|\mathbf{F}|$ denotes the form-degree of $\mathbf{F}$.
One can easily show that the covariant derivative is nilpotent of degree~$2$, i.e. $\DO\circ\DO=0$.
Furthermore the free equations are invariant under the following gauge transformations
\begin{align}
\delta\flomega=\d \flxi - [\Omega,\flxi]_\star\,, \qquad\label{eq:precGauge}
\delta \flC =0\,.
\end{align}
Note that the $\phi$-dependence of $P_{\ga \ga}$ in \eqref{AdSConnection} and the identity $[\phi f, g \psi]=\phi\{ f,g\}\psi$ imply that the covariant derivative $\DO$ acts as follows:
\begin{align}
\label{eq:decomposition}
\DO \left\{ \, g(y,\phi|x) + \tilde{g}(y,\phi|x) \, \psi \, \right\} = \adD g(y,\phi|x) + \tadD \tilde{g}(y,\phi|x) \, \psi \,,
\end{align}
where we have conveniently defined the adjoint and twisted-adjoint covariant derivatives
\begin{align}
\label{covDer}
\adD&=\nabla-\frac12 \, \phi \, h^{\ga\ga}[L_{\ga\ga},\bullet ]_\star=\nabla-\phi h^{\ga\ga}y^{\vphantom{y}}_\ga\pl^y_\ga\,,\\
\tadD&=\nabla-\frac12 \, \phi \, h^{\ga\ga}\{L_{\ga\ga},\bullet \}_\star=\nabla+\frac{i}{2}\phi h^{\ga\ga} (y_\ga y_\ga-\pl^y_\ga \pl^y_\ga)\,,
\end{align}
where $\nabla$ is the usual Lorentz-covariant derivative:
\begin{align}
\label{eq:lorentzDer}
\nabla&=\d\bullet-\frac12 \varpi^{\ga\ga}[L_{\ga\ga},\bullet]_\star=\d-\varpi^{\ga\ga} y^{\vphantom{y}}_\ga \pl^y_\ga\,.
\end{align}
The above differential form of the operators $\adD$ and $\tadD$ can be easily derived using \eqref{eq:moyalProduct}. An important difference between the \emph{adjoint covariant derivative $\adD$} and the \emph{twisted-adjoint covariant derivative $\tadD$} is that the former commutes with the $y_\alpha$-number operator $y^\nu\pl^y_\nu$, i.e. it slices fields into finite-dimensional modules each having a fixed degree in $y_\alpha$, while the latter mixes components with different even (odd) powers of $y_\ga$. Both covariant derivatives are nilpotent as an immediate consequence of $\DO\circ\DO=0$.

Due to \eqref{eq:decomposition} it is useful to decompose $\flomega$ and $\flC$ as follows:
\begin{align}
\label{eq:splittingTwistedPhys}
\flomega(y,\phi,\psi)&=\somega(y,\phi)\psi+\pomega(y,\phi)\,, & \flC(y,\phi,\psi)&=\sC(y,\phi)+\pC(y,\phi)\psi\,.
\end{align}
We refer to the fields $\pomega$ and $\pC$ as {\it physical} and to $\somega$ and $\sC$ as the {\it twisted} sector of the theory.\footnote{In \cite{Prokushkin:1998bq} they were called auxiliary, but we use the term twisted since many of the fields in the physical sector are auxiliary as well. Twisting is related to the type of higher-spin algebra representation they take values in as compared to the physical fields.} Using \eqref{eq:decomposition} the equations of motion and gauge transformations split as
\begin{align}
\tadD\somega &=0\,, &  \delta\somega&=\tadD \sxi \,, & \adD \sC&=0 \,, & \delta \sC = 0\,,\label{shadoweq}\\
\adD\pomega&=0\,, &  \delta \pomega&=\adD \pxi \,, &\tadD \pC &=0 \,, & \delta \pC = 0\,.\label{physeq}
\end{align}
By expanding in $y_\alpha$ we can see that the equations of motion have the following content:

\begin{description}
\item[Higher-Spin frame-like fields]\hspace*{-5pt}, the vielbein and the spin-connection, are contained in $\pomega(y,\phi)$:
\begin{equation}
\label{splittingomega}\pomega(y,\phi)=\sum_s \frac{1}{(2s-2)!}y_{\ga(2s-2)}\left(\omega^{\ga(2s-2)}+\phi e^{\ga(2s-2)}\right)\,,\end{equation}
while splitting the equations \eqref{physeq} with respect to $\phi$ leads to the generalized zero-torsion and zero-curvature conditions on these fields:
\besubeqs
\label{HSTR}
\begin{align}
T^{\ga(n)}&=\nabla e^{\ga(n)}-h\fud{\ga}{\gc}\wedge \omega^{\gc \ga(n-1)}=0\,,\\
R^{\ga(n)}&=\nabla \omega^{\ga(n)}-h\fud{\ga}{\gc}\wedge e^{\gc \ga(n-1)}=0\,,
\end{align}
\esubeqs
which are exactly \eqref{frameequations} in the spinorial language of \eqref{eq:spinref} and are therefore equivalent to the Fronsdal equation \eqref{FronsdalEq} as explained in Subsection \ref{sec:framemetric}.
It is clear from this point of view that the higher-spin fields are topological since \eqref{omegaCeq} and \eqref{eq:precGauge} describe a flat connection. 

\item[Two physical scalar fields] encoded in $\pC$. Indeed, the component form of the equations \eqref{physeq} projected onto the two orthogonal subspaces by $\Pi_\pm=\frac{1\pm \phi}{2}$ are
\begin{equation}\label{scalarunfld}
\nabla \pC_\pm^{\ga(n)}\pm i h^{\ga\ga} \pC^{\ga(n-2)}_\pm \mp \frac{i}{2}h_{\gc\gc} \pC^{\gc\gc\ga(n)}_\pm=0\,,
\end{equation}
and tell us that $\pC^{\ga\ga}_\pm$ parametrizes the first derivative of $\pC_\pm$. 
Contracting \eqref{scalarunfld} for $n=2$ with an inverse vielbein leads to
\begin{equation}
h_{\ga\ga}^\mm (\nabla_\mm \pC_\pm^{\ga\ga}\pm i h^{\ga\ga}_\mm \pC_\pm)=0 \,,
\end{equation}
where the contraction with the vielbein produces a trace of $\pC^{\ga(4)}$, which is identically zero, and which we have therefore left out altogether. Combining the resulting equation with \eqref{scalarunfld} for $n=0$,
\begin{equation}
\nabla \pC_\pm \mp \frac{i}{2} h_{\gc\gc} \pC^{\gc\gc}_\pm=0 \,,
\end{equation}
we recover the Klein--Gordon equation,
\begin{align}
\square \pC_\pm=-\frac34 \pC_\pm\,, 
\end{align}
for two real scalars.\footnote{\label{fut:realityCond} According to \cite{Prokushkin:1998bq} the scalars obey $\pC_+^\dagger=\pC_-$, which follows from the reality conditions $(\flC(\phi))^\dagger = \flC(-\phi)$ where $y_\ga^\dagger = y_\ga$ and $(\phi \psi)^\dagger=\psi \phi$.}

The rest of the equations express the remaining components as derivatives of the scalar:
\begin{align}
\label{eq:compOfPhysC}
\pC_{\ga(2k)}=(4i \, \phi \, h_{\ga\ga}^{\mm}\nabla_{\mm})^k \pC(x)\,.
\end{align}
Therefore, the dynamical content of $\tadD \pC=0$ is given by two scalar fields.\footnote{At this point it is clear that there is no need for doubling of the scalar fields. Indeed a single scalar field on the AdS or a more general higher-spin background can be described along the same lines by taking $\pC$ be a function $\pC(y)$ of $y_\alpha$ and imposing 
\be\d \pC+A_+ \star\pC-\pC\star A_-=0\,,\ee 
where $A_\pm(y)$ are two flat connections of $Aq(2,0)$. This equation is consistent, but how to introduce nonlinearities in $\pC$ therein is not known. The Prokushkin--Vasiliev construction allows to construct consistent nonlinearities for such free equations of motion, but then one does need the $\phi$ element (and in fact also $\psi$).} Let us note that the mass corresponds 
to a conformally coupled scalar, but Prokushkin--Vasiliev theory is not conformal. The value of the mass is given by the $\textrm{sp}(2)$-Casimir operator computed on the given oscillator representation as can be seen by comparing with \eqref{eq:quadrCasimir}.

\item[A twisted zero-form] denoted by $\sC$. The equations for $\sC$ decompose into an infinite set of Killing equations
\begin{align}
\label{eq:twistedComponentEomC}
\nabla \sC^{\ga(n)}_\pm \mp h\fud{\ga}{\gc} \sC_\pm^{\gc \ga(n-1)}=0\,.
\end{align}
This can be seen by observing that the above covariant constancy condition precisely coincides with the condition for a $0$-form gauge parameter $\xi(y,\phi)$ to be a Killing tensor:
\begin{equation}
\delta\pomega(y,\phi)=\DO \xi(y,\phi)= \adD \xi(y,\phi) \equiv 0\,.
\end{equation}
Also, it is obvious that the above component equation \eqref{eq:twistedComponentEomC} does \emph{not}, unlike its physical counterpart \eqref{scalarunfld}, mix different components of $\sC$. 
It is not clear what the physical interpretation of such Killing tensors is and their role within the Gaberdiel--Gopakumar conjecture is unclear. They generically mix with dynamical fields at the interacting level, as we explore in Section \ref{sec:presentation}. Let us note that a non-vanishing value for $\nu$, as defined in Subsection~\ref{sec:HSalgebra}, would lead to the following vacuum value for the twisted zero-form \cite{Prokushkin:1998bq}:
\begin{equation}
\label{eq:nuVEV}
\sC = \nu \,.	
\end{equation}
We will discuss this point in more detail in Section~\ref{sec:presentation}.

\item[A twisted one-form] called $\somega$. One could think of it as the gauge field associated with $\sC$. In this case the corresponding equations look like those for $\pC$, but imposed on one-forms. Moreover, just as for $\sC$, it is not clear what the physical interpretation of this set of fields is~---~they are definitely not related to Fronsdal fields. In particular, their role within the Gaberdiel--Gopakumar duality is unclear.

\end{description}


\subsection{Non-linear Equations of Motion}
\label{sec:unfolded}
In the last subsection we have restricted our attention to the free theory. In this section we will discuss the non-linear equations of motion. By expanding these equations of motion around an $\textrm{AdS}_3$-background and considering linear fluctuations one recovers the free equations of $\flomega$ and $\flC$ discussed in the last subsection. Let us denote the fields of the non-linear theory by $\W$ and $\B$, whose linear order fluctuations are then the fields $\flomega$ and $\flC$. Interactions for these fields can a priori arise from allowing for the most general nonlinearities on the right-hand side of their equations of motion, that is \cite{Vasiliev:1988xc, Vasiliev:1988sa},
\besubeqs
\label{xpaceseqX}
\begin{align}
\d \W&=F^{\W}(\W,\B)\label{xpaceseqXA}\,,\\
\d\B&=F^{\B}(\W, \B)\,.\label{xpaceseqXB}
\end{align}
\esubeqs
Equations of motion of this form are said to be \emph{unfolded} and are further constrained by Frobenius integrability, i.e. by consistency with $\d^2\equiv0$. The structure functions $F^{\W}(\W,\B)$  and $F^{\B}(\W,\B)$ are assumed to be expandable in $\B$:
\besubeqs
\label{xpaceseq}
\begin{align}
F^{\W}(\W,\B)&=\mathcal{V}(\W,\W)+\mathcal{V}(\W,\W,\B)+
\mathcal{V}(\W,\W,\B,\B)+\dots\label{xpaceseqA}\,,\\
F^{\B}(\W, \B)&=\mathcal{V}(\W,\B)+\mathcal{V}(\W,\B,\B)+\mathcal{V}(\W,\B,\B,\B)+\dots\,,\label{xpaceseqB}
\end{align}
\esubeqs
where our notation is that the functions $\mathcal{V}$ are linear in each argument. The first interaction vertices are given explicitly by the higher-spin algebra: 
\begin{align*}
\mathcal{V}(\W,\W)&=\W \, \wedge \star  \, \W\,,
&\mathcal{V}(\W, \B)&=\W \star \B-\B\star \W\,.
\end{align*}
We shall also refer to the vertices $\mathcal{V}$ as cocycles,\footnote{Due to the integrability condition the vertices $\mathcal{V}$ can be also interpreted as Chevalley--Eilenberg cocycles with value in infinite-dimensional modules that $\W$ and $\B$ take values in \cite{Vasiliev:2007yc,Sullivan77}. Since these modules are infinite-dimensional it is difficult to say anything directly. A prescription to write a solution for the structure functions is given by Vasiliev equations, which can be thought of as a tool to generate the required interaction terms.} and they can be extracted from the Prokushkin--Vasiliev equations as we will detail in Section~\ref{sec:Veq}. Notice that the deviation of $\W$ from a flat connection is proportional to $\B$.

As a consequence of Frobenius integrability, the equations enjoy a gauge symmetry with a gauge parameter $\flxi$:
\besubeqs
\label{xpaceseqGS}
\begin{align}
&\delta \W=\d\flxi+\flxi\frac{\delta}{\delta \W}F^{\W}(\W,\B)=\d\flxi-[\W,\flxi]_\star+\mathcal{O}(\B)\label{xpaceseqGSA}\,,\\
&\delta \B=\flxi\frac{\delta}{\delta \B}F^{\B}(\W, \B)=\flxi \star \B-\B \star \flxi+\mathcal{O}(\B^2)\,.\label{xpaceseqGSB}
\end{align}
\esubeqs
We stress that the deformation of the gauge symmetry is governed directly by the higher-spin algebra to the lowest order only. The fully non-linear gauge symmetry algebra is a deformation of the higher-spin algebra in the form of an open algebra with structure 'constants' that depend on the fields themselves (algebroid).

The simplest background solution for these non-linear equations is provided by a flat connection $\Omega$ of the higher-spin algebra at vanishing matter field $\B=0$. We take $\Omega$ to be the $\textrm{AdS}_3$ flat connection of \eqref{AdSConnection} and then expand up to the second order:
\begin{align}
\W = \Omega + \flomega + \flomega^{(2)} + \dots\,,  \qquad \B = \flC + \flC^{(2)} + \dots\,. 
\end{align}

\paragraph{Linear Fluctuations:}

For the first-order perturbations $\flomega$ and $\flC$ one finds, in general,
\besubeqs
\begin{align}
&\d \flomega=\{\Omega,\flomega\}_\star+\mathcal{V}(\Omega,\Omega,\flC) \qquad \longrightarrow\qquad \DO \flomega=\mathcal{V}(\Omega,\Omega,\flC)\label{xpaceseqEA}\,,\\
&\d\flC=\Omega\star \flC-\flC\star \Omega\label{xpaceseqEB} \qquad \hspace*{29pt}\longrightarrow\qquad\DO \flC=0\,.
\end{align}
\esubeqs
We thus see that $\flomega$ may generically not be a flat connection as it can have a non-vanishing source represented by $\mathcal{V}(\Omega,\Omega,\flC)$. For Prokushkin--Vasiliev Theory on the $\textrm{AdS}_3$-background we will find $\mathcal{V}(\Omega,\Omega,\flC)=0$ (up to a field redefinition of $\flomega$, see Section~\ref{sec:Firstorder}). This statement implies the flatness of higher-spin-connections to the first order and is related to the non-propagating nature of higher-spin fields in three dimensions\footnote{In higher dimensions, and in particular for $D=4$, or on more complicated backgrounds the latter cocycle is non-zero.}. The gauge transformations at linear order are given by
\begin{align}
\delta\flomega=\d \flxi - [\Omega,\flxi]_\star\,, \qquad
\delta \flC =0\,.
\end{align}
Splitting the fields into twisted and physical components as in \eqref{eq:splittingTwistedPhys} we obtain the following equations of motion:
\begin{align}
\tadD\somega &=0\,,\qquad  \adD \sC=0 \,, \\
\adD\pomega&=0\,, \qquad\tadD \pC =0\,, \label{eq:linearPhsEOM}
\end{align}
which are the free equations of motion discussed in the previous subsection.

\paragraph{Second-Order Fluctuations:}

The second-order perturbations $\flomega^{(2)}$ and $\flC^{(2)}$, which are our main concern in this paper, obey
a system of equations which contain source terms a priori involving first- and second-order fields:
\besubeqs\label{xpaceseqBQ}
\begin{align}
&\DO \flomega^{(2)}=\flomega \, \wedge \star \, \flomega+\mathcal{V}(\Omega,\Omega,\flC^{(2)})+\mathcal{V}(\Omega,\flomega,\flC)+\mathcal{V}(\Omega,\Omega,\flC,\flC)\,,\label{xpaceseqBA}\\
&\DO \flC^{(2)}=[\flomega,\flC]_\star+\mathcal{V}(\Omega,\flC,\flC)\,.\label{xpaceseqBB}
\end{align}
\esubeqs
In Prokushkin--Vasiliev Theory we can remove $\mathcal{V}(\Omega,\Omega,\flC^{(2)})$ by a field redefinition of $\flomega^{(2)}$, so that the sources on the above right-hand sides depend on the first-order fields only. The gauge transformation of the second-order fields are then given by
\begin{align}
\delta\flomega^{(2)}=D_\Omega \flxi^{(2)} - [\flomega,\flxi]_\star\, + \flxi \frac{ \mathcal{V}(\Omega,\flomega,\flC) }{\delta \flomega}, \qquad
\delta \flC^{(2)} =[\flxi,\flC]_\star\,.
\end{align}
Again, we can split these equations into physical and twisted components. The linear-order equations of motion allow us to consistently set all first-order twisted fields to zero, and doing so we obtain the following set of equations:
\besubeqs\label{xpaceseqBQX}
\begin{align}
&\adD \pomega^{(2)}=\pomega \, \wedge \star \, \pomega +\mathcal{V}(\Omega,\Omega,\pC,\pC) \,, \qquad ( \tadD \pC^{(2)} ) \psi=[\pomega,\pC \psi]_\star \,, \label{eq:secondPhsEOM} \\
& ( \tadD \somega^{(2)}) \psi=\tilde{\mathcal{V}}(\Omega,\pomega,\pC) \,, \qquad \hspace*{68.5pt}\adD \sC^{(2)}= \tilde{\mathcal{V}}(\Omega,\pC,\pC) \,. \label{eq:secondTwEOM}
\end{align}
\esubeqs
Let us stress again that the cocycles depend linearly on all their arguments and that their $\psi$-dependence is also linear. We will study \eqref{xpaceseqBQX} extensively in Section~\ref{sec:presentation}. The cocycle $\mathcal{V}(\Omega,\Omega,\pC,\pC)$, which is bilinear in the scalar fields, yields the matter backreaction that sources the Fronsdal equation to the second order, i.e. it encodes the generalized stress-tensors. We will analyze this term in Section~\ref{sec:PresentationBackreactionFronsdal}. 

Note that by \eqref{eq:secondTwEOM} we can not set the second-order twisted fields to zero consistently. But by performing field redefinitions one might be able to remove the source terms appearing in its equations of motion and afterwards set these fields to zero. We will indeed show that this is possible for Prokushkin--Vasliev Theory. We will discuss field redefinitions in the next section.


\subsection{Field Redefinitions and Cohomologies} 
\label{sec:cohomology}
It is natural to ask whether we can remove terms from the above equations of motion by a field redefinition. As an example let us consider the cocycle $\formJ=\mathcal{V}(\Omega,\Omega,\pC,\pC)$, which is part of the equation of motion for the gauge-connection \eqref{eq:secondPhsEOM}, 
\begin{align}
\label{eq:pomegaEOM}
&\adD \pomega^{(2)} = \pomega \, \wedge \star \, \pomega + \formJ \,.
\end{align}
We can perform a field redefinition of the type
\begin{align}\label{examplegeneral}
\pomega^{(2)}&\rightarrow \pomega^{(2)} +F(\Omega, \pC,\pC)\,,
\end{align}
where $F$ is linear in every argument. Field redefinitions quadratic in $\pC$, such as $F$, contain generically terms of the form
\begin{equation}
\sum_{n,m,l=0}^\infty f_{n,m,l} \, \pC_{\ga(n) \nu(l)} \, \pC_{\ga(m)}{}^{\nu(l)} \, y^{\ga(n+m)} \,, \label{eq:pseudolocalfieldredefs}
\end{equation}
where one has to appropriately contract with $H^{\ga \ga}$ and $h^{\ga \ga}$ for redefinitions of form-degree 1 and 2 respectively.
Following standard (but unfortunate) terminology we will refer to such field redefinitions as \emph{pseudo-local}. By \eqref{eq:compOfPhysC} a pseudo-local field redefinition generically contains an infinite number of derivatives of the physical scalar field for each spin, e.g. $2s=m+n$ for zero-forms. 

If the cocycle $\formJ$ is exact, i.e. $\formJ=\adD F(\Omega,\pC,\pC)$, then it can evidently be removed by a pseudo-local field redefinition $\pomega^{(2)}\rightarrow \pomega^{(2)} +F(\Omega,\pC,\pC)$. On the other hand, the consistency of \eqref{eq:pomegaEOM} with $\adD^2=0$ leads to
\begin{equation}
 \adD \formJ + \adD \pomega \, \wedge \star \, \pomega  - \pomega \, \wedge \star \, \adD \pomega =  0 \,.
\end{equation}
Upon using the first order equation of motion $\adD \pomega=0$ this implies that the current $\formJ$ is also closed, i.e. $D \formJ=0$. Therefore in order to make sure that the cocycle $\formJ$ cannot be removed by a pseudo-local field redefinition we have to check whether it is an element of $\mathbb{H}^2(\adD,\pC \pC)$, the cohomology of the nilpotent operator $\adD$ with respect to pseudo-local field redefinitions of form-degree $1$ which are quadratic in $\pC$. 

This discussion generalizes to the other non-vanishing cocycles in \eqref{xpaceseqBQX} by considering cohomologies for the covariant derivatives $\adD$ and $\tadD$ with respect to field redefinitions that are linear in $\pC$, linear in both $\pC$ and $\pomega$, or quadratic in $\pC$. The notation for the corresponding cohomologies changes in the obvious way. In Appendix~\ref{app:cohomologies} we have analyzed various cohomologies, and the most relevant results of this analysis are summarized in Table~\ref{tab:cohomologies2bulk}. Hereafter we briefly discuss various implications of these results.

\begin{table}[h]
\begin{center}
\begin{tabular}{c|c|c|c}
 degree n & $\mathbb{H}^n(\adD,\pC \pC)$  & $\mathbb{H}^n(\tadD,\pC)$  & $\mathbb{H}^n(\tadD,\pomega\pC)$  \\ \hline
1 & non-empty & non-empty & non-empty \\
2 & empty & empty & non-empty
\end{tabular}
\caption{Cohomologies of various form-degrees $n$ and classes of field redefinitions.\label{tab:cohomologies2bulk}}
\end{center}
\end{table}

\begin{description}
\item[Form-degree-$2$ cohomology $\mathbb{H}^2(\adD,\pC \pC)$] for field redefinitions quadratic in $\pC$: This cohomology is trivial and therefore any $\formJ$ on the right-hand side of \eqref{eq:pomegaEOM} can be removed by a pseudo-local field redefinition of the type \eqref{eq:pseudolocalfieldredefs}. Thus any backreaction of the scalar fields on the higher-spin fields, including the spinorial counterpart of the canonical $s$-derivative current \eqref{standardcurrent}, can be removed by a pseudo-local field redefinition which generically contains an arbitrary number of derivatives of the scalar field for each spin.\footnote{The fact that the canonical $s$-derivative current can be removed by a pseudo-local field redefinition was first shown in \cite{Prokushkin:1999xq} and led to the development of an integration flow\cite{Prokushkin:1998vn}, which maps all physical and twisted fields in a field frame in which they obey the free equations of motion.} Such redefinitions should not correspond to physically allowed ones. A possible interpretation for the fact that an arbitrary backreaction can be removed is that the class of pseudo-local field redefinitions \eqref{examplegeneral} is too broad. Unfortunately a criterion which restricts the class of field redefinitions to the physically allowed ones is not yet known.\footnote{A conjecture regarding this point was put forward in \cite{Vasiliev:2015wma}.} More comments on this important issue can be found in the conclusions to this paper. 

\item[Form-degree-$1$ cohomology $\mathbb{H}^1(\adD,\pC \pC)$] for field redefinitions quadratic in $\pC$: The non-emptiness of this cohomology\cite{Prokushkin:1999xq} allows for sources to the twisted zero-form's equations of motion, 
\begin{equation}
\adD \sC^{(2)}= \tilde{\mathcal{V}}(\Omega,\pC,\pC)\,,
\end{equation}
that cannot be removed by a pseudo-local field redefinition
\begin{equation}
\sC^{(2)} \rightarrow \sC^{(2)} + G(\pC,\pC)\,,
\end{equation}
which would imply that we cannot consistently choose $\pC^{(2)}\equiv 0$.
Beyond the second order the twisted zero-form $\sC^{(2)}$ would therefore generically produce source terms to the physical \eomsWS, i.e. higher-order analogs of \eqref{eq:secondPhsEOM}. We will discuss this in more detail in Section~\ref{sec:presentation}.

\item[Form-degree-$2$ cohomology $\mathbb{H}^2(\adD,\pomega \pC)$] for field redefinitions linear in both $\pomega$ and $\pC$: The equations of motion for the twisted gauge fields to the second order are given by
\begin{equation}
(\tadD \somega^{(2)})\psi=\tilde{\mathcal{V}}(\Omega,\pomega,\pC) \,.
\end{equation}
As in the previous case the non-triviality of this cohomology therefore might prevent us from setting the twisted field $\somega^{(2)}$ to zero consistently. We will return to this point in Section~\ref{sec:presentation}. 
\end{description}

We note that Table \ref{tab:cohomologies2bulk} also lists the results for $\mathbb{H}^{1,2}(\tadD,\pC)$ and $\mathbb{H}^{1}(\tadD,\omega \pC)$ for later reference. 
\subsection{Conservation}
\label{sec:conservation}

As explained in the last subsection the cocycles $\mathcal{V}(\Omega,\Omega,\pC,\pC)$ and $\mathcal{V}(\Omega,\pC,\pC)$ of \eqref{xpaceseqBQ} have to be closed or, differently put, conserved with respect to the covariant derivative $\adD$. This provides an important consistency requirement to cross-check the validity of our calculations. Let us consider $q$-forms, which are bilinears in the free fields $\pC=\pC(y,\phi|x)$:
\begin{equation}
\formJ_q=\formJ_{\mu_1..\mu_q}(y,\phi|x) \, \d x^{\mu_1}\wedge...\wedge \d x^{\mu_q} \,, \qquad \textrm{with} \; q=0,1,2,3 \,. 
\end{equation}
Obviously the cocycles $\mathcal{V}(\Omega,\Omega,\pC,\pC)$ and $\mathcal{V}(\Omega,\pC,\pC)$ correspond to $q=2$ and $q=1$ respectively.
The operator $D$ defines a complex on $q$-forms $\formJ_q$:
\begin{align}
0\longrightarrow \formJ_0 \longrightarrow \formJ_1 \longrightarrow \formJ_2 \longrightarrow \formJ_3 \longrightarrow 0\,,
\end{align}
In the following we will mostly work with Fourier-transformed fields,
\begin{align}
\label{eq:fourier}
\pC(y,\phi)=\int \d\xi \, e^{iy\xi} \, \pC(\xi,\phi|x)\,,
\end{align}
which leads to the Fourier-transformed expressions for $\formJ_q$
\begin{align}
\label{eq:fourierDecomp}
\formJ_q&= \int \d\xi\, \d\eta\, \formK_q(\xi,\eta,y)\, \pC(\xi,\phi|x) \pC(\eta,-\phi|x)\,,
\end{align}
The sign flip in $\phi$ for one of the zero-forms $\pC$ is due to the fact that $\pC$ in the splitting \eqref{eq:splittingTwistedPhys} is associated with the $\psi$-dependent term. The kernel $\formK_q$ is given for the various form-degrees by
\begin{align}
\formK_0&=K(\xi,\eta,y)\,, \qquad 
\formK_1=h^{\ga\ga} K_{\ga\ga}(\xi,\eta,y)\,, \qquad 
\formK_2=H^{\ga\ga} J_{\ga\ga}(\xi,\eta,y)\,, \qquad 
\formK_3=H J(\xi,\eta,y)\,. \notag
\end{align}
Notice that boldfaced $\formK_q$ denote forms whereas non-boldfaced ones such as $K$ denote components. We have used the definitions
\begin{align}
\label{eq:H}
H^{\ga \ga} \equiv h^{\ga}{}_\sigma \wedge h^{\ga \sigma}\,, \qquad H\equiv H_{\ga\ga}\wedge h^{\ga\ga}\,,
\end{align}
which obey the following identities:
\besubeqs
\label{eqhident}
\begin{align}
h^{\ga\gb}\wedge h^{\gc\gd}=\frac14 \epsilon^{\ga\gc} H^{\gb\gd}+\mbox{3 more}\,, \label{eq:twoFormIdent} \\
H^{\ga\gb}\wedge h^{\gc\gd}=\frac16 (\epsilon^{\ga\gc}\epsilon^{\gb\gd}+\epsilon^{\gb\gc}\epsilon^{\ga\gd}) H \label{eq:threeFormIdent}\,. 
\end{align}
\esubeqs
Now with the help of the equations of motion for the Fourier-transformed fields,
\besubeqs
\begin{align}
\nabla \pC(\xi,+\phi|x)&=-\frac{i}2\phi h^{\ga\ga}\left(\xi_\ga\xi_\ga -\pl^\xi_\ga \pl^\xi_\ga \right)\pC(\xi,+\phi|x) \,, \\
\nabla \pC(\eta,-\phi|x)&=+\frac{i}2\phi h^{\ga\ga}\left(\eta_\ga\eta_\ga -\pl^\eta_\ga \pl^\eta_\ga \right)\pC(\eta,-\phi|x) \,,
\end{align}
\esubeqs
and of the identities \eqref{eqhident} we find the following Fourier representations for $D$:
\besubeqs
\begin{align}
\adD K(\xi,\eta,y)&=h^{\ga\ga} O_{\ga\ga}K(\xi,\eta,y)\,, \label{eq:derFourierZeroForm} \\
\adD h^{\ga\ga}K_{\ga\ga}(\xi,\eta,y)&= \frac14 H^{\ga\ga} O_{\ga\nu} K\fdu{\ga}{\nu}(\xi,\eta,y)\,, \\
\adD H^{\ga\ga} J_{\ga\ga}(\xi,\eta,y)&= \frac16 H O^{\ga\ga} J_{\ga\ga}(\xi,\eta,y) \label{eq:derFourierTwoForm}\,,
\end{align}
\esubeqs
where we have defined
\begin{align}
\label{eq:inspector}
O_{\ga\ga}&\equiv \frac{i}{2} \left[\left(\eta_\ga\eta_\ga -\pl^\eta_\ga \pl^\eta_\ga \right)-\left(\xi_\ga\xi_\ga -\pl^\xi_\ga \pl^\xi_\ga \right)+2iy^{\vphantom{y}}_\ga\pl^y_\ga\right]\,.
\end{align}
Similarly the cocycles $\tilde{\mathcal{V}}(\Omega,\pomega,\pC)$ in \eqref{eq:secondTwEOM} need to be conserved with respect to the twisted-adjoint covariant derivative $\tadD$. Let us therefore also consider $p$-forms which are linear in $\pC$ and $\pomega$:
\begin{align}
\formJ_q&= \int \d\xi\, \d\eta\, \left\{ \formL_q(\xi,\eta,y)\, \pC(\xi,\phi|x) \, \pomega(\eta,-\phi|x) + \bar{\formL}_q(\xi,\eta,y)\, \pomega(\xi,\phi|x) \, \pC(\eta,\phi|x) \right\} \,,
\end{align}
where $\formL_q$ and $\bar{\formL}_q$ are given by
\besubeqs
\begin{align}
\formL_1&=L(\xi,\eta,y)\,, && 
\formL_2=h^{\ga\ga} L_{\ga\ga}(\xi,\eta,y)\,, &&
\formL_3=H^{\ga\ga} S_{\ga\ga}(\xi,\eta,y)\,, \\
\bar{\formL}_1&=\bar{L}(\xi,\eta,y)\,, &&
\bar{\formL}_2=h^{\ga\ga} \bar{L}_{\ga\ga}(\xi,\eta,y)\,, && 
\bar{\formL}_3=H^{\ga\ga} \bar{S}_{\ga\ga}(\xi,\eta,y)\,, &&
\end{align}
\esubeqs
Using the equations of motion for $\pomega$ and $\pC$ we again obtain a Fourier representation for $\tadD$:
\begin{align}
\tadD L(\xi,\eta,y) &=  h^{\ga\ga} I_{\ga\ga}L(\xi,\eta,y) \,, \\
\tadD h^{\ga\ga}L_{\ga\ga}(\xi,\eta,y)&= \frac14 H^{\ga\ga} I_{\ga\nu} L\fdu{\ga}{\nu}(\xi,\eta,y)\,, \label{eq:barredKernels}
\end{align}
where we have defined 
\begin{equation}
\label{eq:tadInspector}
I_{\ga \ga} \equiv \frac{i}{2}\left[\left(y_\ga y_\ga -\pl^y_\ga \pl^y_\ga \right)-\left(\xi_\ga \xi_\ga -\pl^\xi_\ga \pl^\xi_\ga \right)+2i \eta^{\vphantom{\eta}}_\ga \pl^\eta_\ga\right]\,.
\end{equation}
Analogous expressions hold for the barred kernels \eqref{eq:barredKernels} upon replacing $I_{\ga \ga}$ with $\bar{I}_{\ga \ga}$ defined as 
\begin{equation}
\label{eq:barredtadDInspector}
\bar{I}_{\ga \ga} \equiv \frac{i}{2}\left[\left(y_\ga y_\ga -\pl^y_\ga \pl^y_\ga \right)-\left(\eta_\ga \eta_\ga -\pl^\eta_\ga \pl^\eta_\ga \right)+2i \xi^{\vphantom{\xi}}_\ga\pl^\xi_\ga\right]\,.
\end{equation}
As will be discussed in Section~\ref{sec:presentation} we checked conservation for all cocycles studied in this paper.

\section{Presentation of Results: Second-Order Backreactions}
\label{sec:presentation}

In this section we will discuss our main results obtained by studying the equations of motion \eqref{xpaceseqBQX}. We will postpone a detailed explanation of how we extracted the various cocycles from Prokushkin--Vasiliev Theory to Section~\ref{sec:Veq}. As explained in Section~\ref{sec:general}, the Prokushkin--Vasiliev Theory at hand contains two (real) physical scalars, encoded in the field $\pC$, as well as one physical (although non-propagating) higher-spin gauge field for every spin, encoded in the connection $\pomega$. In addition, the theory also contains a twisted sector, represented by a twisted zero-form $\sC$, and a twisted gauge connection $\somega$. Let us first focus on the twisted sector in Subsection~\ref{subsec:twistedresults} before discussing the second order analysis of the physical sector in Subsection~\ref{sec:PresentationBackreactionFronsdal}.

\subsection{Twisted Sector Results}
\label{subsec:twistedresults}
In the following we will discuss whether we can find solutions of our theory for which all twisted fields vanish. We are interested in such consistent truncations because the role of the twisted fields within the AdS/CFT-duality and their field-theoretical interpretation is unclear~---~as was discussed in Section~\ref{sec:precursors}. Therefore a trivial solution for these fields seems to be the most natural choice. 

We will first discuss the twisted sector at linear order. We will see that we need to perform a field redefinition in order to set the first-order twisted fields to zero consistently. This field redefinition is not unique and will lead to the appearance of free parameters in the second-order equations of motion, which can in turn be fixed by going to a field frame for which backreactions to the second-order twisted zero-form $\sC^{(2)}$ and to the twisted gauge connection $\somega^{(2)}$ can be removed by a pseudo-local field redefinition.

This process will involve pseudo-local field redefinitions and therefore it is by no means guaranteed that the theory after the redefinitions is equivalent to the theory before because of the non-localities involved in this step.

As we will show in the following there is only one choice for the free parameters which allows for trivial solutions of the twisted sector. This suggests a relation to the integration flow procedure\cite{Prokushkin:1998bq}~---~as we will discuss at the end of Section~\ref{sec:twistedSecondOrder}.

\subsubsection{Linear Order}
\label{sec:twistedResultsLinearOrder}
We mentioned in Section~\ref{sec:unfolded} that the cocycle $\mathcal{V}(\Omega,\Omega,\flC)$ in the linear equations of motion \eqref{xpaceseqEA} for the connection $\flomega$ vanishes only up to a field redefinition. We will explain this field redefinition in more detail now. 

As will be discussed in Section~\ref{sec:Veq} analyzing Prokushkin--Vasiliev Theory leads to the following equations of motion for the twisted sector \cite{Vasiliev:1992gr}:
\besubeqs
\label{eq:firstOrderSource}
\begin{align}
(\tadD\somega)\psi&=\mathcal{V}(\Omega,\Omega,\pC)= \tfrac18 H^{\ga\ga} (y_\ga+i\pl^w_\ga)(y_\ga+i\pl^w_\ga)\pC(w,\phi|x) \psi|_{w=0}\,, \label{eq:firstOrderSourceW} \\
\adD \sC &= 0 \,,
\end{align}
\esubeqs
where $H^{\ga \ga}$ was defined in \eqref{eq:H}. Notice that there is a source term to $\somega$ linear in the scalar field $\pC$ and therefore we are interested in performing a field redefinition of $\somega$ which removes this source term. After having performed such a field redefinition we can set the linear-order twisted fields to zero consistently. As discussed above this is the truncation of the theory we are interested in. The most general solution of the inhomogeneous differential equation \eqref{eq:firstOrderSourceW} is given by a particular solution thereof together with the general solution of the complementary homogeneous equation.  

As was first shown in \cite{Vasiliev:1992gr}, a particular solution of $\somega$ in \eqref{eq:firstOrderSourceW} is $M_1$ with
\begin{align}
\label{eq:M1}
M_1=\frac{1}{4} \phi h^{\ga\ga} \int_0^1 \d t \; (t^2-1) (y_\ga + it^{-1}\pl^y_\ga)(y_\ga + it^{-1}\pl^y_\ga)\pC (yt,\phi) \, .
\end{align}
Now, let us find the solution $R$ of the complementary homogeneous equation, i.e. $\tadD R = 0$. We are interested in this solution only up to gauge transformations thereof. Therefore we want to identify two solutions $R$ and $R'$ which differ only by a gauge transformation, i.e. $R-R'=\tadD \epsilon$. The most general solution of the homogenous equation~---~up to gauge transformations~---~is therefore an element of the cohomology with respect to the nilpotent operator $\tadD$ and linear functionals in $\pC$ of form-degree $1$, i.e. $R \in \mathbb{H}^1(\tadD,\pC)$.

This cohomology is non-empty as can be seen by comparing with Table \ref{tab:cohomologies2bulk}. Indeed we show in Appendix~\ref{app:cohomologies} that it forms a two-dimensional space with a representative given by\footnote{We have a map from $\pC$ which is a direct sum of two non-isomorphic irreducible modules, i.e. $\pC_B$ and $\pC_F$, to the direct sum of the same modules in which $\somega$ takes values. Therefore the space is two-dimensional, which is a simple instance of Schur's lemma.}
\begin{align}
\label{eq:representativeofH}
R \equiv \frac{1}{4} \phi h^{\ga\ga}\int_0^1 \d t\, (t^2-1) \left\{ g_0 \left(y_\ga y_\ga- t^{-2}\pl^y_\ga\pl^y_\ga\right)\pC_\textsc{b}(ty)+ 2 d_0 t^{-1} y^{\vphantom{y}}_\ga \pl^y_\ga \pC_\textsc{f}(ty) \right\} \, \,.
\end{align}
Here $\pC_\textsc{b}$ and $\pC_\textsc{f}$ are the even and odd parts of $\pC$ with respect to $y_\alpha$ whereas $g_0$ and $d_0$ are parameters accounting for the two-dimensional nature of this cohomology. The above representative has been chosen to look almost exactly like the particular solution $M_1$ with the crucial difference that we had to split $\pC$ into bosonic and fermionic components $\pC_\textsc{b}$ and $\pC_\textsc{f}$. As we discussed in Section~\ref{sec:precursors} we consider the bosonic Prokushkin--Vasiliev Theory, for which the odd components of $\pC$ vanish identically, i.e. $\pC_\textsc{f} \equiv 0$, and therefore the cohomology is only one-dimensional. But in the next section we will also briefly discuss the behavior of Prokushkin--Vasiliev Theory without imposing the bosonic truncation and we therefore kept $\pC_\textsc{f}$ in \eqref{eq:representativeofH} for future reference.

In the case of the bosonic theory the general form of the field redefinition removing the source term of \eqref{eq:firstOrderSource} is therefore given by
\begin{equation}
\label{eq:M1prime}
M'_1 \equiv M_1 + R = \frac14 \phi h^{\ga\ga}\int_0^1 \d t\, (t^2-1) \left( \tilde{g}_0 \, y_\ga y_\ga+2i y_\ga t^{-1} \pl^y_\ga-\tilde{g}_0 \, t^{-2}\pl^y_\ga\pl^y_\ga\right)\pC_\textsc{b}(ty) \,,
\end{equation}
where we defined $\tilde{g}_0=1+g_0$. After performing this field redefinition $\somega (y,\phi|x) \rightarrow \somega(y,\phi|x)  + M'_1$ we can consistently choose trivial solutions for the twisted fields
\besubeqs
\label{eq:solTwLin}
\begin{align}
\somega &= 0 \, , && \sC = 0 \,,
\end{align}
\esubeqs
The parameter $\tilde{g}_0$ will play a key role in the following subsection, where we discuss the second-order equations of motion of the twisted fields. Anticipating the results to be discussed therein, the situation is that the second-order twisted fields can be consistently set to zero only at a particular point in the parameter space of $M'_1$, namely $\tilde{g}_0=0$.

\subsubsection{Second Order}
\label{sec:twistedSecondOrder}
In this section we will discuss the equation of motion \eqref{xpaceseqBQX} for the twisted scalar field to second order before analyzing the corresponding equations for the twisted one-form. In the scalar sector the equation of motion is given by
\begin{equation}
\label{eq:twsC2nd}
\adD \sC^{(2)}= \tilde{\mathcal{V}}(\Omega,\pC,\pC) \,.
\end{equation}
We can analyze each $y_\ga$-component of \eqref{eq:twsC2nd} separately as the adjoint covariant derivative $\adD$ commutes with the $y_\ga$-number operator $y^\nu \partial^y_\nu$. We will try to follow a similar approach as for the linear order and check if we can find a field redefinition which removes the source term on the right-hand side of \eqref{eq:twsC2nd}. This question is of particular interest as by \eqref{eq:nuVEV} the $y_\ga$-independent part of the Killing tensor $\sC$ at zeroth order specifies the $\lambda$-parameter of the $\textrm{hs}(\lambda)$ higher-spin theory.\footnote{The interpretation of the $y_\alpha$-independent component of $\sC$ at second and higher orders is less clear.} If the $y_\ga$-independent component of the source term $\tilde{\mathcal{V}}(\Omega,\pC,\pC)$ cannot be removed by a field redefinition then the identity component of $\sC$ is necessarily deformed at second-order in perturbation theory. Note that \eqref{eq:twsC2nd} arises from \eqref{xpaceseqBB}, which we repeat here for convenience:
\begin{equation}
\DO \flC^{(2)}=[\flomega,\flC]_\star+\mathcal{V}(\Omega,\flC,\flC)\,.
\end{equation}
By $\psi$-counting this reduces in the twisted sector to
\begin{equation}
\label{eq:afterPsiCounting}
\adD \sC^{(2)}=[\somega \psi,\pC \psi]_\star+\tilde{\mathcal{V}}'(\Omega,\pC,\pC) \,,
\end{equation}
where we chose the linear solution $\sC \equiv 0$. Using \eqref{eq:M1prime} we can perform the field redefinition $\somega \rightarrow \somega + M'_1$ and afterward consistently set $\somega \equiv 0$, as discussed in the last subsection. Having done so \eqref{eq:afterPsiCounting} will reduce to \eqref{eq:twsC2nd} but the field redefinition $M'_1$ will lead to an additional contribution to the cocycle $\tilde{\mathcal{V}}(\Omega,\pC,\pC)$ which is then given by
\begin{equation}
\tilde{\mathcal{V}}(\Omega,\pC,\pC) = [M'_1\psi,\pC \psi]_\star +\tilde{\mathcal{V}}'(\Omega,\pC,\pC) \,,
\end{equation}
and will therefore depend on $\tilde{g}_0$ in the bosonic theory. From Prokushkin--Vasiliev Theory one extracts the following explicit form of the source term in \eqref{eq:twsC2nd}:
\begin{align}
\widetilde{\mathcal{V}}(\Omega,\pC,\pC)&=\phi h^{\ga\ga} \int \, \d^2 \xi \d^2 \eta \,  K_{\ga\ga}(\xi,\eta,y) \,\pC (\xi,\phi|x)\pC (\eta,-\phi|x)\,,
\end{align}
where the kernel $K_{\ga \ga}$ is given by
\begin{align}
K_{\ga\ga}(\xi,\eta,y)=\int_0^1 \, \d t \, &\left\{ \frac12\,e^{i(y(1-t)-t\eta)\xi} \, \xi_\ga\left({(1-t^2)}(\xi_\ga-\eta_\ga)+{(1-t)^2} y_\ga\right)\notag \right. \\
&\left. -\frac12\,e^{i(y(1-t)-t\xi)\eta} \, \eta_\ga\left({(1-t^2)}(\eta_\ga+\xi_\ga)-{(1-t)^2} y_\ga\right) \right.\notag\\
&\left.+\frac14 (t^2-1)e^{i(y-\eta)(y+ t\xi)}(\tilde{g}_0 \,(y-\eta)_\ga(y-\eta)_\ga-2 (y-\eta)_\ga \xi_\ga + \tilde{g}_0 \, \xi_\ga \xi_\ga )  \right.\notag\\
&\left.+\frac14 (t^2-1)e^{i(y+\xi)( t\eta-y)}(\tilde{g}_0 \,(y+\xi)_\ga(y+\xi)_\ga-2(y-\xi)_\ga \eta_\ga + \tilde{g}_0 \, \eta_\ga \eta_\ga ) \right\} \notag\,.
\end{align}
In order to check whether there is some field frame in which we can set the $y_\ga$-independent component of $\sC$ to zero, we have to check whether $\widetilde{\mathcal{V}}(\Omega,\pC,\pC)|_{y=0}$ is a non-trivial element in the cohomology $\mathbb{H}^1(\adD,\pC \pC)$.
Performing the integration over $t$ in $K_{\ga \ga}(\xi,\eta,y=0)$ we obtain
\begin{equation}
\label{eq:zeroOrderKaty0}
K_{\ga \ga}(\xi,\eta,y=0) =  f(\eta \xi) \, \left( \, (1+\tilde{g}_0) \, \eta_\ga \eta_\ga - \, (1-\tilde{g}_0) \,  \xi_\ga \xi_\ga \right) \,,
\end{equation}
where we have defined $f(x) = 4  (x \cos(x) - x^{-3} \sin(x) )$.

The Fourier representation of $\adD$ in \eqref{eq:derFourierZeroForm} reads $h^{\ga \ga}O_{\ga \ga}$. The operator $O_{\ga \ga}$ does not mix different powers of $y_\ga$-oscillators. Therefore we can deduce that for $\widetilde{\mathcal{V}}(\Omega,\pC,\pC)|_{y=0}$ to be exact its kernel has to be of the form
\begin{equation}
K_{\ga \ga}(\xi,\eta,y=0) \overset{!}{=} O_{\ga \ga} F(\eta \xi) = \frac{i}{2} \, (\eta_\ga \eta_\ga - \xi_\ga \xi_\ga) \left( F(\eta \xi) + F''(\eta \xi) \right) \,,
\end{equation}
where $F(x)$ is an arbitrary function. By \eqref{eq:zeroOrderKaty0} this is only the case if 
\begin{equation}
\label{eq:sweetspot}
 \boxed{\tilde{g}_0=0}
\end{equation}
and one can easily check that there exists a solution for $F(\eta \xi)$ at this point in parameter space which is given in Appendix~\ref{app:susyTwistedDecoupling}. Therefore we can consistently set the $y_\ga$-independent component of $\sC$ to zero only for this choice of $\tilde{g}_0$.

At this stage one might wonder what will happen if we also consider fermionic excitations. In this case the cohomology $\mathbb{H}^1(\tadD,\pC)$ is two-dimensional and the field redefinition\footnote{It can be shown that including fermionic fields does not change the form of $M_1$.} $\tilde{M}_1$ 
\begin{equation}
\begin{aligned}
\label{eq:M1PrimeFerm}
\tilde{M}_1 \equiv M_1 + R = \frac14 \phi h^{\ga\ga}\int_0^1 \d t\,  (t^2-1) \big\{  ( \tilde{g}_0 \, y_\ga y_\ga &+2i y_\ga t^{-1} \pl^y_\ga-\tilde{g}_0 \, t^{-2}\pl^y_\ga\pl^y_\ga)\pC_\textsc{b}(ty)   \\
&+ ( y_\ga y_\ga+2i \, \tilde{d}_0 \, y_\ga t^{-1} \pl^y_\ga- t^{-2}\pl^y_\ga\pl^y_\ga)\pC_\textsc{f}(ty)  \big\}\,,
\end{aligned}
\end{equation}
therefore contains an additional free parameter $\tilde{d}_0=1+d_0$. As shown in Appendix~\ref{app:susyTwistedDecoupling} by performing an analogous analysis as for the bosonic theory the $y_\ga$-independent component of $\sC$ can consistently be set to zero only for the choice 
\begin{equation}
\label{eq:susyParameterValues}
 \boxed{\tilde{g}_0=\tilde{d}_0=0 
}\,.
\end{equation}
In fact we also show in Appendix~\ref{app:susyTwistedDecoupling} that at this point in parameter space all $y_\ga$-components of the cocycle $\widetilde{\mathcal{V}}(\Omega,\pC,\pC)$ are exact and can thus be removed by a pseudo-local field redefinition. Therefore the two parameter ambiguity introduced at the linear level by removing the source term in \eqref{eq:twsC2nd} with a field redefinition $\tilde{M}_1$ is uniquely fixed by choosing a field frame in which we can consistently set $\sC^{(2)} \equiv 0$. 

Having fixed this ambiguity by \eqref{eq:susyParameterValues} we will now analyze the twisted gauge sector. We will also consider fermionic excitations. The equations of motion for the twisted gauge fields to second order were given in \eqref{eq:secondTwEOM} and read
\begin{equation}
\label{eq:somega2AfterRedef}
(\tadD \somega^{(2)})\psi =\tilde{\mathcal{V}}(\Omega,\pomega,\pC) \,.
\end{equation}
However as we stressed around \eqref{eq:secondTwEOM} this equation only holds after a redefinition of $\somega^{(2)}$. We will discuss this field redefinition in more detail now. The equation \eqref{eq:somega2AfterRedef} can be derived by considering the $\psi$-dependent part of \eqref{xpaceseqBA} and using the fact that we set $\sC\equiv 0$, which leads to
\begin{equation}
\label{eq:somega2BeforeRedef}
(\tadD \somega^{(2)})\psi=\tilde{\mathcal{V}}'(\Omega,\pomega,\pC) + \tilde{\mathcal{V}}(\Omega,\Omega,\pC^{(2)})\,.
\end{equation}
From Prokushkin--Vasiliev Theory one obtains the following expression for the second source term:
\begin{equation}
\tilde{\mathcal{V}}(\Omega,\Omega,\pC^{(2)})=\tfrac18 H^{\ga\ga} (y_\ga+i\pl^w_\ga)(y_\ga+i\pl^w_\ga)\pC^{(2)}(w,\phi|x) \psi|_{w=0}
\end{equation}
It is therefore of the same form as the corresponding source term \eqref{eq:firstOrderSourceW} at linear order. Performing a field redefinition, $\somega^{(2)} \rightarrow \somega^{(2)} + \tilde{M}^{(2)}_1$ with  
\begin{equation}
\begin{aligned}
\label{eq:M1PrimeFermSecondOrder}
\tilde{M}^{(2)}_1  = \frac14 \phi h^{\ga\ga}\int_0^1 \d t\,  (t^2-1) \big\{  ( \tilde{g}_1 \, y_\ga y_\ga &+2i y_\ga t^{-1} \pl^y_\ga-\tilde{g}_1 \, t^{-2}\pl^y_\ga\pl^y_\ga)\pC^{(2)}_\textsc{b}(ty)   \\
&+ ( y_\ga y_\ga+2i \, \tilde{d}_1 \, y_\ga t^{-1} \pl^y_\ga- t^{-2}\pl^y_\ga\pl^y_\ga)\pC^{(2)}_\textsc{f}(ty)  \big\}\,,
\end{aligned}
\end{equation}
removes the source term $\tilde{\mathcal{V}}(\Omega,\Omega,\pC^{(2)})$ in \eqref{eq:somega2BeforeRedef}. This can be shown as for the linear case but now this field redefinition, apart from removing the source term $\tilde{\mathcal{V}}(\Omega,\Omega,\pC^{(2)})$, also leads to an additional contribution to $\tilde{\mathcal{V}}'(\Omega,\pomega,\pC)$ due to the fact that the equation of motion for $\pC^{(2)}$ is given by $(\tadD \pC^{(2)})\psi=[\pomega,\pC \psi]_\star$ as opposed to the linear case $\tadD \pC=0$.
Therefore after performing this field redefinition we obtain \eqref{eq:somega2AfterRedef} with its source term $\tilde{\mathcal{V}}(\Omega,\pomega,\pC)$ now depending on the parameters $\tilde{d}_1$ and $\tilde{g}_1$.
We will show in Appendix \ref{app:decouplingTwistedGaugeSector} that only for the choice
\begin{equation}
\tilde{g}_0=\tilde{d}_0=\tilde{g}_1=\tilde{d}_1 \,,
\end{equation}
the source term $\tilde{\mathcal{V}}(\Omega,\pomega,\pC)$ in \eqref{eq:somega2AfterRedef} is exact and can therefore be removed by a pseudo-local field redefinition.

Summarizing, we have shown that \emph{only} for the parameter choice 
\begin{equation}
\boxed{\tilde{g}_0=\tilde{d}_0=\tilde{g}_1=\tilde{d}_1=0} \,,
\end{equation}
there exists a field frame in which we can consistently set all second-order twisted fields to zero:
\begin{align}
\sC^{(2)} = 0 \,, && \somega^{(2)} = 0 \,.
\end{align}
However it is important to stress that it is not at all obvious whether the theory in this field frame is equivalent to the theory before the field redefinitions because of the non-localities involved in this step. Furthermore, it is shown in Appendix~\ref{app:cohomologies} that the cohomologies $\mathbb{H}^0(\adD,\pC\pC)$ and $\mathbb{H}^1(\tadD,\pomega\pC)$ are infinite-dimensional and therefore one would generically expect an infinite number of free parameters to enter the third-order equations of motion due to the redefinitions of $\sC^{(2)}$ and $\pomega^{(2)}$ at second order. However these ambiguities do not enter the second-order equations of motion.\footnote{The cohomologies are infinite dimensional with respect to the $\textrm{AdS}_3$-isometry algebra. However the fact that the tensor product of various $\pC$ fields are irreducible higher-spin algebra modules (up to permutations)\cite{Beisert:2004di} makes the cohomology one-dimensional with respect to the higher-spin algebra. In other words higher-spin symmetry relates various irreducible $\textrm{AdS}_3$-isometry algebra components.} 

Interestingly this can be compared to the integration flow formalism pioneered in \cite{Prokushkin:1998bq}. The integration flow by construction maps all physical and twisted fields in a field frame in which they obey the free equations of motion. This is achieved by a pseudo-local B\"acklund--Nicolai-type mapping\cite{Nicolai:1980jc}. In this formalism one can therefore consistently choose a solution with vanishing twisted fields. The fact that there is only one point in parameter space which allows for a trivial twisted sector suggests that this point should correspond to the integration flow solution of the twisted sector. However integration flow also leads to free equations of motion of the physical fields and therefore corresponds to a different field frame for the physical sector. We will discuss possible interpretations of this observation in the conclusion to this paper.

\subsection{Backreaction on the Fronsdal Sector} 
\label{sec:PresentationBackreactionFronsdal}
In this subsection we will analyze the implications of the cocycle $\formJ=\mathcal{V}(\Omega,\Omega,\pC,\pC)$ of \eqref{eq:secondPhsEOM} and its relation to corrections of the Fronsdal equation \eqref{FronsdalEq} due to the presence of scalar fields. From the metric-like formulation of the theory one expects these corrections to be of the form \eqref{standardcurrent}, which upon combining all spins into a generating functional expressed in terms of $\pC$ leads to\footnote{This correspondence only holds up to improvement terms to make the metric-like current \eqref{standardcurrent} traceless on-shell as \eqref{eq:canonicalCurrentBD} is on-shell traceless. }
\begin{equation}
\label{eq:canonicalCurrentBD}
\sum_s \frac{1}{(2s)!} \, j_{\ga(2s)} \, y^{\ga(2s)}=\pC(y,\phi) \, \pC(y,-\phi) \,,
\end{equation}
corresponding to the two-form
\begin{equation}
\label{eq:canonicalCur}
\formJ^{can} = H^{\ga \ga} \partial_\ga^y \partial^y_\ga \, \pC(y,\phi) \, \pC(y,-\phi) \,.
\end{equation}
The cocycle $\formJ=\mathcal{V}(\Omega,\Omega,\pC,\pC)$ should be related to the canonical current $\formJ^{can}$, \eqref{eq:canonicalCur}, by a pseudo-local field redefinition. But as we discussed in Section~\ref{sec:cohomology} the cocycle $\formJ$ is exact and therefore can be completely removed by a pseudo-local field redefinition. However the physically allowed class of field redefinitions should allow us to relate the current $\formJ$ to the canonical current $\formJ^{can}$, but should not allow for a field redefinition which also removes the canonical current $\formJ^{can}$. This suggests that the class of pseudo-local field redefinitions is too broad. 

In the following we will calculate the cocycle $\formJ$ in the field frame in which we can consistently set the twisted fields to zero. But let us stress that there is no rigorous argument that this choice corresponds to a physically allowed field frame. 

Due to the fact that we do not have control of the physically allowed field redefinitions the following analysis is only meant to illustrate the tools one would have to apply in order to extract the second-order corrections to Fronsdal equations if this class of field redefinitions was known.

At the second order the Fronsdal equations \eqref{FronsdalEq} acquire a source $\curj_{\mm(s)}$,
\begin{align} 
\label{eq:correctedFronsdal}
F_{\mm(s)}&=\square \varphi_{\mm(s)}+...=\curj_{\mm(s)}\,.
\end{align}
We will refer to the source $\curj_{\mm(s)}$ as the Fronsdal current. The double trace of the Fronsdal operator vanishes. In spinorial language the Fronsdal operator therefore decomposes into two components, $F_{\ga(2s)}$ and $F_{\ga(2s-4)}$, which respectively correspond to its traceless and trace part.

The second order equation of motion \eqref{eq:secondPhsEOM} is given by
\begin{align}
\label{eq:eombackreaction}
\adD \pomega^{(2)} = \formJ + \pomega \, \wedge \star \, \pomega \,. 
\end{align}
In this subsection we will not consider the contribution of $\pomega \, \wedge \star \, \pomega$, which is independently conserved by \eqref{eq:linearPhsEOM} and would lead to self-interactions of the Fronsdal field governed by the higher-spin-algebra. We will therefore only focus on the first term corresponding to a backreaction of the scalars in the Fronsdal equation \eqref{eq:correctedFronsdal}.

Extracting $\formJ$ from Prokushkin--Vasiliev Theory is a technically involved task. We postpone the discussion of how we calculated $\formJ$ in Fourier space to Section~\ref{sec:Veq} and only present the result here. In Fourier space the current $\formJ$ is of the general form \eqref{eq:fourierDecomp} and therefore reads
\begin{equation}
\label{eq:backreactionGeneralForm}
\formJ = H^{\ga \ga} \int \, \d \xi \d \eta \, K_{\ga \ga} \, \pC(\xi,\phi|x) \pC(\eta,-\phi|x) \,,
\end{equation}
where the kernel $K_{\ga \ga}$ is given by
\begin{align}
\label{eq:Jkernel}
K_{\ga \ga} &= y_\ga y_\ga \, f_1(\xi \eta, y\xi , y\eta ) + y_\ga \xi_\ga \, f_2(\xi \eta, y\xi , y\eta ) + y_\ga \eta_\ga \, f_3(\xi \eta, y\xi , y\eta )  \notag \\ &+ \xi_\ga \xi_\ga \, f_4(\xi \eta, y\xi , y\eta ) + \eta_\ga \eta_\ga \, f_5(\xi \eta, y \xi , y\eta ) + \xi_\ga \eta_\ga \, f_6(\xi \eta, y\xi , y\eta ) \,,
\end{align}
and $f_{1 \dots 6}$ are functions determined by our calculation in Section~\ref{sec:Veq}. The precise form of the current $\formJ$ as extracted from Prokushkin--Vasiliev Theory is given in Appendix~\ref{app:simplifiedBackreaction}. Let us illustrate the interpretation of the various terms in \eqref{eq:Jkernel} by considering a term in the kernel of the form
\begin{equation}
K_{\ga \ga}= \ldots + \xi_{\ga(N)} \eta_{\ga(M)} y_{\ga(2-N-M)} \, (y \xi)^n (y \eta)^m (\eta \xi)^l + \ldots \,.
\end{equation}
By expanding the corresponding two-form $\formJ$ in its spin-components, i.e. $\formJ=\sum_{k=0}^{\infty} \frac{1}{k!} \formJ_{\ga(k)} y^{\ga(k)}$, one obtains the following tensor structure from this term
\begin{equation}
\formJ_{\ga(2+n+m-N-M)} \sim \ldots + f^{n,m,l}_{N,M} \, H^{\beta(N+M)}{}_{\ga(2-N-M)} \, \pC_{\gb(N)\ga(n)\nu(l)}(\phi) \, \pC^{\nu(l)}{}_{\gb(M)\ga(m)}(-\phi) + \ldots\,. \label{eq:fnml}
\end{equation}
The constant $f^{n,m,l}_{N,M}$ is worked out in Appendix~\ref{app:indexFormFromFourierSpace}. The spin-components of $\formJ_{\ga(k)}$ can uniquely be decomposed in three pieces
\begin{align}\label{gentwoformdec}
\formJ_{\ga(k)}&=H^{\gb\gb}\compA_{\ga(k)\gb\gb}+H\fdu{\ga}{\gb}\compB_{\ga(k-1)\gb}+H_{\ga\ga}\compC_{\ga(k-2)}\,,
\end{align}
where $A,B,C$ are zero-forms which are completely symmetric in all their spinorial indices.
\subsubsection{Independently Conserved Subsectors}
\label{sec:independentlyconservedsectors}
The adjoint covariant derivative $\adD$ commutes with the $y_\ga$-number operator $y^\nu \partial^y_\nu$ and therefore each spin-component of the current $\formJ$ is conserved independently. However, as we will explain in the following, each spin-component splits even further into various independently conserved subsectors. To see this let us define 
\begin{equation}
 \zeta^\pm_\ga=(\xi\pm\eta)_\ga \,.
\end{equation}
In \eqref{eq:Jkernel} the kernel $K_{\ga \ga}$ was parametrized by six functions $f_{1 \dots 6}(\xi \eta,y \xi, y \eta)$. Using these $\zeta^\pm_\ga$ we can define the following contractions
\begin{align}
Z_1 = \frac12 y\zeta^+ \,, && Z_2 = \frac12 y\zeta^-  \,, && Z_3 = \xi \eta \,,
\end{align}
and we can then decompose the kernel $K_{\ga \ga}$ as follows
\begin{align}
K_{\ga \ga}=\sum_{n,m} \frac{1}{(n-1)!(m-1)!}\, K^{(n,m)}_{\ga\ga} Z_1^{n-1} Z_2^{m-1} 
\end{align}
where we defined
\begin{align}
K^{(n,m)}_{\ga\ga}=& y_\ga y_\ga \, k^{(n,m)}_1 (Z_3) +y_\ga \zeta^+_\ga \, Z_2\, k^{(n,m)}_2 (Z_3)+y_\ga \zeta^-_\ga \,Z_1\, k^{(n,m)}_3 (Z_3) \notag \\ 
&+ \zeta^+_\ga \zeta^+_\ga \, Z_2^{2}\, k^{(n,m)}_4(Z_3)+\zeta^-_\ga \zeta^-_\ga \, Z_1^{2}\, k^{(n,m)}_5(Z_3)+ \zeta^+_\ga \zeta^-_\ga \, Z_1 Z_2\, k^{(n,m)}_6(Z_3)\,,
\end{align}
In the expression above any negative power of $Z_i$ is understood to be set to zero. This decomposition has the following nice property: Each kernel $K^{(n,m)}_{\ga \ga}$ is independently conserved with respect to the adjoint covariant derivative $\adD$ as was first shown in \cite{Prokushkin:1999zk} and therefore corresponds to an independent coupling. Among those only one is proportional to the canonical current \eqref{eq:canonicalCur}, while the others are proportional to improvements which do not contribute to the Witten diagram computation. Note that the spin of the kernel $K^{(n,m)}_{\ga \ga}$ is given by $2s=m+n+2$ and therefore this decomposition splits each spin-component further into independently conserved pieces. This splitting crucially relies on the fact that we are expanding around an $\textrm{AdS}_3$-vacuum and does generically not hold on a more general background on which the covariant derivative would mix various spin components.

For bosonic fields the kernel $K_{\ga \ga}$ is invariant under $\eta \rightarrow - \eta$. This symmetry exchanges $Z_1$ with $Z_2$ and therefore the sectors $(n,m)$ and $(m,n)$ are no longer independent for the bosonic truncation of the theory.

\subsubsection{Solving the Torsion Constraint}
We can decompose the covariant derivative as $\adD = \nabla + \phi Q$ with $Q=-h^{\ga\ga}y_\ga\pl^y_\ga$. The cocycle $\formJ$ can be split into $\formJ=\formJ^0+\phi \formJ^1$ and the second-order gauge connection $\pomega^{(2)}$ in its generalized Riemann and torsion components as in \eqref{HSTR}. We can then rewrite \eqref{eq:eombackreaction} as 
\begin{align}
T'^{(2)} \equiv \nabla e^{(2)}+Q\omega^{(2)}&=\formJ^1\,, &  R'^{(2)} \equiv \nabla \omega^{(2)} +Q e^{(2)}&=\formJ^0\,,\label{TRJeq}
\end{align}
where we have dropped the second term on the right-hand side of \eqref{eq:eombackreaction} as discussed in the previous subsection. The explicit form of $\formJ^1$, which can be found in Appendix~\ref{app:simplifiedBackreaction}, shows that the higher-spin theory has non-vanishing torsion. In particular the current $\formJ$ is of the form \eqref{eq:backreactionGeneralForm} and therefore only depends on $\phi$ through the zero-forms $\pC$. Therefore $\formJ^0$ is obtained by considering the symmetric part of $K_{\ga \ga}$ in \eqref{eq:backreactionGeneralForm} with respect to $\xi$ and $\eta$ while $\formJ^1$ is obtained from the anti-symmetric component.

We therefore need to solve the torsion constraint in order to find the source to the Fronsdal equations \eqref{eq:correctedFronsdal}. This can be done by defining
\begin{align}
\omega^{(2)} = \omega^{(2)}(e) &+ Q^{-1}\formJ^1 \,, 
\end{align}
where $Q^{-1}\formJ^1$ is the contorsion two-form and $\omega^{(2)}(e)$ is the solution for $\omega^{(2)}$ in terms of vielbein $e$ at vanishing torsion. Plugging this expression into \eqref{TRJeq} gives
\begin{subequations}
\begin{align}
T^{(2)}=\nabla e^{(2)}+Q\omega^{(2)}(e)&=0 \,, \\
R^{(2)}=\nabla \omega^{(2)}(e)+Q e^{(2)}&= \formj\,, \label{jgenexpr}
\end{align}
\end{subequations}
where $\formj$ is given by
\begin{equation}
\formj=\formJ^0-\nabla Q^{-1}\formJ^1\,. \label{eq:torsionFreeCurrent}
\end{equation}
It is important to note that the operator $Q^{-1}$ is well-defined and in the basis \eqref{gentwoformdec} reads\footnote{$k>0$ is implied in the relation above as there is no torsion constraint to be solved for the case of spin $1$.}
\begin{align}
(Q^{-1}\formJ)_{\ga(k)}&=\frac{2}{k}h^{\gb\gb}\compA_{\ga(k)\gb\gb}-h\fdu{\ga}{\gb}\compB_{\ga(k-1)\gb}-\frac{2}{k+2}h_{\ga\ga}\compC_{\ga(k-2)}\,.
\end{align}
In the following subsection we will study $\formj$ more closely and discuss how it is related to the Fronsdal current \eqref{eq:correctedFronsdal}.

\subsubsection{Obtaining the Fronsdal Current}
In this subsection we will first derive that $\formj$, in a decomposition analogous to \eqref{gentwoformdec}, has a vanishing $B$ component. This observation will allow us to relate this object to the Fronsdal current $\curj_{\mm(s)}$ appearing in \rf{eq:correctedFronsdal}. 

Let us first note that the nilpotence of $\adD$ and the conservation of $\formJ$ imply the following relations: 
\begin{align}
\label{eq:commu}
\adD^2&=0 & \rightarrow \hspace*{19	pt}\qquad\; \{\nabla,Q\}&=0 \,, & \nabla^2+Q^2&=0\,, \\
\adD \formJ&=0 & \rightarrow \qquad \; \nabla \formJ^0 + Q \formJ^1 &=0\,, &  Q \formJ^0 + \nabla \formJ^1 &= 0 \,. \label{eq:cons} 
\end{align}
Using these relations one derives
\begin{align}
\nabla \formj=\nabla R^{(2)}=0\,, && Q\formj=Q R^{(2)}=0\,.
\end{align}
These relations correspond to the differential and algebraic Bianchi identities respectively. The first condition implies that the Fronsdal current $\formj$ is conserved with respect to the Lorentz-covariant derivative $\nabla$. The second condition implies that 
\be
h\fdu{\ga}{\nu} \wedge \formj_{\nu \ga(k-1)}=h\fdu{\ga}{\nu} \wedge R^{(2)}_{\nu \ga(k-1)}\equiv 0\,.
\ee
By using \eqref{eq:threeFormIdent} one can show that this is only guaranteed to hold if and only if $\compB \equiv 0$ in the decomposition \eqref{gentwoformdec} and therefore
\begin{align}
\formj_{\ga(k)}&=H^{\gb\gb}\compj_{\ga(k)\gb\gb}+H_{\ga\ga}\compj'_{\ga(k-2)}\,. \label{eq:vanishingOfBcomp}
\end{align}
The vanishing of $\compB$ therefore provides a consistency check of our calculations and we checked explicitly that our results pass this test. The decomposition \eqref{eq:vanishingOfBcomp} allows us to relate $\formj$ to the Fronsdal current: the two above components of rank $k+2$ and $k-2$ correspond to the trace and traceless parts of the Fronsdal current $\curj_{\mm(s)}$ with $2s = k+2$, in accordance with the mapping between spacetime and twistor indices explained in Section~\ref{sec:framemetric}. These components can be conveniently expressed by
\besubeqs
\begin{align}
\compj_{\ga(2s+2)}&= \sum_l\sum_{n+m=2s} a^{n,m,l} \, \pC_{\ga(n+1)\nu(l)}(\phi) \, \pC\fud{\nu(l)}{\ga(m+1)}(-\phi) \,,  \\ 
\compj'_{\ga(2s-2)}&= \sum_l\sum_{n+m=2s} c^{n,m,l} \, \pC_{\ga(n-1)\nu(l)}(\phi) \, \pC\fud{\nu(l)}{\ga(m-1)}(-\phi) \label{GeneralFrCurrent} \, .
\end{align}
\esubeqs
We summarize some of our explicit results for the coefficients $a^{n,m,l}$ and $c^{n,m,l}$ of $\formj$ in the following subsection. 
\subsubsection{Explicit Results}
The explicit expressions for the full $\formj$ are rather involved. In the following we will therefore only illustrate its form by considering the following two interesting components:
\begin{description}
  \item[Spin 1:] We find a source for the two-form $\d\pomega^{(2)}$ (with $\pomega^{(2)}=\pomega^{(2)}(y=0) \,$), which is pseudo-local and reads
\begin{multline}
\d\pomega^{(2)}= \formj = H^{\gb\gb} \left( \sum_{l \in 2 \mathbb{N}}  a_l \left( \pC_{\gb\gb\nu(l)}(\phi) \, \pC^{\nu(l)}(-\phi)+  \pC_{\nu(l)}(\phi) \, \pC\fud{\nu(l)}{\gb\gb}(-\phi) \right) \right. \\ 
\left. - \sum_{l \in 2 \mathbb{N}+1} a_l \pC_{\gb\nu(l)}(\phi) \, \pC\fud{\nu(l)}{\gb}(-\phi) \right) \notag \,,
\end{multline}
where the coefficients are given by
\begin{equation}
a_l=\frac{i (-i)^{l}}{l!} \frac{1}{(l+2)^2 (l+4)}\,.
\end{equation}
One can decompose this result with respect to $\phi$ to obtain equations of motion for two spin-1 fields. We checked that the coefficients obey the conservation identity \eqref{eq:conservationABCBasis}, which holds if the coefficients of the first two terms are equal while the coefficient of the third term can be arbitrary and does not affect conservation. At the linear level one can choose the connection $\omega$ to take values in $\textrm{hs}(1/2) \oplus \textrm{hs}(1/2)$ and therefore it will not contain any spin-1 field. To the second-order however a source term for the spin-1 field is produced by the scalar fields. This source term can be removed by a pseudo-local field redefinition and might therefore just be a result of our particular choice of field-frame, but unless one has full control of the physically allowed field redefinitions it is difficult to draw any definite conclusions from this result. We will further discuss this point in the conclusions to this paper.

\item[Spin 2:] From our discussion in Subsection~\ref{sec:independentlyconservedsectors} it follows that we have five independently conserved subsectors $(3,-1)$, $(2,0)$, $(1,1)$, $(0,2)$, $(-1,3)$ for $(n,m)$ in the case of spin-2. However we are considering bosonic fields and therefore the sectors $(n,m)$ and $(m,n)$ are not independent as also discussed in Subsection~\ref{sec:independentlyconservedsectors}. Thus the backreaction of the scalar fields splits into three separately conserved components for spin-2:
\begin{align}
R^{(2)}_{\ga\ga}&=\formj_{\ga \ga}=J^{(3,-1)}_{\ga \ga} + J^{(1,1)}_{\ga \ga} + J^{(2,0)}_{\ga \ga} \,,& \text{where} \; \; R^{(2)}_{\ga\ga}&\equiv\nabla \omega^{(2)}_{\ga\ga}+h_{\ga}{}^{\nu} \wedge e^{(2)}_{\nu\ga}\,.
\end{align}
We find the following expressions for these components
\begin{align}
J^{(3,-1)}_{\ga\ga}&=H^{\gb\gb}\compj^{(3,-1)}_{\ga\ga\gb\gb} \,, \\
J^{(1,1)}_{\ga\ga}&=H^{\gb\gb}\compj^{(1,1)}_{\ga\ga\gb\gb}+H_{\ga\ga}\compj^{'(1,1)} \,, \\
J^{(2,0)}_{\ga\ga}&\equiv 0 \,.
\end{align}
For the expressions above we define
\begin{align}
\compj^{(3,-1)}_{\ga(4)}&=\sum_{l \in 2 \mathbb{N}}  a_l \left( \pC_{\ga(4) \nu(l)}(\phi) \, \pC^{\nu(l)}(-\phi) + 3 \, \pC_{\ga(2) \nu(l)}(\phi) \, \pC^{\nu(l)}{}_{\ga(2)}(-\phi) \right) \,, \\
\compj^{(1,1)}_{\ga(4)}&=\sum_{l \in 2 \mathbb{N}} b_l \left( \pC_{\ga(4) \nu(l)}(\phi) \pC^{\nu(l)}(-\phi) - \pC_{\ga(2)\nu(l)} (\phi) \, \pC^{\nu(l)}{}_{\ga(2)}(-\phi) \right) \,,\\
\compj^{'(1,1)} &= \sum_{l \in 2 \mathbb{N}} b'_{l} \, \pC_{\nu(l)} (\phi) \, \pC^{\nu(l)}(-\phi) \,,
\end{align}
where projection on the $\phi$-independent part is implied.
The coefficients are then given by
\begin{align}
a_l &= \frac{i^{l-1}}{4 l!} \left(\frac{1}{1+l}-\frac{6}{2+l}+\frac{9}{(3+l)^2}+\frac{19}{4(3+l)}-\frac{6}{4+l}+\frac{7}{5+l}-\frac{3}{4 (7+l)}\right) \,,\nonumber \\
b_l&=-\frac{i^{l-1}}{4 l!} \left(\frac{1}{2+l}-\frac{1}{(3+l)^2}-\frac{13}{4 \
(3+l)}+\frac{4}{4+l}-\frac{1}{5+l}-\frac{1}{6+l}+\frac{1}{4 (7+l)}\right) \,, \nonumber\\
b'_l&=\frac{i^{l-1}}{l!} \left(\frac{1}{3 (1+l)^2}+\frac{7}{12 (1+l)}-\frac{3}{2+l}+\frac{1}{3+l}+\frac{1}{3 (4+l)}-\frac{1}{4 (5+l)} - \frac16 \delta_{l,0} \right) \,.\nonumber
\end{align}\noindent
As a consistency check we confirmed that the backreaction is conserved by using \eqref{eq:conservationABCBasis}. Let us note that these expressions can be straightforwardly expressed in metric-like language by using \eqref{eq:compOfPhysC}. The canonical current \eqref{eq:canonicalCur} is part of only one sector, namely $J^{(3,-1)}_{\ga \ga}$. Therefore the class of physically allowed field redefinitions should allow us to completely remove the other non-vanishing and independently conserved current $J^{(1,1)}_{\ga \ga}$. Furthermore both currents are generically of pseudo-local form \eqref{eq:pseudolocalfieldredefs}. If we truncate them to some finite value of $l$ in \eqref{eq:pseudolocalfieldredefs} we observe that $J^{(1,1)}_{\ga \ga}$ can be removed by local field redefinitions whereas $J^{(3,-1)}_{\ga \ga}$ can only be removed by a pseudo-local redefinition.

\end{description}

Our calculation shows that the current $\formj$ is pseudo-local, as illustrated by the two examples above. One might think that this is an immediate consequence of the fact that our calculation also results in a pseudo-local cocycle $\formJ$. However extracting $\formj$ from $\formJ$ by solving the torsion constraint as in \eqref{eq:torsionFreeCurrent} might potentially project out all the pseudo-local terms in $\formJ$. In fact one needs to consider a pseudo-local ansatz if one wants to recover the canonical current \eqref{eq:canonicalCur} upon solving the torsion constraint while keeping $\phi \rightarrow -\phi$ symmetry, which is the case for Prokushkin--Vasiliev Theory. We discuss this point in more detail in Appendix~\ref{app:localcurrents} and \ref{app:canonicalCurrentsIntegral}.


\section{Fixing the Cubic Action}
\label{sec:cubic}

In this section we explain how to determine completely the cubic action for the physical sector of Prokushkin--Vasiliev Theory. In the previous section we presented our results concerning the second-order physical equations of motion for the various fields presented in the theory and in particular we obtained the backreaction to the physical gauge connection at order~$2$ in perturbation theory. As explained in Section~\ref{sec:PresentationBackreactionFronsdal}, upon solving the torsion constraint this backreaction is the source for the Fronsdal tensor. From the standpoint of an action principle, these currents correspond to $0$--$0$--$s$-like couplings. In the frame formalism that we have been dealing with so far such couplings would read
\begin{align}
2 \sum_s g_s\int e^{a(s-1)}\wedge \formj_{a(s-1)}\,,
\end{align}
where 
$\formj_{a(s-1)}$ is a conserved two-form, 
 bilinear in the physical scalar field $\pC(y=0)$. In the metric-like picture we deal with currents $j_{\mm(s)}$ with $\formj^{a(s-1)}=-\tfrac12 \formj\fdu{\mm}{a(s-1)} \epsilon\fud{\mm}{\nn\rr} \d x^\nn\wedge \d x^\rr$ being the two-form dual thereof. The corresponding  cubic couplings are known and classified: they read
\begin{align}\label{metriclikecubic}
S_\textsc{currents}&= 2\sum_s \frac{g_s}{s}\int \varphi^{\mm(s)} j_{\mm(s)}\,,
\end{align}
with the corresponding currents given by derivatives of the scalar fields, that is,\footnote{Whereas in the rest of this paper we have set $\Lambda = 1$ in this section we restore it for the purpose of keeping track of the terms which vanish in the flat space-limit.}
\begin{equation}
\label{currentsmetriclike}
j_{\mm(s)}(\Phi) \equiv (-i)^s\Phi^* (\cev{\nabla}_\mm - \vec{\nabla}_\mm)^s\Phi+\Lambda(\ldots)\,,
\end{equation}
for which we refer to \cite{Bekaert:2010hk}. These currents are hermitian and all prefactors are introduced for convenience (although the $i$ is needed in order to make odd-spin currents hermitian too). The second term in the above right-hand side denotes terms proportional to the cosmological constant $\Lambda$ which are needed to make the current conserved on $\textrm{AdS}_3$. We have chosen to express the above currents in terms of one complex scalar field $\Phi$ and its complex conjugate $\Phi^*$, which are to be identified with the $\Pi_\pm$-projected components of $\pC(y=0)$, that is, $\Phi = \Pi_+\pC(y=0)$ and $\Phi^* = \Pi_-\pC(y=0)$. As one can check, odd-spin conserved currents can be written down only if at least two real scalars are involved. As in this section we are interested in dealing with even and odd spins altogether, the above thus constitutes the minimalistic option involving one complex scalar, which corresponds to the truncation of Prokushkin--Vasiliev Theory we consider. 
As one can verify, the above expressions for the currents indeed yield cubic couplings in \rf{metriclikecubic} which are unique up to field redefinitions and boundary terms.\footnote{As we are about to explain these currents deform the gauge transformations of the scalars. Evidently the currents are unique only as equivalence classes in the space of such couplings, for improvements which do not deform the gauge transformations can always be constructed.} 

The form of the spin-$s$ coupling is thus known for all spins. However, to the best of our knowledge the relative coefficients $g_s$ of \rf{metriclikecubic} have never been determined before. Indeed these are left arbitrary at the cubic level, where the cubic cross couplings are invariant independently. These $g_s$ coefficients in fact constitute the last piece of information needed to determine completely the cubic action for the physical sector of Prokushkin--Vasiliev Theory at $\lambda=\frac12$. Indeed, the kinetic pieces are known and the higher-spin self-couplings are also known: they can be extracted from the Chern--Simons action which describes the pure gauge sector \cite{Achucarro:1987vz,Blencowe:1988gj} (see below).  

Presumably, the relative coefficients $g_s$ could be read off by comparing \rf{metriclikecubic} with the Prokushkin--Vasiliev backreaction. Such is, however, a non-trivial task, because the Fronsdal currents that are so produced still contain infinite pseudo-local tails of derivatives and it is not clear which class of field redefinitions one should use in order to map these tails to canonical form \eqref{currentsmetriclike} (see previous section). Another possibility is to start from the consistent cubic action
and proceed with the quartic Noether analysis. If a quartic completion exists thereof, quartic terms will be found which make the action gauge invariant to quartic order, and the relative coefficients $g_s$ are expected to be fixed in this manner.\footnote{Also one could think of using more modern methods such as the BRST--Antifield ones, which are particularly suited for addressing quartic-order issues. They are reviewed e.g. in \cite{henneauxbook} and in Chapter~$4$ of \cite{Gomez:2014dwa}.} However there is a simpler way of determining the value of the relative coefficients $g_s$, which we now detail. The idea is to look at the deformation of the gauge transformations for the scalar $\Phi$, so that we can write down the complete cubic action including the relative coefficients without having to go through the full quartic Noether analysis but only employing the known solutions to the so-called admissibility condition \cite{Konshtein:1988yg,Konstein:1989ij} (see \cite{Joung:2014aba} for an example in which admissibility condition was used to this effect in a simpler context). Note that we are not going to repeat the analysis of the admissibility condition from scratch. The most general solution for the theory at hand has already been discussed in the literature. We only match the metric-like result with the known solution to the effect of fixing the metric-like action. \\

The coupling \rf{metriclikecubic} is on-shell gauge invariant to the lowest order, that is, under $\delta^{(0)}\Phi = 0$ and $\delta^{(0)}\varphi_{\mm(s)} = \nabla_\mm \xi_{\mm(s-1)}$ we have
\begin{equation}
\delta^{(0)} S_\textsc{currents} \approx 0\,,
\end{equation}
where $\approx$ denotes an on-shell equality and we have neglected boundary terms as we will do through the rest of this section. This interaction term is abelian but deforms the gauge transformation rules for the scalar field. Differently put, in order to make the term off-shell gauge invariant we need to assign transformation rules to the scalar field, so that the terms in the above right-hand side are canceled by the gauge variation of the scalar kinetic piece. Here we are simply expanding the full invariance condition $\delta S = 0$ to order~$1$ in perturbation theory, that is,
\begin{equation}
\label{noetherbasic}
\delta^{(0)}S^{(1)} + \delta^{(1)}S^{(0)} = 0\,.
\end{equation}
In the above, $S^{(0)}$ is the kinetic piece:
\begin{equation}
S^{(0)} \equiv \int \textrm{det}|h| \,  \left(\nabla_\mm \Phi^* \nabla^\mm\Phi +m^2\Phi^*\Phi\right) + S^{(0)}_\textsc{cs} \equiv S^{(0)}_\textsc{scalars}+S^{(0)}_\textsc{cs} \,,
\end{equation}
where $S^{(0)}_\textsc{cs}$ is the quadratic piece of the full Chern--Simons action for a higher-spin gauge connection valued in $\textrm{hs}(\tfrac{1}{2})$. Recall that
\begin{align}
S_\textsc{cs}&\equiv \frac{k}{4\pi}\int \textrm{tr}\Big(\pomega \wedge \d\pomega-\frac23\pomega\wedge\pomega\wedge \pomega \Big)\,,
\end{align}
where $k$ is the Chern--Simons level, $\pomega=\pomega(y,\phi)$ is the higher-spin connection and we do not take twisted fields into account, thereby dropping $\psi$. 
The trace\footnote{To be precise, $f(0)$ is a super-trace \cite{Vasiliev:1999ba}, but since we consider bosonic higher-spin fields $f(y)=f(-y)$, it reduces to a trace.} is $\textrm{tr} f(y)=f(0)$. The quadratic and cubic pieces are extracted from the above action by perturbing around the $\textrm{AdS}_3$ vacuum $\Omega$ of \rf{AdSConnection}, that is, performing $\pomega\rightarrow \Omega+\pomega$:
\begin{equation}
\label{csexpansion}
S^{(0)}_\textsc{cs} + S^{(1)}_\textsc{cs} = \frac{k}{4\pi}\int \textrm{tr}\Big(\pomega \wedge \d\pomega-2\Omega\wedge \pomega \wedge\pomega -\frac23 \pomega\wedge\pomega\wedge\pomega \Big)\,,
\end{equation}
where $S^{(1)}_\textsc{cs}$ is part of $S^{(1)}$ in \rf{noetherbasic}, which thus contains two terms: $S^{(1)} = S^{(1)}_\textsc{cs} + S_\textsc{currents}$. 

The way one determines the deformation of the gauge transformations for the scalar is identifying terms proportional to the equations of motion for the scalar in \rf{noetherbasic}. This goes as follows: the Chern--Simons cubic self-coupling in \rf{csexpansion} is off-shell invariant on its own under the zeroth-order gauge transformations for the higher-spin connection. Thus $\delta^{(0)}S^{(1)}$ roughly reads
\begin{equation}
\label{zerovariations1}
\delta^{(0)}S^{(1)} = \delta^{(0)}S^{(1)}_\textsc{currents} \equiv \int \xi \times \mathcal{C}(\mathcal{E}, \Phi) \,,
\end{equation}
where $\mathcal{C}(\mathcal{E}, \Phi)$ is the expression obtained by taking the divergence of the currents \rf{currentsmetriclike} when integrating by parts, so that as indicated it is linear in both the scalar field $\Phi$ and the equations of motion $\mathcal{E} = \mathcal{E}(\Phi)$ thereof. By definition $\mathcal{E} \approx 0$ and hence $\mathcal{C} \approx 0$. On the other hand, the gauge transformations for the higher-spin gauge connection do not get deformed by the above cubic cross-coupling. This is evident by noticing that only the equations of motion for the scalar appear in the above right-hand side. Therefore $\delta^{(1)}S^{(0)}$ yields the following expression:
\begin{equation}
\label{firstordervariationS0}
\delta^{(1)}S^{(0)} = \delta^{(1)}S^{(0)}_\textsc{scalars} = \int \mathcal{E}(\Phi)\delta^{(1)}\Phi\,.
\end{equation}
The variations $\delta^{(1)}\Phi$ are linear in $\xi$ and in $\Phi$ itself. Now integrating by parts in \rf{zerovariations1} in order to write its integrand as $\mathcal{E}(\Phi)\times (\dots)$ and comparing with the above right-hand side one can read off the searched-for variations $\delta^{(1)}\Phi$. They depend on the relative couplings $g_s$, since they depend on the current. The 'trick' we will now use is, instead of solving some consistency condition for the quartic Lagrangian, to solve some consistency condition for the first-order gauge transformations of the scalar field. This workaround will prove to be much quicker in determining completely the relative coefficients $g_s$.\\ 

Let us consider the following consistency condition, which is part of the Noether procedure:
\begin{equation}
[\delta_\xi,\delta_\epsilon] \Phi \approx \delta_{[\xi,\epsilon]} \Phi\,.
\end{equation}
Expanding this equation in perturbation theory and retaining the piece of order~$2$ we obtain
\begin{equation}
[\delta^{(1)}_\xi,\delta^{(1)}_\epsilon]\Phi \approx \delta^{(0)}_{[\xi,\epsilon]^{(1)}}\Phi + \delta^{(1)}_{[\xi,\epsilon]^{(0)}} \Phi - ([\delta^{(0)}_\xi,\delta^{(2)}_\epsilon] - \xi \leftrightarrow\epsilon)	\Phi \,.
\end{equation}
Solving the above consistency condition for $\delta^{(1)}\Phi$ should fix the relative coefficients $g_s$ which it depends on. In general, doing so is as hard as solving the corresponding consistency condition for the cubic interaction term involving the currents, for one needs to find a quartic completion $\delta^{(2)}\Phi$ such that the above condition is fulfilled. The trick is to restrict one's attention to Killing tensors, that is, to gauge parameters $\xi$ and $\epsilon$ such that $\delta^{(0)}\varphi_{\mm(s)} = \nabla_\mm \xi_{\mm(s-1)} =0$ and similarly for $\epsilon$. In such a case the last term in the above right-hand side is zero, because $\delta^{(2)}\Phi$ is proportional to the higher-spin field and hence $\delta^{(0)}_\xi \delta^{(2)}_\epsilon\Phi$ is zero on Killing tensors by simply using the chain rule. Further noticing that $\delta^{(0)}_{[\xi,\epsilon]^{(1)}}\Phi = 0$ we find
\begin{equation}
\label{admissibility}
[\delta^{(1)}_\xi,\delta^{(1)}_\epsilon]\Phi \approx \delta^{(1)}_{[\xi,\epsilon]^{(0)}} \Phi \qquad \textrm{on Killing tensors}\;\xi\;\textrm{and}\;\epsilon . 
\end{equation}
This condition is necessary but non-sufficient in order for the variations $\delta^{(1)}\Phi$ to be consistent at order~$2$ in the Noether analysis. The advantage of this procedure is now clear: we are solving (part of) a second-order consistency condition in which no second-order quantity enters. Note that the above requirement also goes under the name of \emph{admissibility condition} for the scalar couplings \cite{Konshtein:1988yg,Konstein:1989ij}. In words, it says that the first-order gauge transformations should close to an algebra on the scalar field when restricting to rigid parameters, i.e. the scalar field needs to sit in a representation of the higher-spin algebra of rigid symmetries. 

Remarkably, there is a well-known solution to the above condition on $\delta^{(1)}\Phi$: the gauge transformations for the scalar derived from the Prokushkin--Vasiliev theory are known to pass the above admissibility condition.\footnote{In some sense such laws of transformation are the unique ones solving the admissibility condition \cite{Vasiliev:1997ak}.} These gauge transformation are given here below, and we observe that there are no free coefficients therein. According to Section~\ref{sec:unfolded} we have\footnote{For ease of notation the $\delta \flC^{(2)}$ of Section~\ref{sec:unfolded} is here denoted just by $\delta^{(1)}\flC$.}
\begin{equation}
\label{appeq:precGauge}
\delta^{(1)} \flC = [\pxi,\flC]_\star\,,
\end{equation}
where $\pxi = \pxi (y,\phi)$ is the first-order piece of $\boldsymbol{\xi}$ appearing in Section~\ref{sec:unfolded}.\footnote{We set to zero consistently the $\psi$-dependent part of $\pxi$ since we choose $\somega=0$.} The above transformation rules can indeed be checked to satisfy the admissibility condition \rf{admissibility}. Recalling that $\pC$ is embedded into $\flC$ as $\flC = \pC\psi + \sC $ and splitting the gauge parameter as $\xi=\xi_\omega+\phi\xi_e$, the interplay between $\phi$ and $\psi$ is seen to lead to 
\begin{align}
\delta^{(1)} \pC &= [\xi_\omega,\pC]_\star+\phi \{\xi_e,\pC\}_\star\,.
\end{align} 
From Prokushkin--Vasiliev Theory the transformations of the scalar field $\Phi = \Pi_+\pC (y=0)$ thus read
\begin{align}\label{twistedgtr}
\delta \Phi =\textrm{tr}\{\xi_e,\pC_+\}_\star = 2\sum_s \frac{(-1)^{s-1}}{(2s-2)!}\xi_e^{\ga(2s-2)} \pC_{+\,\ga(2s-2)} = \sum_s \frac{(2i)^s}{(2s-2)!}\xi_e^{\mm(s-1)} \nabla_{\mm(s-1)}\Phi\,,
\end{align} 
where we have used \rf{eq:compOfPhysC} in order to express $\pC_{\ga(2s-2)}$ as derivatives of $\pC(y=0)$ and have defined $\xi_e = i \sum_s \tfrac{1}{(2s-2)!} \xi_e^{\ga(2s-2)} y_{\ga(2s-2)}$.\footnote{The reality conditions, $\pomega^\dagger = - \pomega$, require the gauge parameter $\xi$ to be imaginary \cite{Prokushkin:1998bq}.} Comparing the last expression above with the one obtained from cubic action cross-couplings (see beginning of this section),
\begin{equation}
\label{gaugetransfscubic}
\delta^{(1)} \Phi= \sum_s (2i)^s g_s \xi_{\mm(s-1)}\nabla^{\mm(s-1)}\Phi\,,
\end{equation}
we read off the relative $g_s$ coefficients. They are the following:
\begin{equation}
\label{gsatlast}
\boxed{g_s=\frac{1}{(2s-2)!}} \,.
\end{equation}
Let us note that the restriction to Killing tensors also implies that \rf{twistedgtr} is, in fact, the only Lorentz-invariant combination one could write which is linear in the scalar and the gauge parameter~---~up to the relative factors. This is the solution to the admissibility condition \rf{admissibility} at the Lagrangian level. The complete cubic action for the physical sector of Prokushkin--Vasiliev Theory thus reads
\begin{equation}
\begin{aligned}
\label{eq:fullcubicaction}
S_\textsc{cubic} &= S^{(0)}_\textsc{cs} + S^{(1)}_\textsc{cs} + S^{(0)}_\textsc{scalars} +  S_\textsc{currents}\\
&= \frac{k}{4\pi}\int \textrm{tr}\left(\pomega \wedge \d\pomega-2\Omega\wedge \pomega \wedge\pomega -\frac23 \pomega\wedge\pomega\wedge\pomega \right)\hspace*{-10pt} &&+ \int \textrm{det}|h| \,   \left(\nabla_\mm \Phi^* \nabla^\mm\Phi +m^2\Phi^*\Phi\right) \\
&{}&&+ 2 \sum_s g_s\int e^{a(s-1)}\wedge \formj_{a(s-1)}\,,
\end{aligned}
\end{equation}
where the above cross-couplings can be rewritten in the metric-like language of \eqref{metriclikecubic} since at cubic order the identification \eqref{eq:metricframeid} holds \cite{Fredenhagen:2014oua,Campoleoni:2012hp}. Note also that the Lagrangian which solves the admissibility condition will depend on the chosen HS algebra. For different values of $\lambda$ in $hs(\lambda)$ the coupling constant are hence expected to be different (see \cite{Charlotte} for the corresponding analysis). Below we give explicit expressions for some low-spin currents. \\

Some comments are in order. Firstly we note that the above scalar transformation rules generically hold for Killing tensors only. For generic gauge parameters $\xi$ the right-hand side of \rf{gaugetransfscubic} would include terms with derivatives of the gauge parameter, produced by integrating by parts in \rf{firstordervariationS0} to isolate $\mathcal{E}(\Phi)$.  Such terms can always be removed by redefining the scalar field, which will supplement the currents \rf{currentsmetriclike} with improvements. A unique combination of improvements is required to uplift \rf{gaugetransfscubic} beyond killing tensors. The above procedure thus fixed the deformed gauge variations of the scalar up to field redefinitions. Requiring the gauge transformations to contain no derivatives of the gauge parameter determines the field frame to be the Prokushkin--Vasiliev one. The choice of redefinitions that recovers the higher-spin algebra structure constants is perhaps more natural, as it is (among other things) the one associated with the usual stress tensor in the spin-2 sector, as we detail below. 

Secondly let us stress that in so fixing the $g_s$ coefficients, although we have determined completely the cubic action, it is not implied that a quartic completion thereof exists. Indeed, the condition we have solved is necessary but non-sufficient. A priori, there might be no consistent quartic completion, a unique one, or many. It could be argued that the very existence of Prokushkin--Vasiliev Theory indicates that such a quartic completion does exist. However, our cubic action is free of twisted fields, whereas we have only proven that the latter can be consistently set to zero in Prokushkin--Vasiliev Theory to order~$2$ in perturbation theory. We thus consider it an open issue whether or not one can achieve full consistency starting from our cubic action.

The above result and its simplicity are to be contrasted with the pseudo-local nature of the Prokushkin--Vasiliev backreaction, in which the above simple coefficients are well hidden and hard to extract. It is important to stress, however, that the gauge transformations are blind to the addition of off-shell conserved currents on top of the above ones. In principle those can be pseudo-local. For instance, one can obtain conserved currents of spin~$s$ as\footnote{The subleading terms in $\Lambda$ can be conveniently extracted from the corresponding ambient space form but we do not specify them in the following.}
\begin{equation}
\label{eq:improvTerms}
j_{\mm(s)}=(g_{\mm\mm}\Box-\nabla_{\mm}\nabla_\mm)^k j_{\mm(s-2k)}+\Lambda(\ldots)\,.
\end{equation}

It is however conceivable that the higher-derivative tail which is seen to arise from Prokushkin--Vasiliev's equations boils down to a pseudo-local contribution to the canonical currents, precisely in the same fashion as the canonical stress tensor differs from the spin-$2$ current given below by terms of the form \eqref{eq:improvTerms}.

Another important comment is that one can write down the cubic cross-couplings corresponding to the above ones before the torsion constraint has been solved for.	 Such an action term would read
\be \int \textrm{tr}\left(\pomega(y,\phi)\star \wedge \formJ^\textsc{fr}(y,\phi)\right)\,,\ee
where $\pomega(y,\phi)$ takes values in the higher-spin algebra and contains both vielbeins and spin-connections, and $\formJ^\textsc{fr}$ is the backreaction that has the property that the Fronsdal current it yields upon solving the torsion constraint is the canonical $s$-derivative one (see Appendix~\ref{app:localcurrents} and \ref{app:canonicalCurrentsIntegral}).\footnote{The Fronsdal current that it corresponds to is however traceless and differs from \rf{currentsmetriclike} by improvement terms.} This way of writing the coupling is more natural from the Prokushkin--Vasiliev vantage point. The coefficients $g_s$ are the same. \\

It is instructive to give explicit forms for the spin-2 and spin-3 currents for which the deformation of the scalar gauge transformation does not involve derivatives of the gauge parameter. For the sake of generality we restore the cosmological constant $\Lambda$ and do not fix the mass term of the scalar field $m^2=\mu \Lambda$ entering the mass-shell equation $(\Box-m^2)\Phi=0$, where $\mu=\lambda^2-1$ and in our case $\lambda=\frac12$ 

\paragraph{Spin-2 current.}
In the case of spin-2 we can construct a current that differs from the canonical stress tensor by a trivial improvement term of the type  \eqref{eq:improvTerms}.
Such a spin-2 current reads
\begin{multline}
\hspace*{21pt}-j^{\mm\mm}=  \Phi \nabla^{\mm }\nabla^{\mm}\Phi^*  -  2 \nabla^{\mm}\Phi^*\nabla^{\mm}\Phi+  \Phi^* \nabla^{\mm }\nabla^{\mm}\Phi\\ +2\Lambda ( 1-   \mu) g^{\mm \mm} \Phi^* \Phi  -   g^{\mm \mm}\Phi  \Box \Phi^* -   g^{\mm \mm} \Phi^* \Box \Phi\,,\hspace*{20pt}
\end{multline}
and the induced gauge transformations are as anticipated; without derivatives of the gauge parameter:
\begin{equation}
\delta \Phi=-2\xi^\mm\nabla_\mm \Phi\,.
\end{equation}

\paragraph{Spin-3 current.} 
In the spin-3 case one builds a current differing from the canonical form by improvement terms and the result is given by
\begin{align}
-i j^{\mm(3)}&=  \Phi \nabla^{\mm}\nabla^{\mm}\nabla^{\mm}\Phi^* -    \Phi^* \nabla^{\mm}\nabla^{\mm}\nabla^{\mm}\Phi + 3  \nabla^{\mm}\Phi^*\nabla^{\mm}\nabla^{\mm}\Phi -  3 \nabla^{\mm}\nabla^{\mm}\Phi^*\nabla^{\mm }\Phi\\
&+ 2 \Lambda (4 - 3 \mu)  g^{\mm\mm}\,\Phi \nabla^{\mm}\Phi^*  - 2 \Lambda (4 - 3 \mu) g^{\mm\mm}\, \Phi^* \nabla^{\mm}\Phi\nonumber\\
& -  \tfrac{9}{2}  g^{\mm\mm}\,\Box \Phi \nabla^{\mm}\Phi^*+ \tfrac{9}{2} g^{\mm\mm}\, \Box \Phi^* \nabla^{\mm}\Phi-  \tfrac{3}{2}  g^{\mm\mm}\,\Phi \nabla^{\mm}\Box \Phi^* + \tfrac{3}{2}  g^{\mm\mm}\,\Phi^* \nabla^{\mm}\Box \Phi\,.\nonumber
\end{align}
The corresponding induced gauge transformation on the scalar again do not display any derivative acting on the gauge parameter due to the above field-redefinition terms, so that we have:
\begin{equation}
\delta \Phi=-\frac{i}{3}\,\xi^{\mm\mm}\nabla_{\mm}\nabla_{\mm} \Phi\,,
\end{equation}
which is given for $g_3=\tfrac{1}{24}$. 

Summarizing, we have used admissibility condition to fix the last piece of arbitrariness in the cubic action for the physical sector of Prokushkin--Vasiliev Theory. The solution to this consistency condition is the transformation rules for the Prokushkin--Vasiliev scalar, precisely. The terms \rf{eq:fullcubicaction} form the unique cubic action for the physical sector of Prokushkin--Vasiliev Theory\footnote{Note that we do not need to repeat from scratch the analysis of admissible HS algebra which is already present in the literature. The original result here is to match the metric-like result against to the structure constant of the known admissible HS algebras to the effect of fixing the Lagrangian of the theory to this order.}. 

\paragraph{Frame-like action.} 

There is yet another way in which one may think of constructing the cubic cross-couplings discussed hereabove. One can write down a quadratic action for scalar fields in the following way:
\begin{equation}
S_\textsc{rr}= \sum_k a_{2k}\int  h_{\nu\nu} R^{\ga(2k-1)\nu } R\fud{\nu}{\ga(2k-1)} \,,\label{RRaction}
\end{equation}
where the free curvatures are defined as $R=(\tadD \pC) \psi$. For generic values of the $a_{2k}$ coefficients the above term involves all components $\pC^{\ga(k)}$ of the physical scalar. However, as explained in \cite{Shaynkman:2000ts} one can tune the coefficients so that all components $\pC^{\ga(k)}$ with $k>2$ drop out, i.e. only the first two (bosonic) components, $\pC(y = 0) \equiv \Phi$ and $\pC^{\alpha\alpha}$ are involved. The corresponding coefficients are $a_n=\frac{1}{(n+2)n!}$. Up to boundary terms the above expression is the first-order action for scalar fields, and yields the standard kinetic term upon solving for $\pC^{\alpha\alpha}$ as in \rf{eq:compOfPhysC}. It is important to stress that such an action can be written in $RR$-like form in $\textrm{AdS}$-space or for massive fields only, and not for massless fields in flat space, since it relies on the presence of the $yy$-piece in the $h^{\ga\ga}(y_\ga y_\ga-\pl_\ga\pl_\ga)$-part of $\tadD$. Indeed reintroducing the cosmological constant $\Lambda$ we see that it multiplies this term as $\Lambda h^{\ga\ga}y_\ga y_\ga$ so that it degenerates in the flat limit. 

One can now turn on interactions by simply replacing the background derivative $\tadD$ with $\adD_\omega=\tadD-[\pomega,\bullet]_\star=\d-[\Omega+\pomega,\bullet]_\star$, which is similar to the Yang--Mills interactions considered in \cite{Shaynkman:2000ts}. The action 
is then found to be consistent up to the cubic level following the standard arguments of the Fradkin--Vasiliev approach \cite{Fradkin:1986qy,Fradkin:1987ks}. Indeed the variation is proportional to the free equations of motion:
\begin{align}
\delta S=2\sum_k a_{2k} \int h_{\gc\gc} [\xi,R]^{\ga(2k-1)\gc}_\star R\fdu{\ga(2k-1)}{\gc} \,,
\end{align}
and therefore vanishes on the free mass-shell $R=0$. We note, however, that the interacting action also contains quartic terms, which we neglect at cubic order. This action must be the cubic action we have constructed in this section, as it is gauge invariant under the same deformed gauge transformations $\delta \pC\psi=[\pxi,\pC\psi]$. The $RR$-like action is however pseudo-local, since it involves all components of $\pC(y)$ even if we restrict to a particular spin in $\pomega$, and differs from the local cubic action constructed above by a boundary term and further by a bulk term proportional to $F=\d\pomega-\pomega\star\wedge \pomega$. It would be interesting to see which of the two actions it is easier to extract correlation functions from, as they are computed in \cite{Chang:2011mz,Chang:2011vka}. It is also interesting to point out that in the cubic action constructed via the Noether procedure the coefficients which we determine parametrize the interactions, and it is a requirement about consistency of the interactions which fixes them. 

As a final comment let us note that the $RR$-like action is formally consistent to the cubic order over any background that is described by a flat connection $\Omega$ of the higher-spin algebra, e.g. a higher-spin black-hole. If $\Omega$ has non-vanishing components beyond the spin-$2$ sector the action in the free approximation will depend on higher components $\pC^{\ga(2k)}$ with $k>1$, which brings in higher derivatives\footnote{This indicates a difference between the Cauchy problem where data is given at $t=0$ and the Taylor-like problem that arises within the unfolded approach (the components $\pC^{\ga(k)}$ parametrize on-shell derivatives of the scalar field at a point). While the solution to the unfolded problem is always given by some $\pC(y|x_0)$, the Cauchy problem can change.} in the equations of motion as in \cite{Cabo-Bizet:2014wpa}. While for the simplest background, which is \textrm{AdS}, the $s$--$0$--$0$ vertices are gauge invariant for any $s$ separately, it is not so on more general backgrounds. On those gauge invariance requires a relative normalization of different vertices to be fixed in terms of the trace of the higher-spin algebra. It should be stressed that the mass of the scalar field that can be consistently coupled is also fixed by the representation theory of the higher-spin algebra in order to compensate the variation of the scalar-current coupling. Let us also recall that a scalar cannot be coupled to a Chern--Simons theory for an $\textrm{sl}(N)$ algebra with $N>2$, and having a consistent coupling requires the $\textrm{hs}(\lambda)$ algebra.


\section{Extraction of the Physical Equations}
\label{sec:Veq}

In this section we explain how the equations of motion \rf{eq:linearPhsEOM} and \rf{eq:secondPhsEOM} for the physical fields presented and discussed in Section~\ref{sec:presentation} are extracted from the master equations of the Prokushkin--Vasiliev theory. In a nutshell, the procedure for doing so goes as follows: one considers master equations for master fields. The master fields depend on a doubled set of oscillators, that is, on the $y_\alpha$'s of Section~\ref{sec:presentation} but also on some $z_\alpha$'s which obey analogous commutation relations (see below). The components of the master fields along the new $z_\alpha$ oscillators are purely auxiliary, and the role of some of the master equations is to allow one to solve for them in terms of the physical fields (those that multiply $y_\alpha$ oscillators only). The other master equations become the higher-spin equations of motion once we plug the master fields with their $z_\alpha$-dependent part solved for ($z_\alpha$-on-shell forms). As one can prove, the obtained equations no longer depend on $z_\alpha$. In the rest of this section we detail this procedure and obtain the first- and second-order equations of motion for the physical higher-spin gauge connections and scalar fields. 

\subsection{Master Fields and Master Equations}
\label{subsec:masterfieldsandeqs}

The Prokushkin--Vasiliev master equations are expressed in terms of three master fields
\begin{align}
\MW&=\MW_\mm(y,z,\phi,\psi|x)\, \d x^\mm\,, \qquad \MB=\MB(y,z,\phi,\psi|x)\,, \qquad\MS_\ga=\MS_\ga(y,z,\phi,\psi|x)\,.
\end{align}
The master field $\MW$ is a spacetime one-form which includes the higher-spin gauge connections and dreibeins as well as auxiliary components. The zero-form master field $\MB$ includes the (complex) scalar field and also auxiliary components. The master field $\MS_\ga$ is completely auxiliary in the sense that it can be completely expressed in terms of the zero-form $\MB$, as will be explained below. All master fields are functions of the spacetime coordinates $x^\mm$, the Clifford factors $\phi$ and $\psi$ introduced in \eqref{eq:cliffordElements}, and two sets of (mutually) commuting oscillators $y_\ga$ and $z_\ga$, i.e. they obey
\begin{align}
y_\ga y_\gb = y_\gb y_\ga \,, \qquad z_\ga z_\gb = z_\gb z_\ga \,, \qquad y_\ga z_\gb = z_\gb y_\ga\, .
\end{align}
The $y_\alpha$ oscillators are those of Section~\ref{sec:HSalgebra} which are involved in the star-product \rf{eq:moyalProduct}, whereas the $z_\alpha$ oscillators are new ones, satisfying the following commutation relations:
\begin{equation}
[y_\alpha,y_\beta]_\star = 2i\epsilon_{\alpha\beta}\,,\qquad [z_\alpha,z_\beta]_\star = -2i\epsilon_{\alpha\beta}\,.
\end{equation}
The corresponding star-product, generalizing \rf{eq:moyalProduct}, reads
\begin{align}
\label{eq:starproduct}
\mathbf{f}(y,z)\star \mathbf{g}(y,z)=\frac{1}{(2\pi)^2}\int \d^2u\, \d^2v\, \, \mathbf{f}(y+u,z+u) \, \mathbf{g}(y+v,z-v) \exp{(iv^\ga u_\ga)}\,,
\end{align}
where $v^\ga=\epsilon^{\ga \gb}v_\gb$ and the antisymmetric epsilon tensor obeys $\epsilon^{\ga \gamma} \epsilon_{\gb \gamma}=\delta^{\ga}_\gb$. All our conventions are summarized in Appendix~\ref{app:notation}. In the rest of this section all star-products will refer to this 'enlarged' star-product. Evidently, upon considering functions of $y_\alpha$ only in the above formula one recovers the $y_\alpha$-star-product of \rf{eq:moyalProduct}. 

The physical fields of Section~\ref{sec:general}~and~\ref{sec:presentation}~---~the (higher-spin) gauge connections and dreibeins $\pomega$ as well as the scalar $\pC$~---~are embedded into the above master fields via their $z_\alpha$-independent components. That is, 
\begin{equation}
\begin{aligned}
\MB =&\;\flC + \MB^{(2)} + \MB^{(3)} + \dots = \B(y) + \mathcal{O}(z)\\
&\;\flC = \flC(y) = \pC \psi + \sC  \\
&\hspace*{70pt}\pC = \Pi_+\pC + \Pi_-\pC \equiv \pC_+ + \pC_- \,,
\end{aligned}
\end{equation}
so that, as explained in Subsection \ref{sec:precursors}, the identity component of $\pC$ is the physical scalar. For the connection we have
\begin{equation}
\begin{aligned}
\MW =&\;\flomega + \MW^{(2)} + \MW^{(3)} + \dots = \W(y) + \mathcal{O}(z)\\
&\;\flomega = \flomega(y) = \pomega + \somega \psi \\
&\hspace*{70pt}\pomega = \omega + \phi e\,.
\end{aligned}
\end{equation}
We have also displayed the 'twisted' fields, which are discussed in Section~\ref{sec:precursors}. As is explained in Section~\ref{sec:presentation}, one of the main points of this paper is to study the possibility of consistently setting them to zero, at order~$2$ in perturbation theory. The actual gauge connections and dreibeins $\omega(x)_\mm^{\alpha_1\dots \alpha_{2s}}$ and $e(x)_\mm^{\alpha_1\dots \alpha_{2s}}$ are extracted as explained in \rf{splittingomega}. Also note that $\pC_\pm$ are the projected components of $\pC$ with respect to the projectors $\Pi_\pm$, that we have used in \rf{scalarunfld}. \\

The main prescription of Vasiliev-like theories (including Prokushkin--Vasiliev Theory) is to use part of the master equations to solve for the $z_\alpha$-dependent part of the master fields in terms of the physical sector. One then plugs these $z_\alpha$-on-shell forms into the dynamical master equations thereby extracting the physical equations of motion for the physical fields. As will be seen, the dynamical master equations are linear covariant constancy conditions in the full $y_\alpha$ and $z_\alpha$ space, and plugging the master fields with their $z_\alpha$-part solved for therein is really what produces interactions. The Prokushkin--Vasiliev master equations read as follows:\footnote{\label{myfootinyourass}For the original Prokushkin--Vasiliev theory the master equations formally read the same as Equations \rf{threedVasiliev}, although for master fields $\PVW$, $\PVB$ and $\PVS_\alpha$ which depend on two extra Clifford-like elements $\rho$ and $k$, and the Kleinian $\varkappa$ in \rf{threedVasiliev} is replaced by $k\varkappa$. However we may project out these two extra elements by declaring $\PVW= \MW(y,z,\phi,\psi)$, $\PVB=\MB(y,z,\phi,\psi)$ and $\PVS_\ga=\rho\, \MS_\ga(y,z,\phi,\psi)$, yielding the Vasiliev theory \cite{Vasiliev:1992ix}. Then sitting at $\lambda = \tfrac12$ ($\nu = 0$) corresponds to the theory we study, and which we keep naming Prokushkin--Vasiliev Theory although it is really a truncation thereof \cite{Vasiliev:1992gr,Prokushkin:1998bq}.}
\allowdisplaybreaks{\besubeqs\label{threedVasiliev}
\begin{align}
\d \MW&=\MW \wedge\star \MW\,,  \\
\d \MB \star\varkappa&=[\MW, \MB \star\varkappa]_\star\,,\\
\d \MS_\ga&= [\MW,\MS_{\ga}]_\star\,,\\
0&=\{\MB \star\varkappa, \MS_\ga\}_\star \, ,\label{threedVasilievBS} \\
[\MS_\ga,\MS_\gb]_\star &= - 2i \epsilon_{\ga \gb} (1 + \MB \star \varkappa) \label{threedVasilievSS}\,,
\end{align}
\esubeqs}\noindent 
where the last three equations above are those that allow one to solve for the $z_\alpha$-dependent part of the master fields and the first two will then generate the physical equations for $\omega$, $e$ and $\pC$ (and for the twisted sector as well). Here above the Kleinian $\varkappa \equiv \exp{(i y^\ga z_\ga)}$ was introduced, which has the properties $\varkappa \star \varkappa = 1$ and 
\begin{align}
\varkappa \star \mathbf{f}(y,z) = \mathbf{f}(-y,-z) \star \varkappa \label{kleinian} \,.
\end{align}
The above Prokushkin--Vasiliev equations are invariant under the following gauge transformations parametrized by the master gauge parameter $\boldsymbol{\xi} = \boldsymbol{\xi}(y,z,\psi,\phi|x)$:
\besubeqs\label{gaugevariation}
\begin{align}
\delta \MW &=\d\boldsymbol{\xi}-[\MW,\boldsymbol{\xi}]_\star\,,\\
\delta \MB \star\varkappa &=[\boldsymbol{\xi},\MB \star\varkappa]_\star\,,\\
\delta \MS_\ga &= [\boldsymbol{\xi},\MS_\ga]_\star \label{Sgaugevariation}\,.
\end{align}
\esubeqs
Let us note that the meaning of the last two master equations above is perhaps more easily understood when they are rewritten in the following manner \cite{Alkalaev:2014nsa}: 
\besubeqs
\begin{align}
\d \MW&=\MW \wedge \star \MW\,, \\
\d \mathbf{T}_{\ga\gb}&=[\MW,\mathbf{T}_{\ga\gb}]_\star\,,\\
\d \MS_\ga&= [\MW,\MS_{\ga}]_\star\,,\\
\tfrac{i}4\{\MS_\ga, \MS_\gb\}_\star&=\mathbf{T}_{\ga\gb}\label{DefT}\,,\\
[\mathbf{T}_{\ga\gb}, \MS_{\gc}]_\star&=\MS_\ga \epsilon_{\gb\gc}+\MS_\gb \epsilon_{\ga\gc} \label{TSCom} \,.
\end{align}
\esubeqs
Here (\ref{DefT}) defines the zero-from $\mathbf{T}_{\ga \gb}$ that together with $\MS_\ga$ constitutes the five generators of $\textrm{osp}(1|2)$, two odd plus three even ones, as can be seen by inspecting (\ref{DefT}) and (\ref{TSCom}) which are the defining relations of the $\textrm{osp}(1|2)$ algebra. 
One then recovers the system (\ref{threedVasiliev}) by setting
$\MB \star \varkappa = -\frac{i}{2} (\MS^\ga \star \MS_\ga + 1)$. 

Before moving to perturbation theory of the above master equations we should point out that the original Prokushkin--Vasiliev system of equations is more general \cite{Prokushkin:1998bq}, and the theory that we are interested in and which is introduced hereabove is really a (consistent) truncation thereof. As explained in Section~\ref{sec:HSalgebra}, the original Prokushkin--Vasiliev theory is really a one-parameter family of theories, each of them based on the algebra $\textrm{hs}(\lambda)$. The parameter $\lambda$ is then related to the vacuum value of $\sC$, which is denoted by $\nu$. The truncation of interest to us is that which corresponds to setting $\nu = 0$. In some sense this theory is technically simpler, since we can make use of the explicit realization of the star product \rf{eq:starproduct}, which we cannot do for generic values of $\lambda$ (or $\nu$). Also, with respect to the original theory proposed in \cite{Prokushkin:1998bq} we address the so-called reduced version thereof (see footnote~\textsuperscript{\ref{myfootinyourass}}), although we expect it to have features similar to those of the more general theory. 

\subsection{Vacuum Values and $\boldsymbol{z}_\mathbf{\alpha}$-dependence} 

The above Prokushkin--Vasiliev master equations are background independent. However, we will be interested in perturbative field excitations propagating on the (pure) $\textrm{AdS}_3$ vacuum solution. The vacuum of Section~\ref{sec:general} is given by
\begin{align}\label{AdSconn}
\Omega&=\frac12\varpi^{\ga\ga} L^\textsc{y}_{\ga\ga} +\frac12 \phi h^{\ga\ga} L^\textsc{y}_{\ga\ga}\, \qquad \text{with} \; \; L^\textsc{y}_{\ga\ga}=-\frac{i}4 \{ y_\ga,y_\ga \}_\star\,.
\end{align}
In order to start the perturbative analysis we further need to define the background values for the two other master fields. As we are interested in pure AdS$_3$ we choose the following \cite{Prokushkin:1998bq}:
\begin{align}
\MW=\Omega \,,\qquad \MB=0 \,, \qquad \MS_\ga=z_\ga \,, \label{vaccum}
\end{align}
and it can be easily checked that (\ref{threedVasiliev}) are satisfied using  $[z_\ga, \bullet]_\star=-2i \pl^z_\ga\bullet$. We further stress that $\MW$ is taken to be equal to the above $\Omega$ at zeroth-order, so that there are no higher spins turned on, and the scalar field is set to zero, so that we work on a vacuum with no matter. Lastly, the auxiliary master field $\MS_\alpha$ takes the simplest non-zero form. In particular this form is consistent with the fact that the components multiplying $z_\alpha$ oscillators are auxiliary as $\MS_\alpha$ is to be purely auxiliary. Note also that, as explained in Section~\ref{sec:HSalgebra} we work at $\nu=0$ which is why the twisted scalar field has a zero vacuum value. 

For practical purposes we will rewrite the master equations in terms of new master fields, shifted by their background values as
\begin{align}
\label{shiftMasterField}
\MS_\ga&\rightarrow z_\ga+2i \MA_\ga\,, & \MW&\rightarrow \Omega+\MW\,, & \MB&\rightarrow 2i\MB\,,
\end{align}
where the extra factors of $2i$ are included for convenience. We shall be working in the bosonic theory, which is implemented by declaring $\MW$ and $\MB$ to be of even degree in the total number of $y_\alpha$ and $z_\alpha$ oscillators, while $\MS_\alpha$ is taken to be of odd degree. This is consistent with the aforementioned background values. In terms of the Kleinian operator $\varkappa$ of \rf{kleinian}, the bosonic projection can be rephrased as follows: \begin{equation}
\varkappa\star\MB\star \varkappa=\MB\,, \qquad \varkappa\star \MW\star \varkappa=\MW\,, \qquad \varkappa\star \MA_\ga\star \varkappa=-\MA_\ga\,.
\end{equation} 
The new, background-shifted and bosonic master equations now take the form
\besubeqs\label{VasilievEqAB}
\begin{align}
\DO \MW&=\MW\wedge \star \MW\,,  \label{VasilievEqBA}\\
\DO \MB&=[\MW, \MB]_\star\,,\label{VasilievEqBB}\\
\pl^z_\ga \MW&= \DO \MA_{\ga}-[\MW,\MA_{\ga}]_\star\,,\label{VasilievEqBC}\\
\pl^z_\ga \MB&=[\MA_\ga,\MB]_\star\,,\label{VasilievEqBD}\\
\pl^z_\ga \MA^{\ga}&=\MA_{\ga}\star\MA^{\ga}+\MB\star \varkappa\,,\label{VasilievEqBE}
\end{align}
\esubeqs
where we are using the $\textrm{AdS}_3$ covariant derivative of \rf{covDerNonLLTDef}. 

The prescription for extracting the physical equations of motion from the above master equations is now as follows: one solves the last three master equations for the $z_\alpha$-dependent part of the three master fields in terms of their physical, $z_\alpha$-independent parts (for $\MS_\alpha$, which will be seen to be proportional to $z_\alpha$, we solve in terms of the physical components of the other master fields). More precisely, we obtain such $z_\alpha$-on-shell forms by making use of the following integration formulas:
\besubeqs
\label{homotopyIntegrals}
\begin{align}
\pl^z_\ga \mathbf{f}^\ga(z,y)=\boldsymbol{g}(z,y)  \hspace*{6pt}\; &\rightarrow \; \mathbf{f}_\ga(z,y) = \partial^z_\ga \boldsymbol{\epsilon}(z,y) + z_\ga \homo{1}{\boldsymbol{g}(z,y)} \, , \label{homotopycontracted}\\
\pl^z_\ga \mathbf{f}(z,y)=\boldsymbol{g}_\ga(z,y) \; &\rightarrow \;\hspace*{5pt} \mathbf{f}(z,y)=\epsilon(y) + z^\ga \homo{0}{\boldsymbol{g}_\ga(z,y)} \,,
\end{align}
\esubeqs
where the homotopy integrals are defined as \footnote{For $n\neq m$ the nested homotopy integrals can be resolved as $\Gamma_n\circ \Gamma_m=-(\Gamma_n-\Gamma_m)/(n-m)$. For $n=m$ one needs $\int \d t\, t^n \log t$, etc.}
\begin{align}
\homo{n}{f(z)}=\int_0^1  \d t\,t^n\, f(zt)\, .
\end{align}
One thereby obtains
\besubeqs
\label{pert}
\begin{align}\label{pertAA}
\MB&=\B(y,\phi,\psi)+z^\ga \homo{0}{[\MA_\ga,\MB]_\star}\,,\\
\MA_{\ga}&=\pl^z_\ga \boldsymbol{\epsilon}(y,z,\phi,\psi)+z_\ga \homo{1}{\MA_{\nu}\star\MA^{\nu}+\MB\star \varkappa}\,,\label{pertB}\\
\MW&=\W(y,\phi,\psi)+ z^\ga\homo{0}{\DO \MA_{\ga}-[\MW,\MA_{\ga}]_\star}\,.\label{pertC}
\end{align}
\esubeqs 
The 'initial data' $\B(y,\phi,\psi)$ and $\W(y,\phi,\psi)$ are the physical, $z_\alpha$-independent fields, encoding the higher-spin gauge connections and dreibeins as well as the scalar fields to all orders, as described more precisely in the previous subsection. The arbitrary function $\boldsymbol{\epsilon}$ is commented on below. Now, upon plugging the above $z_\alpha$-on-shell forms into the first two master equations (\ref{VasilievEqBA},\ref{VasilievEqBB}) one can, without loss of generality, evaluate them at $z=0$, yielding
\begin{align}
\label{dynEoMs}
\DO \MW|_{z=0}&=\left.\MW\wedge \star \MW\right|_{z=0}\,, \\
\DO \MB|_{z=0}&=\left.[\MW, \MB]_\star\right|_{z=0}\,,
\end{align}
and in the rest of this work we will always assume the equations to be evaluated at $z=0$, even when not explicitly stated. The reason one can take the above equations at $z_\alpha = 0$ is simply that, as one can prove, once we have plugged the solutions \rf{pert} therein these equations no longer depend on $z_\alpha$. This fact is non-trivial, and for its proof we refer to \cite{Vasiliev:1999ba,Sezgin:2002ru,Didenko:2014dwa}. Once we know the equations are $z_\alpha$-independent, putting $z_\alpha$ to zero is not a loss of generality, but makes the following computations easier as we can neglect terms that otherwise would have canceled each other in non-trivial ways. 

In Section~\ref{subsec:lorentzcovariantperturbation} we expand these two equations order by order in perturbation theory, thereby extracting physically meaningful equations of motion for the first-order $\pC(y,\phi)$ and $\pomega(y,\phi)$ and then taking the procedure to order~$2$, describing $\pC^{(2)}$ and $\pomega^{(2)}$. Let us note that such a procedure will also yield equations of motion for the twisted sector, formed by $\sC$ and $\somega$ as well as for their second-order versions $\sC^{(2)}$ and $\somega^{(2)}$. However, before proceeding with perturbation theory there is one more step to perform, which is related to Lorentz invariance and to the arbitrary initial data $\pl^z_\ga\boldsymbol{\epsilon}(y,z,\phi,\psi)$ found in \rf{pertB}. This is discussed in the following subsection. 

\subsection{Lorentz Invariance in The Schwinger--Fock Gauge}
\label{sec:LorentzCovariance}

In the previous subsection we have explained how the physical equations of motion are obtained by solving for the $z_\alpha$-dependence of the master fields as in \rf{pert} and then plugging the obtained expressions into the first two master equations \rf{dynEoMs} (and further evaluating at $z_\alpha = 0$). However, in solving for the master field $\MS_\alpha$ we find (the derivative of) an arbitrary function $\pl^z_\ga \boldsymbol{\epsilon}(y,z,\phi,\psi)$ in the solution, as is displayed in \rf{pertB}. This master field, however, should be kept completely auxiliary, that is, completely determined in terms of the other fields of the theory. The usual way of removing the arbitrariness in $\boldsymbol{\epsilon}(y,z,\phi,\psi)$ is to impose the Schwinger--Fock gauge:
\begin{equation}
\label{SFgauge}
z^\alpha \MS_\alpha = 0\,. 
\end{equation}
As is easy to check, this gauge choice implies $\boldsymbol{\epsilon}(y,z,\phi,\psi) = \boldsymbol{\epsilon}(y,\phi,\psi)$, and hence the first term in the right-hand side of \rf{pertB} vanishes identically.\footnote{Indeed, given that the second term in the right-hand side of \rf{pertB} is proportional to $z_\alpha$, the gauge  $z^\alpha \MS_\alpha = 0$ implies $z^\alpha \pl^z_\ga \boldsymbol{\epsilon}(y,z,\phi,\psi) = 0$, and noticing that $z^\alpha \pl^z_\ga$ is the number-of-oscillators operator in $z_\alpha$-space, we conclude that $\boldsymbol{\epsilon}$ cannot depend on $z_\alpha$~---~unless it is non-analytic in $z_\alpha$.} Evidently, going to such a gauge leaves one with only a subset of the original gauge transformations. As we will see below, at order~$1$ in perturbation theory the residual gauge parameters are simply the $\boldsymbol{\xi}$'s of \rf{Sgaugevariation} which are independent of $z_\alpha$. At higher orders the $z_\alpha$-dependent part of $\boldsymbol{\xi}$ will be non-zero but expressed in terms of the $z_\ga$-independent components thereof. Differently put, there is a gauge freedom in the solution for $\MS_\alpha$ and one chooses to fix the gauge~---~leaving unaffected the part of the gauge freedom endowing the physical fields, that is the $z_\alpha$-independent part of $\boldsymbol{\xi}$. As we will be working in the above Schwinger--Fock gauge for the master field $\MS_\alpha$, expression \rf{pertB} becomes
\begin{equation}
\label{pertBbis}
\MA_{\ga}=z_\ga \homo{1}{\MA_{\nu}\star\MA^{\nu}+\MB\star \varkappa}\,.
\end{equation}

The issue with Lorentz invariance is now that the generators $L^\textsc{y}_{\ga\ga}$ of the original $\star$-product or also their naive extension to the $y_\alpha,z_\alpha$ Weyl algebra $L^\textsc{yz}_{\ga \ga}=-\tfrac{i}{2}(y_\ga y_\ga -z_\ga z_\ga)$ do not preserve the above condition, i.e. $z^\alpha \delta_\Lambda \MS_\alpha \neq 0$, where $\delta_\Lambda \MS_\alpha$ is the gauge variation of $\MS_\alpha$ from \rf{Sgaugevariation} with $\Lambda = \tfrac{1}{2} \Lambda^{\ga\ga} L^{\textsc{yz}}_{\ga\ga}$ (explicit computations can be found in Appendix~\ref{app:llt}). One then concludes that, in this gauge, neither $L^\textsc{y}_{\ga\ga}$ nor $L^{\textsc{yz}}_{\ga\ga}$ provide us with a proper realization of the Lorentz generators on all the master fields present in Prokushkin--Vasiliev's equations. One might be tempted to instead conclude that there is a tension between the Schwinger--Fock gauge and Lorentz invariance. However, as we will see below one can identify other, field-dependent generators that realize the Lorentz symmetry.

Any proper set of Lorentz generators should satisfy the following requirements: (i) they ought to transform all fields covariantly and the corresponding gauge variations of the fields should close to the Lorentz algebra,\footnote{We will, however, allow for generators which realize the algebra \emph{only} with respect to the fields' variations, i.e. we allow for algebroids.} (ii) they have to preserve the Schwinger--Fock gauge. Fortunately, one can find generators which satisfy both of them, and they read \cite{Vasiliev:1997ak}:
\begin{equation}
\label{correctedlgen}
 L^\textsc{s}_{\alpha\beta} \equiv L^\textsc{yz}_{\alpha\beta}  -\frac{i}{4}\{\MS_\alpha,\MS_\beta\}_\star= L^\textsc{yz}_{\alpha\beta}-\mathbf{T}_{\ga\gb}\,.
\end{equation}
Using \rf{threedVasilievBS} and \rf{threedVasilievSS} one proves straightforwardly that $z^\alpha \delta_\Lambda \MS_\alpha = 0$ for $\Lambda = \tfrac{1}{2}\Lambda^{\alpha\beta}  L^\textsc{s}_{\alpha\beta} \equiv \Lambda^\textsc{s}$. As for the closure of the algebra, there is a subtlety: the above generators do not close to the Lorentz algebra per se. The Lorentz algebra is only recovered when computing commutators of the fields' gauge variations, while the above commutators obey
\begin{equation}
\label{Lorentznonclosure}
[ L^\textsc{s}_{\ga\ga}, L^\textsc{s}_{\gb\gb}]_\star=\epsilon_{\ga\gb} L^\textsc{s}_{\ga\gb}+\frac{\delta  L^\textsc{s}_{\ga\ga}}{\delta \B}[L^\textsc{y}_{\gb\gb},\B]_\star-\frac{\delta  L^\textsc{s}_{\gb\gb}}{\delta \B}[L^\textsc{y}_{\ga\ga},\B]_\star\,.
\end{equation}
The above issues are presented in further detail in Appendix~\ref{app:llt}. 

It might be helpful to point out that, at zeroth order in perturbation theory $\MS_\alpha = z_\alpha$ and hence $ L^\textsc{s} = L^\textsc{y}$. This fits nicely with the fact that for $\MS_\alpha = z_\alpha$ the condition $z^\alpha \MS_\alpha = 0$ is trivially satisfied and preserved under the `naive' Lorentz generators $L^\textsc{y}$. Note that this also implies that at zeroth order $\MW$ indeed is the chosen background $\Omega$ as in \rf{vaccum}. At first and higher orders the expression \rf{correctedlgen} acquires a dependence on the auxiliary field $\MS_\alpha$, and the correct Lorentz generators are no longer the naive ones. \\

Finding the correct Lorentz generators \rf{correctedlgen} is not the end of the story: we need to define the spin-connection accordingly\,! Indeed, the spin-connection is naturally defined to be the coefficient of the Lorentz generators in $\MW$, and the spin-connection thus enters via the following equation: 
\begin{equation}
\label{gaugeconnectionredef}
\MW \equiv \frac{1}{2}\omega^{\alpha\beta} L^{s}_{\alpha\beta} + \mathsf{W}\,,
\end{equation}
where $\mathsf{W}$ is assumed to be independent of $\omega^{\alpha\beta}$. In terms of this (correct) spin-connection the perturbation theory looks a little different from that obtained by (wrongly) declaring the coefficient of $L^\textsc{y}$ to be the spin-connection. This relabeling amounts to a (pseudo-local) field redefinition from the point of view of the physical theory in terms of $\pC$ and $\pomega$. The above is the correct object to be called a spin-connection when in the Schwinger--Fock gauge.

Our last point before considering the perturbative analysis in the next subsections is to comment on the Lorentz-transformation rules of the fields, now with respect to the corrected generators. By spin-connection we mean the one defined by \rf{gaugeconnectionredef}. Under a local Lorentz transformation, in the Schwinger--Fock gauge the master fields are rotated as follows:
\besubeqs
\label{covariantrotations}
\begin{align}
\delta \Big(\mathsf{W}+\frac{1}{2}\omega^{\ga\ga}L^\textsc{s}_{\ga\ga}(\B)\Big)&=\frac12\left(\d\Lambda^{\ga\ga} -\omega\fud{\ga}{\nu}\Lambda^{\nu \ga}\right)L^\textsc{s}_{\ga\ga}-[\mathsf{W},\frac12\Lambda^{\ga\ga}L^\textsc{yz}_{\ga\ga}]_\star\,,\label{newgaugevariation}\\
\delta \MB\star \varkappa&=\frac12\Lambda^{\gb\gb}[L^\textsc{s}_{\gb\gb},\MB\star \varkappa]_\star=\frac12\Lambda^{\gb\gb}\frac{\delta{\MB\star\varkappa}}{\delta \B}[L^\textsc{y}_{\gb\gb},\B]_\star\,,\\
\delta \MS_\ga&=\frac12\Lambda^{\gb\gb}[L^\textsc{s}_{\gb\gb},\MS_\alpha]_\star=\frac12\Lambda^{\gb\gb}\frac{\delta{\MS_\ga}}{\delta \B}[L^\textsc{y}_{\gb\gb},\B]_\star\,,
\end{align}
\esubeqs
where we assumed that $\MB$ and $\MS_\ga$ are expressed in terms of $\B$ according to \rf{pertAA} and \rf{pertBbis}, and the left-hand side of the first line is understood as
\begin{align}
\delta \Big(\mathsf{W}+\frac{1}{2}\omega^{\ga\ga}L^{\textsc{s}}_{\ga\ga}(\B)\Big)&=\delta\mathsf{W}+
\frac{1}{2}\delta\omega^{\ga\ga}L^{\textsc{s}}_{\ga\ga}(\B)+\frac{1}{2}\omega^{\ga\ga}\delta L^{\textsc{s}}_{\ga\ga}(\B) \,.
\end{align}
As anticipated, the tensors are rotated covariantly by the Schwinger--Fock Lorentz generators, even though the generators themselves do not close to the Lorentz algebra~---~see \rf{Lorentznonclosure}. Hence, requirement (i) announced at the beginning of this subsection is fulfilled. Let us point out, finally, that in order to derive the above transformation rules one crucially uses two facts: for the last two lines one uses the (anti-)commutation relations of $\MS_\alpha$ with itself and with $\MB$~---~see \rf{VasilievEqAB}~---~, while for the first line it was sine qua non to correctly identify the spin-connection as in \rf{gaugeconnectionredef} and to use the Lorentz algebra. Note, also, that the above transformation rules for $\MW$ and $\MB$ are the same we would obtain from \rf{gaugevariation} upon using $\boldsymbol{\xi} = \tfrac{1}{2} \Lambda^{\alpha\alpha}L^\textsc{yz}_{\alpha\alpha}$, except for the fact that the spin-connection (respectively its variation) is contracted with $L^\textsc{s}_{\alpha\alpha}$ on the left-hand side (respectively right-hand side) of \rf{newgaugevariation}, not with $L^\textsc{yz}_{\alpha\alpha}$. Again, some more details about the issue of Lorentz invariance in the Schwinger--Fock gauge can be found in Appendix~\ref{app:llt}.

\subsection{Manifest Lorentz-Covariant Perturbation Theory}
\label{subsec:lorentzcovariantperturbation}

Having identified the correct Lorentz generators (\ref{correctedlgen}) to be used in the Schwinger--Fock gauge \rf{SFgauge}, in this subsection we develop a manifestly Lorentz-covariant perturbative expansion of (\ref{threedVasiliev}). 
We want to perform a redefinition of $\MW$ for the practical purpose of making manifest Lorentz covariance with respect to the background. We do so in a two-step fashion. First we perform the following redefinition of the master gauge connection which makes manifest the covariance with respect to the spin-$2$ sector, as well as removing the vielbein $e$ from $\MW$:
\begin{equation}
\label{manifestredefgeneral}
\MW\quad\rightarrow\quad\MW + \frac{1}{2}\omega^{\ga\ga}L^{\textsc{s}}_{\ga\ga}+\frac{1}{2}\,\phi\,e^{\ga\ga} L^{\textsc{Y}}_{\ga\ga} \equiv \MW + \omega + e \,.
\end{equation}
For notational convenience in the above right-hand side, the new, redefined $\MW$ will still be denoted as $\MW$. 
In terms of that new $\MW$, and with all other fields shifted as in \eqref{shiftMasterField}, the master equations (\ref{threedVasiliev}) read as follows:
\besubeqs\label{VasilievEqABP}
\begin{align}
\mathsf{D}^{\textsc{yz}}\MW&=\MW\wedge\star \MW-\frac12 R^{\ga\ga}  L^\textsc{s}_{\ga\ga}-\frac12 \phi T^{\ga\ga} L^\textsc{y}_{\ga\ga}-\frac{i}8 e\fud{\ga}{\nu}\wedge e^{\nu\ga} \{\MS_\ga,\MS_\ga\}_\star\,,\label{VasilievXEqBA}\\
\mathsf{D}^{\textsc{yz}}\MB&=[\MW,\MB]_\star\,,\label{VasilievXEqBB}\\
\pl^z_\ga \MW&=-[e+\MW,\MA_\ga]_\star+\frac{\delta\MA_\ga}{\delta\MB}[e+\MW,\MB]_\star
\label{VasilievXEqBC}\,,\\
\pl^z_\ga \MB&=[\MA_\ga,\MB]_\star\,,\label{VasilievXEqBD}\\
\pl^z_\ga \MA^{\ga}&=\MA_{\ga}\star\MA^{\ga}+\MB\star \varkappa\,,\label{VasilievXEqBE}
\end{align}
\esubeqs
where we have introduced the curvature tensor $R$ and Torsion $T$,
\begin{equation}
\label{RandT}
R^{\ga\ga}\equiv\d\omega^{\ga\ga}-\omega\fud{\ga}{\nu}\wedge \omega^{\nu\ga}-e\fud{\ga}{\nu}\wedge e^{\nu\ga}\,,\qquad T^{\ga\ga}\equiv\d e^{\ga\ga}-2\omega\fud{\ga}{\nu}\wedge e^{\nu\ga}\,,
\end{equation} 
and the new covariant derivative is given by
\begin{align}
\label{eq:covDev}
\mathsf{D}^{\textsc{yz}}&=\d\bullet-\frac12\omega^{\ga\ga}[L^{\textsc{yz}}_{\ga\ga},\bullet]_\star-\frac12e^{\ga\ga}[\phi L^{\textsc{y}}_{\ga\ga},\bullet]_\star\,,
\end{align}
in which the difference with (\ref{covDerNonLLTDef}) is in that the spin-connection is now contracted with $L^\textsc{yz}_{\ga\ga}$ instead of $L^\textsc{y}_{\ga\ga}$. It is important to note that in the first master equation here above we have dropped a term proportional to $z_\alpha$, which we know will not contribute to the corresponding $z_\ga$-independent equation, for by definition the latter is obtained by evaluating the master equation at $z_\alpha = 0$. \footnote{Evidently, terms proportional to $z_\alpha$ appear in other master equations in \rf{VasilievEqABP}, e.g. in the third one \rf{VasilievXEqBC}. However one should remember that such is not one of the master equations that will yield a physical equation of motion. Rather, as we already noted the last three master equations allow one to solve for the $z_\alpha$-dependence of the three master fields, and only the first two master equations are to be evaluated at $z_\alpha = 0$~---~after plugging therein the master fields with their $z_\alpha$-dependence solved for.}

It is now evident that in the above master equations there are no spin-connections appearing outside of covariant derivatives. Hence the master equations are manifestly Lorentz covariant when we correctly identify the spin-connection as in \rf{gaugeconnectionredef}, and the 'price to pay' for making that property manifest is to have some extra terms in the equations \rf{VasilievXEqBA}, \rf{VasilievXEqBC}.\\

We are close to being able to formulate the perturbative master equations. Our last step is the following: the redefinition \rf{manifestredefgeneral} makes manifest Lorentz covariance with respect to the whole spin-$2$ sector. However, as we are interested in perturbation theory we choose to make it manifest with respect to the background only, that is, we set $\omega$ and $e$ in \rf{manifestredefgeneral} to $\varpi=\frac{1}{2}\varpi^{\ga\ga}L^{\textsc{s}}_{\ga\ga}$ and $h=\frac{1}{2}\,\phi\,h^{\ga\ga} L^{\textsc{Y}}_{\ga\ga}$, the $\textrm{AdS}_3$ background spin-connection and dreibein. Because the curvature $R$ and torsion $T$ of \rf{RandT} vanish for this background the resulting equations read
\besubeqs
\label{newDynMasterEoM}
\begin{align}\label{newWeq}
\bar{\mathsf{D}}^{\textsc{yz}}\MW &=\MW\wedge\star \MW-\frac{i}8 h\fud{\ga}{\nu}\wedge h^{\nu\ga} \{\mathbf S_\ga,\mathbf S_\ga\}_\star\,,\\
\bar{\mathsf{D}}^{\textsc{yz}}\MB &=
[\MW,\MB]_\star\,,\label{newBeq}
\end{align}
\esubeqs
and all other equations in (\ref{VasilievEqABP}) remain unchanged. Note that we have introduced the background version of the covariant derivative $\mathsf{D}^{\textsc{yz}}$, that is
\begin{align}
\label{covDerivativeLLTCov}
\bar{\mathsf{D}}^{\textsc{yz}}&=\d\bullet-\frac12\varpi^{\ga\ga}[L^{\textsc{yz}}_{\ga\ga},\bullet]_\star-\frac12h^{\ga\ga}[\phi L^{\textsc{y}}_{\ga\ga},\bullet]_\star\,.
\end{align}
Now using (\ref{homotopyIntegrals}) we can again determine the $z_\alpha$-dependence of the master fields by integrating (\ref{VasilievXEqBC})-(\ref{VasilievXEqBE}), which leads to a slightly different result than (\ref{pert}):
\besubeqs
\label{zdepenMasterField}
\begin{align}
\MB&=\B(y,\phi,\psi)+z^\ga \homo{0}{[\MA_\ga,\MB]_\star}\,,\\
\MA_{\ga}&=z_\ga \homo{1}{\MA_{\nu}\star\MA^{\nu}+\MB\star \varkappa}\,,\\
\MW&=\W(y,\phi,\psi)- z^\nu\homo{0}{[h, \MA_\nu]_\star+[\MW,\MA_{\nu}]_\star}\,,
\end{align}
\esubeqs
where this result already takes into account the  Schwinger--Fock gauge \rf{SFgauge}, which is why the last term in \eqref{VasilievXEqBC} has been dropped, on the account that $z^\nu\homo{0}{\frac{\delta\MA_\nu}{\delta\B}\dots}=0$. Note that in the $z_\alpha$-dependent part of the last equation hereabove the background spin-connection is not present. Indeed such a term would break manifest Lorentz covariance and would arise if we had not identified the Lorentz generators correctly in the Schwinger--Fock gauge, as can be seen by comparing with (\ref{pert}).

\subsection{Order-1 Perturbations}
\label{sec:Firstorder}

As discussed earlier, one of the goals of the present work is to explore the backreactions on the different fields of the Prokushkin--Vasiliev theory at order $2$ in perturbation theory. However, in order to do so we found it was needed to carefuly analyze the first-order perturbation theory first. This will also provide a warm-up exercise in view of the next subsection. As we have explained already, the procedure is to plug the solutions \rf{zdepenMasterField} into the master equations \rf{newDynMasterEoM} and evaluate the result at $z_\alpha = 0$. Also, let us stress once again that the order of the operations plays a crucial role here: if one evaluates the expressions \rf{zdepenMasterField} at $z_\alpha = 0$ \emph{first} and \emph{then} plugs the result in \rf{newDynMasterEoM}, the dynamics is lost. The interactions come from the $z_\alpha$-dependence precisely\,!

At order $1$ it should be evident that the right-hand sides of \rf{newDynMasterEoM} are just zero, which simply stems from the fact that the fields start at order $1$ now, as they have been shifted by their background values (and those right-hand sides are quadratic in the master fields). We thus have $\bar{\mathsf{D}}^{\textsc{yz}} \MW^{(1)}=0 $, $ \bar{\mathsf{D}}^{\textsc{yz}} \MB^{(1)}=0 $, so that the physical first-order equations of motion read
\begin{align}
\label{linEoMMF}
\left.\bar{\mathsf{D}}^{\textsc{yz}} \MW^{(1)}\right|_{z=0}=0 \,, \qquad \left.\bar{\mathsf{D}}^{\textsc{yz}} \MB^{(1)}\right|_{z=0}=0 \,,
\end{align}
where it is implicit that the master fields now stand for the corresponding $z_\alpha$-on-shell forms of \rf{zdepenMasterField}. The first-order versions of \rf{zdepenMasterField} then are 
\besubeqs
\label{linZDepMasterFields}
\begin{align}
\MB^{(1)}&=\pC(y,\phi)\psi+\sC(y,\phi) \,,\label{MB1ZDep}\\
\MA^{(1)}_\ga&=z_\ga \int_0^1 \d t\, t\,\pC (-zt,\phi) e^{ity z} \, \psi+z_\ga \int_0^1 \d t \, t\,\sC(-zt,\phi)e^{ity z}\,,\label{MA1ZDep}\\
\MW^{(1)}&=\pomega(y,\phi)+\somega(y,\phi) \psi+\int_0^1 \d t \,(1-t)t \,\phi \, h^{\ga\ga} z_\ga z_\ga \sC(-zt,\phi)e^{ity z}+M_2 \psi \label{eq:MW1Z}\,,
\end{align}
\esubeqs
where $M_2$ is given by 
\begin{equation}
M_2=-\frac12 \int_0^1 \d t \, (1-t) \, \phi \, h^{\ga\ga}z_\ga e^{ity z} \left(y_\ga (1-t)-i(1+t)t^{-1}\pl^z_\ga\right)\pC(-zt,\phi) \label{eq:M2}\,.
\end{equation}
Note that we have split $\flC(y,\phi,\psi)$ and $\flomega(y,\phi,\psi)$ in their twisted and physical component (see Section~\ref{sec:general} as well as \ref{subsec:masterfieldsandeqs}). 
After some algebra, substituting \eqref{linZDepMasterFields} into \eqref{linEoMMF} yields the following result:
\besubeqs\label{firstordereq}
\begin{align}
\adD\pomega&=0\,,\\
\tadD\somega&= \tfrac18 H^{\ga\ga} (y_\ga+i\pl^w_\ga)(y_\ga+i\pl^w_\ga)\pC(w,\phi)|_{w=0}\,,\label{shadowA}\\
\adD\sC&=0\,,\\
\tadD \pC&=0\,,
\end{align}
\esubeqs
where $H^{\ga\ga} \equiv h^\alpha_{\;\nu}\wedge h^{\nu\alpha}$ and the physical-space covariant derivatives $\adD$ and $\tadD$ are defined in (\ref{covDer}). 

The equations for $\pomega$, $\pC$ and $\sC$ are exactly as in (\ref{shadoweq}) and (\ref{physeq}). The equation of motion \eqref{shadowA} for the twisted one-form $\somega$ is the one displayed in \rf{eq:firstOrderSource}. It differs from (\ref{shadoweq}) by a source term involving the physical scalar fields. As explained in Section~\ref{sec:twistedResultsLinearOrder} we wish to consider solutions of (\ref{firstordereq}) for which the twisted fields $\somega$ and $\sC$ are zero. As is further detailed in Section~\ref{sec:twistedResultsLinearOrder}, the above source term can be removed by performing the field redefinition $\somega (y,\phi|x) \rightarrow \somega(y,\phi|x) + M_1$. 
In other words we are finding a particular solution to the inhomogeneous first-order equation for $\somega$. Hence the equations of motion in the twisted sector are exactly given by \eqref{shadoweq} after performing this field redefinition,
i.e.
\begin{align}
\label{redeftwomegalin}
\tadD \somega=0\,.
\end{align}
Therefore we can consistently consider the trivial solution for the redefined fields \eqref{redeftwomegalin} and \eqref{shadowA}, that is, $\somega = 0 $, $ \sC = 0$, which we assume in the following. 

After having performed the field redefinition of $\somega$ by \rf{eq:M1} the one-form $\MW$ at linear order is modified: instead of \eqref{eq:MW1Z} it is now given by
\begin{align}
\label{MW1ZDep}
\MW^{(1)}&=\pomega(y,\phi)+M_1\psi+M_2 \psi \equiv \pomega(y,\phi) + M \psi \,,
\end{align}
where the solutions \eqref{eq:solTwLin} have been used to eliminate the terms involving twisted fields in the right-hand side of \rf{eq:MW1Z}. However the above field redefinition is not unique: as exposed in Section~\ref{sec:cohomology}, the generic zero-mode for the homogeneous equation \rf{redeftwomegalin} 
is parametrized by an arbitrary parameter $g_0$ (in bosonic theory). This means that the generic form of $\MW^{(1)}$ after performing the field redefinition that removes the source term of \rf{shadowA} is the following:
\begin{equation}
\MW^{(1)} = \pomega(y,\phi) + M \psi + R \psi \,.
\end{equation}
This fact will play a crucial role in the following subsection, where we address the second-order backreactions on the twisted fields. Recalling the results presented in Section~\ref{sec:PresentationBackreactionFronsdal}, the situation is that the twisted fields can be consistently set to zero at second-order only at a particular point in the parameter space describing the zero-mode \rf{eq:representativeofH}. 

\subsection{Order-2 Perturbations}

In spirit, the second-order analysis much resembles the first-order one: we solve for the $z_\alpha$-dependence of the second-order master fields, plug the result in the first two master equations and evaluate the latter at $z_\alpha = 0$, thereby obtaining the physical-space second-order equations of motion for the fields. As can be expected, however, the details are much more intricate, and as we will see here below at order~$2$ the Prokushkin--Vasiliev theory truly becomes non-trivial, namely the fields start to interact. However, as the computational procedure has been made clear in the previous subsection and we wish to keep the presentation concise we shall skip some specifics of the calculations and shall not display the obtained expressions explicitly. The latter are to be found in Appendix~\ref{subsec:fourierspace}. 

The $z_\alpha$-dependence of the second-order excitations is again easily computed from \eqref{zdepenMasterField} and is found to be
\besubeqs
\label{eq:ZDepM2nd}
\begin{align}
\MB^{(2)}&=\pC^{(2)}(y,\phi)\psi+\sC^{(2)}(y,\phi)+z^\ga \homo{0}{[\MA^{(1)}_\ga,\MB^{(1)}]_\star}\,,\\
\MA^{(2)}_{\ga}&=z_\ga \homo{1}{\MA^{(1)}_{\nu}\star\MA^{(1)\nu}}+z_\ga \homo{1}{\MB^{(2)}\star \varkappa}\,,\\
\MW^{(2)}&=\pomega^{(2)}(y,\phi)+\somega^{(2)}(y,\phi)\psi- z^\nu\homo{0}{[h, \MA^{(2)}_\nu]_\star+[\MW^{(1)},\MA^{(1)}_{\nu}]_\star}\,,
\end{align}
\esubeqs
where we split again the $z_\alpha$-independent parts of the master fields into their physical and twisted components. To obtain the physical equations of motion one now has to insert \eqref{eq:ZDepM2nd} as well as \eqref{MW1ZDep}, \eqref{MB1ZDep} and \eqref{MA1ZDep} into the first two master equations at second-order,
\besubeqs
\begin{align}\label{newWeq2}
\bar{\mathsf{D}}^{\textsc{yz}}\MW^{(2)}|_{z=0}&= (\MW^{(1)}\wedge\star \MW^{(1)}- i H^{\ga\ga} \MA_\ga^{(1)}\star\MA_\ga^{(1)})|_{z=0}\,,\\
\bar{\mathsf{D}}^{\textsc{yz}}\MB^{(2)}|_{z=0}&=
[\MW^{(1)},\MB^{(1)}]_\star|_{z=0}\,.\label{newBeqX}
\end{align}
\esubeqs
In the following the evaluation at $z_\alpha=0$, which is always meant after all star-products have been performed, will no longer be indicated explicitly. It is important to note that we will only consider the case where we have chosen vanishing solutions for the linear twisted fields as in \eqref{eq:solTwLin}. 

After some algebra the above equations are turned into
\begin{align}
\bar{\mathsf{D}}^{\textsc{yz}}\big( \pC^{(2)}\psi+\sC^{(2)} \big)&=-\mathsf{D}^{\textsc{yz}}z^\ga \homo{0}{[\MA_\ga^{(1)},\pC\psi]_\star}+[\pomega+M\psi,\pC\psi]_\star\,,\\
\bar{\mathsf{D}}^{\textsc{yz}}\left( \pomega^{(2)}+\somega^{(2)}\psi\right)&= \mathsf{D}^{\textsc{yz}}z^\nu\homo{0}{[h, \MA^{(2)}_\nu]_\star}+
\mathsf{D}^{\textsc{yz}}z^\nu\homo{0}{[\pomega+M\psi,\MA^{(1)}_{\nu}]_\star}\nonumber\\&+(\pomega+M\psi)\wedge\star (\pomega+M\psi)-i H^{\ga\ga}\MA^{(1)}_{\ga}\star\MA^{(1)}_\ga\,.
\end{align}
Splitting again these equations in their physical and twisted components we arrive at the following equations of motion:
\besubeqs
\label{eq:secOrderEoMUnfolded}
\begin{align}
(\tadD \pC^{(2)})\psi&=\mathcal{V}(\pomega,\pC)\,,\\
{D} \sC^{(2)}&=\widetilde{\mathcal{V}}(\Omega,\pC,\pC)\,,\\
(\tadD\somega^{(2)})\psi&=\widetilde{\mathcal{V}}(\Omega,\pomega,\pC)\,,\\
D\pomega^{(2)}&=\mathcal{V}(\pomega,\pomega)+\mathcal{V}(\Omega,\Omega,\pC,\pC)\label{eq:omegasecord}\,,
\end{align}
\esubeqs
with the physical cocycles found to be
\besubeqs
\label{eq:secOrderCocyclesPhys}
\begin{align}
\mathcal{V}(\Omega,\Omega,\pC,\pC)&=(M\psi) \wedge\star (M \psi)- i H^{\ga\ga} \MA_\ga^{(1)}\star\MA_\ga^{(1)}+\bar{\mathsf{D}}^{\textsc{yz}}z^\nu\homo{0}{[M \psi,\MA^{(1)}_{\nu}]_\star}+\label{eq:cocycleOOCC}\\
&+\bar{\mathsf{D}}^{\textsc{yz}}z^\ga\homo{0}{[h, z_\ga \homo{1}{\MA^{(1)}_{\nu}\star\MA^{(1)\nu}}]_\star}+\bar{\mathsf{D}}^{\textsc{yz}}z^\ga\homo{0}{[h, z_\ga \homo{1}{\MB^{(2)}\star \varkappa}]_\star}\,, \nonumber \\
\mathcal{V}(\pomega,\pomega)&=\pomega\wedge\star\pomega\,,\\
\mathcal{V}(\pomega,\pC)&=[\pomega,\pC\psi]_\star\,,
\end{align}
\esubeqs
and those pertaining to the twisted sector reading
\besubeqs
\label{eq:secOrderCocyclesTw}
\begin{align}
\widetilde{\mathcal{V}}(\Omega,\pC,\pC)&=-\bar{\mathsf{D}}^{\textsc{yz}} z^\nu \homo{0}{[\MA^{(1)}_\nu,\pC\psi]_\star}+[M\psi,\pC\psi]_\star\,,\\
\widetilde{\mathcal{V}}(\Omega,\pomega,\pC)&=\{\pomega, M\psi\}_\star+\bar{\mathsf{D}}^{\textsc{yz}}z^\nu \homo{0}{[\pomega,\MA^{(1)}_\nu]_\star}\,.
\end{align}
\esubeqs
Obtaining an explicitly $z_\alpha$-independent expression thereof is a task of considerable technical difficulty and we will outline the main techniques we used for performing this calculation in the next subsection. The final form of the various cocycles, with no $z_\alpha$'s involved anymore, is given in Section~\ref{sec:twistedSecondOrder} where we present the corresponding results and comment on them, whereas the explicit expressions for some of them are collected in Appendix~\ref{app:technicaldetailsBackreaction}.

\subsection{Explicit Evaluation of Cocycles}
\label{subsec:explicit}
As commented on at the end of the previous subsection, evaluating the cocycles displayed there is not an easy task. In order to do so we have developed some methods for computing, which we now illustrate on the following example:
\begin{equation}
\label{M2wedgeM1}
\left. (M_2 \psi) \wedge \star (M_1 \psi) \right|_{z=0} \,,
\end{equation}
which is found in \eqref{eq:cocycleOOCC}. Each of the $M_i$'s hereabove contains a scalar field $\pC$, and it turns out to be computationally advantageous to consider the Fourier transformations thereof, given by \eqref{eq:fourier}. We will furthermore adopt the convention that the wave vector of the first $\pC$ field is denoted by $\xi$ and that of the second field (for the above piece the one in $M_1$) by $\eta$. This is important as for each term in the cocycle \eqref{eq:cocycleOOCC} we will obtain an expression of the form
\begin{equation}
\int \d\xi \d\eta \, f(y,\xi,\eta) \, \pC(\xi,\phi|x) \psi \, \pC(\eta,\phi|x) \psi
=\int \d\xi \d\eta \, f(y,\xi,\eta) \, \pC(\xi,\phi|x) \, \pC(\eta,-\phi|x) \,,
\end{equation}
so that this convention amounts to associating a wave vector $\eta$ with the master field that comes with a flipped sign for $\phi$. 

Now using the bosonic version of \eqref{eq:M1} and \eqref{eq:M2} for $M_1$ and $M_2$ we can rewrite \eqref{M2wedgeM1} hereabove using the integral representation of the star product \eqref{eq:starproduct} as
\begin{align}
\label{eq:M2M1cont}
-\frac{1}{32\pi^2} h^{\ga \ga} &\wedge h^{\gb \gb} \left\{ \int \, \d t \, \d q \, \d\xi \, \d\eta \, \d^2 u \, \d^2v \; (1-t)(q^2-1) e^{i q  (y+v)  \eta-it y  (\xi-u) + i v u}\right.\\ & \left.  \times\, u_\ga [ (y+u)_\ga (1-t)-(1+t)\xi_\ga ] (y+v - \eta)_\gb (y + v - \eta)_\gb \vphantom{\int x}\right\} \pC(\xi,\phi|x) \, \pC(\eta,-\phi|x) \,. \nonumber
\end{align}
After shifting $u_\ga \rightarrow u_\ga - q \eta_\ga$  and $v_\ga \rightarrow v_\ga - t (y+\xi)_\ga$ the above expression becomes 
\begin{align}
-\frac{1}{32\pi^2} h^{\ga \ga} &\wedge h^{\gb \gb} \left\{ \int \, \d t \, \d q \, \d\xi \, \d\eta \, \d^2 u \, \d^2v \; (1-t)(q^2-1) R^2  e^{i v u} \right.\\ & \left.  \times\, (u-q\eta)_\ga [ (y_\ga+u_\ga-q\eta_\ga) (1-t)-(1+t)\xi_\ga ] \right. \nonumber \\&\left. \times\, (y_\gb+v-t(y+\xi)_\gb - \eta_\gb) (y_\gb + v_\gb - t(y+\xi)_\gb - \eta_\gb)\vphantom{\int x} \right\} \pC(\xi,\phi|x) \, \pC(\eta,-\phi|x) \,, \nonumber
\end{align}
where we have defined $R^2\equiv\exp i \left(q (y-t(\xi+y))\eta \right)$.
We can now evaluate the integrals over $u$ and $v$ by using the following identities:
\besubeqs
\begin{align}
&\frac{1}{(2\pi)^2} \int \d^2 u \, \d^2v \; e^{iv  u} = 1 \,, \\
&\frac{1}{(2\pi)^2} \int \d^2 u \, \d^2v \, e^{iv  u} \; u_\ga v_\gb = i \epsilon_{\ga \gb}\, ,
\end{align}
\esubeqs
whereas this type of integral vanishes if the number $u_\alpha$'s is different from the number of $v_\alpha$'s. Using these identities we arrive at our final result for \eqref{M2wedgeM1}, that is
\begin{align}
\int \d t \, \d q & \, \d^2\xi \, \d^2\eta \;  (q^2-1) R^2  \\ & \times\, \left\{
\frac{-i}{4} \,  H^{\ga \ga} \left(T^2_\ga S^2_\ga + q (1-t)^2 \eta_\ga S^2_\ga \right) +
\frac{1}{8} \, h^{\ga \ga} \wedge h^{\gb \gb} \;\left( q \; \eta_\ga T^2_\ga S^2_\gb  S^2_\gb \right) \right\} \pC(\xi,\phi|x) \, \pC(\eta,-\phi|x) \nonumber \,,
\end{align}
where we have denoted certain combinations of $y_\ga$, $\eta_\ga$ and $\xi_\ga$ by $S^2$ and $T^2$, whose definitions are found in \eqref{eq:shortcuts}. One could simplify this expression further by using the basic identity \rf{appeq:twoFormIdent} for two-forms, but evaluating the resulting expression is rather cumbersome and can be done most easily using a computer algebra program. The piece which we have explicitly evaluated in this subsection is part of the cocycle \eqref{eq:cocycleOOCC}, which is by far the hardest one to compute. In Appendix~\ref{subsec:fourierspace} we display, however, the simplest form we obtain for the whole of it.

\paragraph{Consistent truncation to the physical sector:} The above computation is an example of how to explicitly evaluate a piece on the right-hand side of equations of motion for the physical gauge connection. However, hereabove we do so in the field-frame corresponding to the redefinition of $\MW$ by \rf{eq:M1}. We now explain how to obtain the expression for the cocycle studied above in the field frame corresponding to using the redefinition $\tilde{M}_1$ at $\tilde{g}_0 = \tilde{d}_0 = 0$, that is, the frame for which the second-order twisted fields can be trivialized consistently. 

At $\tilde{g}_0 = \tilde{d}_0 = 0$ the redefinition $\tilde{M}_1$ is given by 
\begin{align}
\tilde{M}_1&=\frac{i}4 \phi h^{\ga\ga} \int_0^1 \d t \, (t^2-1) \, (y-\xi)_\ga (y-\xi)_\ga \, \sin{(ty\xi)} \, \pC (\xi,\phi|x) =\frac12\left(M_1-\left.M_1\right|_{t\rightarrow-t}\right)\,.
\end{align}
This form suggests that the expressions for the redefinition $\tilde{M}_1$ can be obtained by anti-symmetrizing over $t$, over $q$ or over both $t$ and $q$ for terms that are of the form $(M_1 \psi)  \star X$, $X \star (M_1 \psi)$ or $(M_1 \psi) \wedge\star (M_1 \psi)$ respectively. For the example \eqref{M2wedgeM1} we would therefore need to anti-symmetrize with respect to $q$. The resulting expression is a bit more intricate as it involves additional types of exponentials. It is therefore advantageous to calculate the various cocycles with respect to $M_1$ and then impose appropriate antisymmetrization.


\section{Discussion and Outlook}
\label{sec:conclusions}

Briefly put, our results are the following:
\begin{description}
\item[Twisted Fields:] the second-order Prokushkin--Vasiliev theory at $\lambda = \frac12$, with first-order twisted fields set to zero, possesses free real parameters. Only at one point in this parameter space can one consistently set all second-order twisted fields to zero without redefining the physical fields. 
\item[Physical Sector:] the backreactions on the second-order physical fields in this theory have been computed explicitly in the Schwinger--Fock gauge in a manifestly Lorentz-covariant manner, in particular at the point in parameter space mentioned hereabove.
\item[Cubic Action:] we have determined completely the cubic action describing the physical sector of the theory. The relative coupling constants $g_s$ parametrizing each spin-$s$ canonical current were fixed by solving the admissibility condition, which is part of the Noether procedure. 
\end{description}
\noindent
Along the way, we have also shed light on how to formulate perturbation theory in a manifestly local Lorentz-covariant way in the Schwinger--Fock gauge and have also systematically computed all cohomologies relevant for our second-order analysis. Hereafter we comment and expand on the above results. \\

Let us first comment on twisted fields. When truncating the theory to linear order in perturbation theory, it was known since the work of Vasiliev\cite{Vasiliev:1992ix} that one can indeed set these fields to zero after a field redefinition of the twisted gauge connection $\somega$. In Section~\ref{subsec:twistedresults} we establish that there is a two-parameter ambiguity in this field redefinition. The twisted gauge connection $\somega$ will enter the equations of motion when we expand the theory to second order in perturbation theory, as in \rf{eq:afterPsiCounting}. Changing the two parameters $\tilde{d}_0$ and $\tilde{g}_0$ in the redefinition \eqref{eq:M1PrimeFerm} thus modifies the twisted scalar's equation of motion accordingly. As for the second-order twisted gauge connection $\somega^{(2)}$ one first removes the backreaction depending on $\pC^{(2)}$ along the same lines as for the corresponding first-order equation. The remaining backreaction depends on two other parameters $\tilde{d}_1$ and $\tilde{g}_1$. The most general second-order theory is thus parametrized by four parameters (two in the bosonic case). As we explain in Section~\ref{subsec:twistedresults}, \emph{there is a unique point in parameter space where the second-order backreaction on the twisted scalar and twisted gauge connection can be removed by a pseudo-local field redefinition}. That point corresponds to the values $\tilde{d}_{0,1} = \tilde{g}_{0,1} = 0$, for the parameters defined in \rf{eq:M1prime} and \rf{eq:M1PrimeFermSecondOrder}. Only for this single point in parameter space there exists a truncation of Prokushkin--Vasiliev Theory to its physical sector to second order in perturbation theory. For any other set of values for the parameters the backreaction on either $\sC^{(2)}$ or $\somega^{(2)}$ is non-trivial in cohomology, and hence cannot be removed by any pseudo-local field redefinition. 

Let us stress once again that we analyzed Prokushkin--Vasiliev Theory only up to order~$2$ in perturbation theory, and we have nothing definite to say about higher-order perturbation theory. The field redefinitions which allow us to consistently set the second order twisted fields to zero are not unique as our cohomological analysis of Appendix~\ref{app:cohomologies} shows. In fact they form an infinite-dimensional parameter space of possible field redefinitions, whereas at order~$1$ the redefinitions of $\somega$ form a two-dimensional parameter space. However this infinite set of parameters presumably boils down to only one independent parameter by higher-spin covariance. Beyond this we can only say that two scenarios are possible: it could be that at order~$3$ the truncation to the physical sector leaves us with free parameters or not. 

One should also note that there exists a so-called \emph{non-local integration flow} for the Prokushkin--Vasiliev Theory \cite{Prokushkin:1998bq,Prokushkin:1998vn}. In brief, this is a pseudo-local, non-perturbative field redefinition that maps the original theory to a free one. In particular it achieves the decoupling of the twisted fields at all orders. In comparison with our findings, this means that the first-order part of the integrating flow's field redefinition on $\somega$ corresponds to the redefinition \eqref{eq:M1PrimeFerm} of $\somega$ at $\tilde{d}_0 = \tilde{g}_0 = 0$. However the integration flow also leads to free equations in the physical sector. This is compatible with the fact that, as we prove in Appendix~\ref{app:cohomologies}, the relevant cohomology is trivial and therefore any backreaction to the physical equations of motion of the second-order gauge field can be removed.

All of our discussion highlights the urgent need for a better understanding of field redefinitions in Prokushkin--Vasiliev Theory and their locality properties. Allowing for any kind of field redefinition, including pseudo-local ones, should not be physically allowed as one can then remove any backreaction of the matter fields to the gauge fields at order 2. It is easy to implement the requirement that field redefinitions should be local and not pseudo-local. However, this is not the correct criterion, as can be seen e.g. by noticing that pseudo-local tails in our physical backreaction cannot be removed by local field redefinitions. Another possible criterion might be the asymptotic behavior of the pseudo-local field redefinitions. But as we checked in Appendix~\ref{appsec:Prop}, field redefinitions that should be physically not allowed seem to have the same asymptotic behavior as physically allowed ones --- at least to leading order. It is also important to note that the field redefinition performed in \rf{manifestredefgeneral} which enforces manifest Lorentz covariance (with respect to the background) \emph{is} of the pseudo-local type. Similarly, the field redefinition of $\sC^{(2)}$ which removes the backreaction thereof at $\tilde{g}_0 = \tilde{d}_0=0$ is also pseudo-local. It is unclear what the functional class corresponding to acceptable pseudo-local field redefinitions is. Ultimately, 'acceptable' means that the said class of field redefinitions does not change the observables as for example correlation functions. Those should correspond to redefinitions allowing us to remove everything from the backreaction but the canonical currents of Section~\ref{sec:cubic} (once the torsion constraint is solved for). Such a requirement is conceptually clear, but is nevertheless difficult to translate explicitly in terms of restrictions on the functional class of field redefinitions one should allow for.\footnote{See e.g. \cite{Taronna:2011kt} for a discussion of flat-space non-localities in the context of amplitudes computation.}

The interpretation one should have of the twisted fields is unclear to us. These include Killing-like tensor fields, sitting in finite-dimensional representations of the $\textrm{AdS}_3$ isometry algebra. From the perspective of Minimal Model Holography\cite{Gaberdiel:2012uj}, there does not seem to be any natural boundary dual for them. Thus, having in mind this duality one could conjecture the existence of a non-perturbative formulation of a higher-spin theory in dimension~$3$ involving no twisted fields and defined at any value of the $\lambda$ parameter. In particular it would be interesting to find out whether one can reformulate Prokushkin--Vasiliev Theory without twisted fields at the non-perturbative level. 

In fact there exists another three-dimensional, matter-coupled higher-spin theory which involves no twisted fields: the Vasiliev $D$-dimensional theory\cite{Vasiliev:2003ev,Bekaert:2005vh,Alkalaev:2014nsa} is defined without the twisted sector and can be consistently considered at $D=3$, as we discuss in Appendix~\ref{app:Ddimensionaltheory}. One can \emph{choose} to couple a twisted sector to this $D$-dimensional Vasiliev Theory at $D=3$ and in fact also for any $D$, even though it is not required by consistency and its original formulation does not include it. As we discuss in Appendix~\ref{app:Ddimensionaltheory}, in comparison to Prokushkin--Vasiliev Theory the $D$-dimensional theory has different features, and a manifestation thereof is the equations of motion for the spin-$1$ sector and the behavior of the twisted sector. Furthermore the $D$-dimensional theory at $D=3$ corresponds to $\lambda = 1$ and at present it is not known how to embed this point into one parameter family of theories.

Lastly, let us comment on our findings regarding the cubic action for the physical sector of the Prokushkin--Vasiliev theory, which are presented in Section~\ref{sec:cubic}. Let us stress again that we have fixed completely the cubic action, and we have done so by solving a necessary but non-sufficient quartic-order condition \rf{admissibility}, the so-called admissibility condition. This means that our cubic action is not guaranteed to be consistent at quartic order. Differently put, it is not necessarily true that we can find quartic terms to add on top of our cubic action such that it is consistent at that order. In general, it may be that the spectrum needs to be enlarged in order to achieve full consistency starting from our cubic action. In particular, it would be most interesting to find out whether twisted fields are required in order to achieve consistency to all orders for generic values of $\lambda$. 

Let us further comment on the possibility of removing the entire physical backreaction by means of a pseudo-local field redefinition. Although it signals a lack of control at the level of the equations of motion, it could make sense at the level of the corresponding action. The boundary terms produced by an exact current $J=\adD U$ should be kept at the action level while they are neglected for the equations of motion. In a rather daring fashion, one might then think of this feature as realization of the AdS/CFT lore, since by performing a (pseudo-local) field redefinition we are producing a left-over boundary term. 

Let us close by highlighting some open questions and possible continuations of the present investigations.
\begin{itemize}
\item An obvious generalization of our results would be that of considering the Prokushkin--Vasiliev theory at generic values of $\lambda$, which is especially interesting from the AdS/CFT perspective. 
\item Understanding the role of the twisted sector, if any, in the context of Minimal Model Holography seems a prime issue. In light of the  Gaberdiel--Gopakumar duality one would like to either construct a theory involving no twisted fields or try to make sense of twisted fields from the boundary perspective. 
\item A related issue is that of completing the cubic action \eqref{eq:fullcubicaction} to quartic order, if possible at all. More particularly, one would wish to see whether a completion thereof exists. 
\item Another possible direction of investigation, although potentially intricate technically, is to explore the equations of motion at order~$3$ in perturbation theory, paying special attention to the possibility of setting the twisted fields to zero consistently and making sure the whole parameter space of the theory is taken into account. 
\item Last but not the least, we mention the most pressing issue of correctly characterizing the functional class of field redefinitions one should allow for in the context of Prokushkin--Vasiliev Theory and in other higher-spin theories more generally. These should leave the correlation functions invariant and we expect them to yield the canonical currents \eqref{standardcurrent} starting from the backreaction computed in this work. 
\end{itemize}

\section*{Acknowledgments}
\label{sec:Aknowledgements}

We would like to thank Glenn Barnich, Slava Didenko, Matthias Gaberdiel, Carlo Iazeolla, Joris Raeymaekers, Rakibur Rahman, Per Sundell, Stefan Theisen and Mikhail Vasiliev for many useful discussions. We are also very much indebted to Andrea Campoleoni and Stefan Fredenhagen for crucial collaboration in the initial stages of this project and for useful discussions. We are also grateful to the organizers of the Saint Nicolas Workshop on String Field Theory, Higher Spins and Related Topics in Prague, for hospitality during the realization of part of this work. The research of G.\,L.\,G. was supported by the Grant Agency of the Czech Republic under the grant 14-31689S as well as by the Alexander von Humboldt Foundation. The research of E. Skvortsov and M. Taronna was supported by the Russian Science Foundation grant 14-42-00047 in association with Lebedev Physical Institute.


\begin{appendix}
\renewcommand{\thesection}{\Alph{section}}
\renewcommand{\theequation}{\Alph{section}.\arabic{equation}}
\setcounter{equation}{0}\setcounter{section}{0}


\section{Notation and Conventions}
\label{app:notation}
\setcounter{equation}{0}

Our symmetrization convention and index notation go as follows: indices denoted by the same letter \emph{without further subindices} are assumed to be symmetrized without extra factors. For example, $X_\alpha Y_\alpha$ is understood as $X_{\alpha_1}Y_{\alpha_2} + X_{\alpha_2}Y_{\alpha_1}$, without further normalization. If symmetric indices sit on the same object we further contract the notation as follows: a symmetric rank-$n$ tensor will be denoted as $T_{\alpha(n)}$, which means the tensor components $T_{\alpha_1\dots\alpha_n}$ are completely symmetric in the exchange of any two indices. Note that this can lead e.g. to expressions of the form $X_\alpha Y_{\alpha(n-1)}$, which should be understood as $X_{\alpha_1}Y_{\alpha_2\dots\alpha_n} + \; (n-1) \; \textrm{terms}$. 

The master fields entering the master equations \rf{threedVasiliev} are $\MW=\MW_\mm(y,z,\phi,\psi|x) \d x^\mm$,  $\MB=\MB(y,z,\phi,\psi|x)$ and $\MS_\ga=\MS_\ga(y,z,\phi,\psi|x)$. We work at $\lambda = \tfrac12$ so that $\MB$ has zero vacuum value. $\MW$ is shifted by the $\textrm{AdS}_3$ background connection $\Omega \equiv \tfrac12\varpi^{\alpha\alpha}L_{\alpha\alpha} + \tfrac12 \phi h^{\alpha\alpha}L_{\alpha\alpha}$ as $\MW \rightarrow \Omega + \MW$. The auxiliary field $\MS_\alpha$ is shifted according to $\MS_\alpha \rightarrow z_\alpha + 2i\MA_\alpha$. The shifted master fields obey the master equations \rf{VasilievEqAB}. The breakdown of master fields into field components is as follows: for the scalar one has
\begin{equation}
\begin{aligned}
\MB =&\;\flC + \MB^{(2)} + \MB^{(3)} + \dots = \B(y) + \mathcal{O}(z)\\
&\;\flC = \flC(y) = \pC \psi + \sC  \\
&\hspace*{70pt}\pC = \Pi_+\pC + \Pi_-\pC \equiv \pC_+ + \pC_- \,.
\end{aligned}
\end{equation}
For the connection it reads
\begin{equation}
\begin{aligned}
\MW =&\;\flomega + \MW^{(2)} + \MW^{(3)} + \dots = \W(y) + \mathcal{O}(z)\\
&\;\flomega = \flomega(y) = \pomega + \somega \psi \\
&\hspace*{70pt}\pomega = \omega + \phi e\,.
\end{aligned}
\end{equation}
For all these fields the full $y_\alpha$, $z_\alpha$-space star-product is
\begin{equation}
\mathbf{f}(y,z)\star \mathbf{g}(y,z)=\frac{1}{(2\pi)^2}\int \d^2u\, \d^2v\, \, \mathbf{f}(y+u,z+u) \, \mathbf{g}(y+v,z-v) \exp{(iv^\ga u_\ga)}\,.
\end{equation}
In particular we have $[y_\alpha,y_\beta]_\star = 2i\epsilon_{\alpha\beta}$, $[z_\alpha,z_\beta]_\star = -2i\epsilon_{\alpha\beta}$. 

We use various covariant derivatives through the text. The $\textrm{AdS}_3$ covariant derivative is given by
\begin{align}
\label{appcovDerNonLLTDef}
\DO \mathbf{F}&=\d \mathbf{F}-\Omega \wedge \star   \mathbf{F} + (-1)^{|F|} \mathbf{F}  \wedge  \star \Omega \,,
\end{align} 
where $|\mathbf{F}|$ denotes the form degree of $\mathbf{F}$. The covariant derivative acts as $\DO (X + \tilde{X}\psi) =\adD X + (\tadD X)\psi$, where
\begin{subequations}
\begin{align}
\adD&=\nabla-\frac12 \phi h^{\ga\ga}[L_{\ga\ga},\bullet ]_\star=\nabla-\phi h^{\ga\ga}y^{\vphantom{y}}_\ga\pl^y_\ga\,,\\
\tadD&=\nabla-\frac12 \phi h^{\ga\ga}\{L_{\ga\ga},\bullet \}_\star=\nabla+\frac{i}{2}\phi h^{\ga\ga} (y_\ga y_\ga-\pl^y_\ga \pl^y_\ga)\,,\\
&\hspace*{16pt}\nabla=\d\bullet-\frac12 \varpi^{\ga\ga}[L_{\ga\ga},\bullet]_\star=d-\varpi^{\ga\ga} y^{\vphantom{y}}_\ga \pl^y_\ga\,.
\end{align}
\end{subequations}
In the process of extracting physical equations from master equations in Section~\ref{sec:Veq} we also use the following covariant derivative:
\begin{align}
\label{appeq:covDev}
\mathsf{D}^{\textsc{yz}}&=\d\bullet-\frac12\omega^{\ga\ga}[L^{\textsc{yz}}_{\ga\ga},\bullet]_\star-\frac12e^{\ga\ga}[\phi L^{\textsc{y}}_{\ga\ga},\bullet]_\star\,,
\end{align}
as well as its background version
\begin{align}
\label{appcovDerivativeLLTCov}
\bar{\mathsf{D}}^{\textsc{yz}}&=\d\bullet-\frac12\varpi^{\ga\ga}[L^{\textsc{yz}}_{\ga\ga},\bullet]_\star-\frac12h^{\ga\ga}[\phi L^{\textsc{y}}_{\ga\ga},\bullet]_\star\,.
\end{align}
Our conventions for index contraction and raising\,/\,lowering are as follows:
\begin{equation}
y_\alpha = y^\beta\epsilon_{\beta\alpha}\,,\qquad y^\alpha = \epsilon^{\alpha\beta}y_\beta\,,\qquad \epsilon^{\alpha\beta}\epsilon_{\alpha\gamma} = \delta^\beta_\gamma\,,
\end{equation}
so that $A_\alpha B^\alpha = -A^\alpha B_\alpha \equiv -AB = BA$ and $\epsilon_{12} = \epsilon^{12} \equiv 1$. Our derivatives have indices which are raised and lowered in the usual way, so that $\partial_\alpha = \partial^\beta\epsilon_{\beta\alpha}$, $\partial^\alpha = \epsilon^{\alpha\beta}\partial_\beta$ and we have
\begin{equation}
\partial_\alpha y_\beta = \epsilon_{\alpha\beta}\,,\qquad \partial^\alpha y^\beta = \epsilon^{\alpha\beta}\,,\qquad \partial_\alpha y^\beta = \delta^\beta_\alpha\,,\qquad \partial^\alpha y_\beta = -\delta^\alpha_\beta\,, 
\end{equation}
and analogously for the $z_\alpha$-oscillators. Last but not least we recall the following identities for the background vielbein: 
\begin{equation}
\label{appeq:twoFormIdent}
h^{\alpha\alpha}_\mu h_{\alpha\alpha}^\nu = -\frac{1}{2}\delta^\nu_\mu \qquad h^{\alpha\alpha}_\mu h_{\beta\beta}^\mu = - \frac14  \delta^\alpha_\beta \delta^\alpha_\beta \,.
\end{equation}

\section{Backreactions}
\label{app:technicaldetailsBackreaction}
\setcounter{equation}{0}
In this Appendix we will summarize our results concerning the various backreactions in Prokushkin--Vasiliev Theory. We will first focus on the backreaction to the Fronsdal fields in Section~\ref{subsec:fourierspace} and then discuss the twisted scalar and gauge sector in Section~\ref{app:twistedBackreactions}.

\subsection{Fronsdal Sector}
\label{subsec:fourierspace}
In the following Section~\ref{app:rawexpressions} we will collect the raw expressions for all the contributions to $\mathcal{V}(\Omega,\Omega,\pC,\pC)$ in Fourier space and then in Section~\ref{app:simplifiedBackreaction} summarize our strategy to simply relate these expressions by partial integration and Fierz identities.

\subsubsection{Raw Expressions for the Backreaction on the Fronsdal Sector}
\label{app:rawexpressions}
In the following we will first summarize the result for the backreaction $\mathcal{V}(\Omega,\Omega,\pC,\pC)$ when using the field redefinition $M_1$, as discussed in Section~\ref{sec:twistedResultsLinearOrder}, and then explain how the results for the redefinition $M'_1$ can be obtained from it.

It is convenient to introduce the following exponents that appear in various structures below
\besubeqs\label{ExponentsAll}
\begin{align}
Q&=\exp i \left(tq (\eta+y)(y+\xi)\right)\,, &
P&=\exp i \left(t (\eta+y)(y+\xi)\right)\,, \\
K&=\exp i (y-q \eta)(y+t\xi)\,, &
R^1&=\exp i \left(t (y-q(\eta+y))\xi \right)\,,\\
R^2&=\exp i \left(q (y-t(\xi+y))\eta \right)\,. 
\end{align}
\esubeqs
The Fourier images of the functions involved in the computation when setting to zero the linearized twisted sector are:
\begin{align}
\aA_\ga&=z_\ga \int_0^1 \d t \, t e^{it(y+\xi)z}\pC (\xi,\phi|x)\psi\,,\\
M_1&=\frac14 \phi h^{\ga\ga} \int_0^1 \d t \, (t^2-1) (y-\xi)_\ga (y-\xi)_\ga e^{ity\xi}\pC (\xi,\phi|x)\,,\\
M_2&= \frac12 \phi h^{\ga\ga} \int_0^1 \d t \, (-z_\ga y_\ga (1-t)^2 +(1-t^2) z_\ga \xi_\ga) e^{it(y+\xi)z}\pC (\xi,\phi|x)\,.
\end{align}
A lengthy but straightforward computation of all the terms of the backreaction, in complete analogy with the example discussed in Section~\ref{subsec:explicit}, yields the following result. The terms therein are to be added up as they are and integrated over the homotopy parameters $t$, $q$ and the wave-twistors $\xi$, $\eta$ after multiplying with $\pC(\xi,\phi)\pC(\eta,-\phi)$:
\begin{subequations}
\begin{align}
\label{PstructureA}
\bar{\mathsf{D}}^{\textsc{yz}}z^\ga\homo{0}{[h, z_\ga \homo{1}{\aA_{\nu}\star\aA^{\nu}}]_\star} &=  \frac{-i}{2} \;  H^{\ga \ga} \; t^3\; P  \;  (\eta - \xi)_\ga (\eta - \xi)_\ga \,,\\
\label{PstructureB}
\bar{\mathsf{D}}^{\textsc{yz}}z^\ga\homo{0}{[h, z_\ga \homo{1}{\MB^{(2)} \star \varkappa}]_\star} &= \frac{i}{2} \;  H^{\ga \ga} \; t^2\; P  \;  (\eta - \xi)_\ga (\eta - \xi)_\ga\,,\\
\bar{\mathsf{D}}^{\textsc{yz}}z^\nu\homo{0}{[M_1 \psi,\aA_{\nu}]_\star} &= \frac{i}{2}   \; H^{\ga \ga} \; q(1-q)t \; \left\{  S^1_\ga \xi_\ga \;  R^1 + \xi \leftrightarrow \eta  \right\}\,,\\
\bar{\mathsf{D}}^{\textsc{yz}}z^\nu\homo{0}{M_2 \psi \star \aA_{\nu}} &= \frac{i}{2}    \;  H^{\ga \ga} \; \left( -q^2t \; C^1_\ga(\xi-\eta)_\ga-q^3 t (1-t)^2 \; (\eta+y)_\ga (\xi-\eta)_\ga \right. \\ \nonumber &-\left. tq^2 C^1_\ga (\xi+y)_\ga + t q^2 (1-t)^2 (1-q) (\eta+y)_\ga (\xi+y)_\ga  \right)Q\\
&+ \frac{1}{2}  \; h^{\ga \ga} \wedge h^{\gb \gb} \;\left(q^3 t^2\;(\xi+y)_\ga (\xi-\eta)_\ga (\eta+y)_\gb C^1_\gb \right) Q\,,\\
\bar{\mathsf{D}}^{\textsc{yz}}z^\nu\homo{0}{ \aA_{\nu} \star M_2 \psi} &= \frac{i}{2}    \;  H^{\ga \ga} \; \left(qt^2 (\xi-\eta)_\ga C^2_\ga + qt^3(1-q)^2 (\xi+y)_\ga (\xi-\eta)_\ga \right. \\ \nonumber &-\left. t^2 q C^2_\ga (\eta+y)_\ga + t^2 (1-t) q (1-q)^2 (\xi+y)_\ga (\eta+y)_\ga \right)Q\\
&+ \frac{1}{2}  \; h^{\ga \ga} \wedge h^{\gb \gb} \;q^2 t^3\; \left( (\eta+y)_\ga (\xi-y)_\ga (\xi+y)_\gb C^2_\gb \right)Q\,,
\end{align}
\vspace*{-26pt}
\begin{align}
\label{Lorentzredef}
-i H^{\ga\ga}\aA_{\ga}\star\aA_{\ga} &= i   \; H^{\ga \ga} \; t^2 q^2 Q \; (y+\xi)_\ga (y+\eta)_\ga\,,\\
M_2\psi\star\wedge M_2 \psi  &= \frac{i}{4}   \; H^{\ga \ga} \; \left\{ C^1_\ga C^2_\ga +  (1-q)^2 t (\xi+y)_\ga C^1_\ga + (1-t)^2 q (\eta+y)_\ga C^2_\ga \right. \\    \nonumber &+\left. q t (1-t)^2 (1-q)^2 (\eta+y)_\ga (\xi+y)_\ga \right\}Q \\
&+ \frac{1}{4}   \; h^{\ga \ga}\wedge \; h^{\gb \gb} \;\left( tq \; (\eta+y)_\ga C^1_\ga (\xi+y)_\gb C^2_\gb \right) Q\,,\\
M_2 \psi\star \wedge M_1\psi &= \frac{-i}{4}   \; (q^2-1) H^{\ga \ga} \left(T^2_\ga S^2_\ga + q (1-t)^2 \eta_\ga S^2_\ga \right) R^2\\
 &+ \frac{1}{8}   \; h^{\ga \ga} \wedge h^{\gb \gb} \;\left( (q^2-1)q \; \eta_\ga T^2_\ga S^2_\gb S^2_\gb \right) R^2\,,\\
M_1 \psi \star \wedge M_2 \psi &= \frac{-i}{4}   \;  H^{\ga \ga} \; (t^2-1) \; \left(- T^1_\ga S^1_\ga + t (1-q)^2 S^1_\ga \xi_\ga \right) R^1\\
 &+ \frac{1}{8}   \;h^{\ga \ga} \wedge h^{\gb \gb} \; \left( (t^2-1)t \; S^1_\ga S^1_\ga \xi_\gb T^1_\gb \right) R^1\,,\\
M_1 \psi \star\wedge M_1 \psi &=  \frac{-1}{16}   \;  H^{\ga \ga} \; (t^2-1)(q^2-1) \; K \; \left(U^1_\ga U^2_\ga (U^1 U^2) + 4i U^1_\ga U^2_\ga  \right)\,, \label{Mtermlast}
\end{align}
\end{subequations}
where we denoted
\besubeqs
\label{eq:shortcuts}
\begin{align}
C^1_\ga &= (1-t^2)\xi_\ga - (1-t)^2 \left( (1-q)y_\ga - q \eta_\ga \right)\,, & U^1_\ga &= (y+t \xi - \eta)_\ga\,,\\
C^2_\ga &= (1-q^2)\eta_\ga - (1-q)^2 \left( (1-t)y_\ga - t \xi_\ga  \right)\,, &U^2_\ga &= (y-q \eta - \xi)_\ga\,, \\
T^1_\ga &= (1-q^2)\eta_\ga - (1-q)^2 (y+t\xi)_\ga\,,  & S^1_\ga &= (1-q)y_\ga - q \eta - \xi\,,\\
T^2_\ga &= (1-t^2)\xi_\ga - (1-t)^2 (y - q \eta)_\ga\,, & S^2_\ga &= (1-t)y_\ga - t \xi- \eta\,.
\end{align}
\esubeqs
In evaluating various cocycles the following formulas were useful:insertions being carefully tracked we find
\besubeqs
\begin{align}
\left.\bar{\mathsf{D}}^{\textsc{yz}}z^\nu\homo{0}{f_\nu(y,z)}\right|_{z=0}&=i\phi h^{\ga\ga}\pl^y_\ga f_\ga(y,0)\,,\\
\left.\bar{\mathsf{D}}^{\textsc{yz}}z^\nu\homo{0}{[h, z_\nu f(y,z)]}\right|_{z=0}&=-H^{\ga\ga}\pl^y_\ga \pl^y_\ga f(y,0)\,,\\
\left.\bar{\mathsf{D}}^{\textsc{yz}}z^\nu\homo{0}{[h, z_\nu \homo{1}{f(y,z)\star \varkappa}]}\right|_{z=0}&=-\frac12 H^{\ga\ga}\pl^y_\ga \pl^y_\ga f(0,-y)\,.
\end{align}
\esubeqs
It is the sum of the above expressions, i.e. \eqref{PstructureA}-\eqref{Mtermlast}, that we checked against the conservation identity \eqref{eq:derFourierTwoForm} and found to consist of several independently conserved quantities. 

\paragraph{Result for the Redefinition $M'_1$:} If we perform a pseudo-local redefinition \eqref{eq:M1prime} that allows for the removal of the backreactions to the twisted-sector at second order then we need to modify the formulas above. The bosonic projection implies that $M'_1$ is given by 
\begin{align}\label{newM1alg}
M'_1&=\frac{i}4 \phi h^{\ga\ga} \int_0^1 \d t \, (t^2-1) (y-\xi)_\ga (y-\xi)_\ga  \, \sin{(ty\xi)}\pC (\xi,\phi|x)=\frac12\left(M_1-\left.M_1\right|_{t\rightarrow-t}\right)\,.
\end{align}
This form implies that the corrected backreaction can be obtained by anti-symmetrizing over $t$, over $q$ or over both $t$ and $q$ the terms of \eqref{PstructureA}-\eqref{Mtermlast} that are of the form $(M_1 \psi) \star X$, $X \star (M_1 \psi)$ or $(M_1 \psi) \wedge \star (M_1 \psi)$ respectively.

\subsubsection{Simplified Backreaction on the Fronsdal Sector}
\label{app:simplifiedBackreaction}
Due to Fierz identities that can be combined with integration by parts over $t$ or $q$ there is no unique way of presenting the final result. One of the simplest forms is summarized below. Our general strategy was to get rid of four-fermion terms by trying to represent them as derivatives of the exponents with respect to $t$ and $q$, times a prefactor that is only bilinear in spinors.

The four-fermion structures in front of $R^1$ and $R^2$ can be reabsorbed by total derivatives. For the contributions containing the $Q$ exponential there are certain four-fermion terms left, which is much less than the 15 coefficients of the most general ansatz. As for the K-terms, the four-fermion terms can be removed up to a single term.

Finally, the full expression for the backreaction splits into the following two independently conserved components
\begin{align}\label{Simplifiedbackreaction}
\formJ&^\textsc{pv}=\formJ^\textsc{redef}+\formJ^\textsc{phys}\,, 
\end{align}
where we have defined
\begin{align}
\formJ^\textsc{phys}&=H^{\ga\ga}\int \d t\,\d q\,\d\xi\, \d\eta\, (\compJ^Q_{\ga\ga}+\compJ^P_{\ga\ga}) \,\pC (\xi,\phi)\pC(\eta,-\phi)\,, \\
\formJ^\textsc{redef}&=H^{\ga\ga}\int \d t\,\d q\,\d\xi\, \d\eta\, (\compJ^K_{\ga\ga}+\compJ^{R_1}_{\ga\ga}+\compJ^{R_2}_{\ga\ga}) \,\pC (\xi,\phi)\pC(\eta,-\phi)\,,
\end{align}
and the various kernels are given by 
{\allowdisplaybreaks
\begin{align*}
\compJ^Q_{\ga \ga}&=  \;\{ d_1 y_\ga y_\ga + d_2 \xi_\ga \xi_\ga  + d_3 \eta_\ga \eta_\ga + d_4 \xi_\ga \eta_\ga + d_5 \xi_\ga y_\ga + d_6 \eta_\ga y_\ga  \\ &\qquad\qquad  - d_7 ( y_\ga \eta_\ga (y \eta) - \xi_\ga y_\ga (y\xi ) - \xi_\ga \eta_\ga (\xi \eta) ) \} Q \,, \nonumber\\
\compJ^K_{\ga\ga}&=-\frac{1}{8} i (1-q) (1+t)((q+t)(\eta_\ga\eta_\ga-\xi_\ga\xi_\ga)-(q+1) (t+1)y_\ga\eta_\ga -(q-1) (t-1)y_\ga\xi_\ga\\&\qquad\qquad+(1+qt)y_\ga y_\ga-(q-1) (t+1) (2 q t-q+t)\xi_\ga\eta_\ga)K\\
&\qquad\qquad-\frac{1}{16} (q-1)^3 (q+1) (t-1) (t+1)^3 (\eta\xi) \xi_\ga \eta_\ga  K\,,\\
\compJ^P_{\ga \ga}&= \;\{ p_1 \xi_\ga \xi_\ga + p_2 \eta_\ga \eta_\ga +  p_3 (\, t \, \xi_\ga \eta_\ga + \xi_\ga y_\ga + \eta_\ga y_\ga) \}P\,,\\
\compJ^{R^1}_{\ga \ga} &=  \;\{ \rho_1 \xi_\ga \xi_\ga + \rho_2 \eta_\ga \eta_\ga + \rho_3 y_\ga \xi_\ga + \rho_4 y_\ga \eta_\ga + \rho_5 \xi_\ga \eta_\ga\} R^1\\
 &\qquad\qquad+\frac{1}{16} i   \; \left(t^2-1\right)\{ \xi_\ga \eta_\ga  (t+2) -  y_\ga\eta_\ga \} K_t\,,\\
\compJ^{R^2}&=- \compJ^{R^1}\left(\substack{t\rightarrow -q\\ q\rightarrow t\\ \xi\leftrightarrow\eta}\,,\,\,\substack{R^1\rightarrow R^2\\ K_t\rightarrow K_q}\right)\,, 
\end{align*}
}\noindent
where the functions $K_t$ and $K_q$ are given by
\begin{align}
K_q&=\exp i q(y \eta) =K|_{t=0}\,,\\   
K_t&=\exp i t(y\xi)=K|_{q=0}\,, &   
\end{align}
and the coefficients are functions of $t$, $q$ and are given by
{\footnotesize \allowdisplaybreaks
\begin{align*}
d_1 &= \frac{i}{8} (-q + 4 q^2 - 3 q^3 + 4 q t - 9 q^2 t + 4 q^3 t + 8 q^2 t^2 + q^3 t^2)\,, &&\rho_1 = \frac{i}{4} t (-1 + q)  (1 + q + t)\,,\nonumber \\ 
d_2 &= -\frac{i}{8} (-3 q + 3 q^3 + 4 q t + q^2 t - 8 q^3 t + 3 q^3 t^2)\,, & &\rho_2 = -\frac{i}{4} (-1 + q) q \nonumber\,, \\
d_3 &= -\frac{i}{4}  (-q + 2 q t + q^2 t)\,, &&\rho_3 = \frac{i}{4}  t (-1 + q)^2  (1 + q + t)\,, \nonumber \\
d_4 &= \frac{i}{4} (3 q - 2 q^2 - 2 q t - 3 q^2 t - 2 q^3 t + 10 q^2 t^2 + 2 q^3 t^2)\,, & &\rho_4 = -\frac{i}{4} (-1 + q)^2\nonumber\,, \\
d_5 &= -\frac{i}{4} (-2 q^2 + 3 q^3 - 2 q t + 2 q^2 t - 6 q^3 t + 2 q^2 t^2 + q^3 t^2)\,, & &\rho_5 = \frac{i}{4} (-1 + q) (-1 + q^2 t + q t (1 + t))\nonumber\,, \\ 
d_6 &=  \frac{i}{4} (q - 2 q^2 + 2 q t + 3 q^2 t - 2 q^3 t - 2 q^2 t^2 + 2 q^3 t^2)\,, \nonumber \\
d_7 &= \frac{1}{4} (-q t + 2 q^2 t - 2 q^2 t^2 - 2 q^3 t^2 + 3 q^3 t^3) \,,
\end{align*}
\vspace*{-18pt}
\begin{align*}
p_1 &= -\frac{i}{4} t(1-t)^2\,, &
p_2 &= -\frac{i}{4} (-t + t^3)\,, &
p_3 &= \frac{i}{2} (-t + t^2) \,.
\end{align*} } \noindent
\vspace*{-35pt} 
\paragraph{Result for Redefinition $M'_1$:} As explained around \eqref{newM1alg}, it is easy to navigate to the point where the backreaction for the twisted zero-forms vanishes. The terms $\compJ^Q$ and $\compJ^P$ are left untouched. For the $R^1$ and $R^2$ structures we apply
\begin{align}\label{RRrules}
\compJ^{R^1}&\rightarrow\frac12\left(\compJ^{R^1}-\left.\compJ^{R^1}\right|_{t\rightarrow-t}\right)\,, &
\compJ^{R^2}&\rightarrow\frac12\left(\compJ^{R^2}-\left.\compJ^{R^2}\right|_{q\rightarrow-q}\right)\,.
\end{align}
For the $\compJ^K$ we apply both $t$ and $q$ anti-symmetrization. The only subtlety is about the $K_t$ and $K_q$ terms of $\compJ^{R^1}$ and $\compJ^{R^1}$, since they are a combination of the boundary terms produced by $K$,  $R^1$ and $R^2$ but it turns out that the rules \eqref{RRrules} can be applied to them as well.

\subsection{Backreactions on the Twisted Sector}
\label{app:twistedBackreactions} 
In the following we will summarize various aspects of the backreactions arising in the twisted sector of Prokushkin--Vasiliev Theory. We will first focus on the backreaction to the twisted scalar field in Appendix~\ref{app:susyTwistedDecoupling} and then analyze the gauge sector in Section~\ref{app:decouplingTwistedGaugeSector}.

\subsubsection{Scalar Sector}
\label{app:susyTwistedDecoupling}
Without the bosonic projection a computation similar to the one discussed in Section~\ref{sec:twistedSecondOrder} but with the $\varkappa$ insertions being carefully tracked yields
\be
\label{eq:susycocycle} 
\widetilde{\mathcal{V}}(\Omega,\pC,\pC)=-\bar{\mathsf{D}}^{\textsc{yz}} z^\nu \homo{0}{\aA_\nu\star \pC\psi+\pC\psi\star \pi(\aA_\nu)]}+(M \psi) \star \pC\psi - \pC\psi \star\pi(M \psi)\,,\ee
where $M=M_1 + M_2$ and $\pi(f(y,z))=\varkappa\star f\star \varkappa=f(-y,-z)$. This results in
\begin{align}
\widetilde{\mathcal{V}}(\Omega,\pC,\pC)&=\phi h^{\ga\ga} \int \, \d^2 \xi \d^2 \eta \,  K_{\ga\ga}(\xi,\eta,y) \,\pC (\xi,\phi|x)\pC (\eta,-\phi|x)\,,
\end{align}
with the kernel $K_{\ga \ga}$ given by
\begin{align}
K_{\ga\ga}=\int_0^1 \d t \, \Big\{ &\frac12\,e^{i(y(1-t)+t\eta)\xi} \, \xi_\ga\left({(1-t^2)}(\xi_\ga+\eta_\ga)+{(1-t)^2} y_\ga\right)\notag\\
-&\frac12\,e^{i(y(1-t)-t\xi)\eta} \, \eta_\ga\left({(1-t^2)}(\eta_\ga+\xi_\ga)-{(1-t)^2} y_\ga\right)\label{shadowOCCSUSY}\\
+&\frac14 (t^2-1)e^{i(y-\eta)(y+ t\xi)}(y-\xi-\eta)_\ga(y-\xi-\eta)_\ga\notag\\
+&\frac14 (t^2-1)e^{i(y+t\eta)( y+\xi)}(y+\xi+\eta)_\ga(y+\xi+\eta)_\ga \Big\} \notag\,.
\end{align}
The contribution to the Killing constant is
\begin{align}\label{SUSYconst}
\adD \sC^{(2)}|_{y=0}=\d\sC^{(2)}(y=0)=\int \d^2\xi \d^2\eta \, H^{\ga \ga }K_{\ga \ga}(y=0) \,\pC (\xi,\phi)\pC(\eta,-\phi)\,,
\end{align}
where we have used \eqref{covDer} and the kernel $K_{\ga \ga}(y=0)$ is given as
\begin{align}
\label{eq:SUSYkernel}
K_{\ga \ga}(y=0)=&\phi \left\{ f(\eta\xi)(\eta_\ga\eta_\ga+\xi_\ga \xi_\ga+2\xi_\ga \eta_\ga) +(\eta_\ga\eta_\ga-\xi_\ga\xi_\ga)(f(\eta\xi)+g(\eta\xi)/2)\right\}\,,
\end{align}
with the coefficient functions given as
\begin{equation}
 f(x)=\left(\cos x -\frac{\sin x}x\right)\frac{1}{x^2} \,, \qquad g(x)=-\frac{i \left(2+x^2-2 \cos x-2 x \sin x\right)}{x^3}\notag \,.
\end{equation}
Taking the most general zero-form that can contribute to the Killing constant
\begin{align}
\int \d^2\xi \d^2\eta \, F(\eta\xi)\pC (\xi,\phi)\pC(\eta,-\phi)\,,
\end{align}
the derivative \eqref{eq:derFourierZeroForm} generates a single tensor structure 
\be
\label{eq:exactSUSY}
(\eta_\ga\eta_\ga-\xi_\ga\xi_\ga)(F(\eta\xi)+F''(\eta\xi))\,.
\ee 
Therefore the first term in \eqref{eq:SUSYkernel} cannot be represented as an exact form. However this term is precisely canceled if we take the ambiguity in $M_1$, discussed in Section~\ref{sec:twistedResultsLinearOrder}, into account. As we explained in Section~\ref{sec:twistedResultsLinearOrder} the ambiguity is given by \eqref{eq:representativeofH}, which in Fourier space reads
\begin{align*}
R&= \frac14 \int \d^2\xi \d^2\eta \int_0^1 \d t \, (t^2-1) \phi h^{\ga\ga}\left\{ g_0 (y_\ga y_\ga +\xi_\ga\xi_\ga) \pC_\textsc{b}(\xi,\phi|x) - 2
d_0 y_\ga \xi_\ga \pC_\textsc{f}(\xi,\phi|x)\right\} \cos{(ty\xi)}\,.
\end{align*}
This ambiguity will lead to an additional contribution to \eqref{eq:susycocycle} which is given by $R \psi \star \pC \psi - \pC \psi \star \pi(R \psi)$. Its $y_\alpha$-independent component will modify \eqref{eq:SUSYkernel} by
\begin{align}
g_0 \phi h^{\ga\ga} (\xi_\ga \xi_\ga +\eta_\ga \eta_\ga)f(\eta \xi) -d_0 \phi h^{\ga\ga} (\xi_\ga \eta_\ga)f(\eta \xi) \,.
\end{align}
Combining it with \eqref{eq:SUSYkernel} and comparing with \eqref{eq:exactSUSY} we see that it can be made exact at $g_0=-1$, $d_0=-1$ with
\be 
F(x)=\frac{i \left(-1+e^{i x}\right)}{2 x} +A \cos x +B \sin x\,,\label{susyscalredef}
\ee 
where $A$ and $B$ are integration constants corresponding to fermionic and bosonic components of the super-trace belonging to $\mathbb{H}^0(\adD,\pC \pC)$. The choices $g_0=-1$ and $d_0=-1$ obviously correspond to $\tilde{g}_0=1+g_0=0$ and $\tilde{d}_0=1+d_0=0$ in \eqref{eq:M1PrimeFerm}. Therefore, the ambiguity in making a redefinition given by elements $R \in \mathbb{H}^1(\tadD,\pC)$ is fixed by requiring exactness of \eqref{eq:susycocycle}, the corresponding kernel \eqref{shadowOCCSUSY} having the form
\begin{align}
K_{\ga\ga}&=\frac12\,e^{i(y(1-t)+t\eta)\xi} \, \xi_\ga\left({(1-t^2)}(\xi_\ga+\eta_\ga)+{(1-t)^2} y_\ga\right)\notag\\
&-\frac12\,e^{i(y(1-t)-t\xi)\eta} \, \eta_\ga\left({(1-t^2)}(\eta_\ga+\xi_\ga)-{(1-t)^2} y_\ga\right)\label{shadowOCCSUSYA}\\
&+\frac{i}4 (t^2-1)(y-\xi-\eta)_\ga(y-\xi-\eta)_\ga e^{iy\eta} \sin{t(y-\eta)\xi}\notag\\
&+\frac{i}4 (t^2-1)(y+\xi+\eta)_\ga(y+\xi+\eta)_\ga e^{iy\xi} \sin{t\eta( y+\xi)}\notag\,.
\end{align}
\paragraph{Exactness of $\boldsymbol{K_{\ga \ga}}$ beyond $\boldsymbol{y_\ga = 0}$:} In order to check whether all the $y_\alpha$-components of the above backreaction, and not only the Killing constant, are exact it is quite useful to find the corresponding generating function that represents the above backreaction in the basis of Appendix~\ref{app:Basis}. One can easily extract generating functions for the coefficients of the corresponding tensor structures in this basis. This is done using the following representation of the identity under the contour integral sign of Appendix~\ref{app:Basis}:
\begin{equation}
f(x,y,z)\sim\oint\frac{\omega}{1-\omega}\frac{\beta}{1-\beta}\frac{\gamma}{1-\gamma}\,f(\omega^{-1} x,\beta^{-1} y,\gamma^{-1} z)\,,
\end{equation}
where $\omega=\tau^{-1}$, $\beta=X^{-1}$ and $\gamma=Y^{-1}$. The equivalence relation '$\sim$' is defined in \eqref{eq:equivalenceRel}.
We then arrive to the following contribution for the scalar sector where integration in $t$ is implicit:

\begin{equation}
\vec J^{(1)}=\begin{pmatrix}
\frac{\omega  (g_0+1) \left(t^2-1\right)}{2-2 \omega^2 t^2}\\
\frac{\omega  \left(t^2-1\right) (4 d_0-2 (g_0+2)) \left(\omega  t^2-1\right)}{4-4 \omega^2 t^2}\\
\frac{\omega  \left(t^2-1\right) (4d_0+2 g_0) \left(\omega  t^2+1\right)}{4-4 \omega ^2 t^2}\\
-\frac{\omega  \left(t^2-1\right) (2d_0-2 (g_0+2)) \left((2 \omega -1) t^2-1\right)}{8 \omega ^2 t^2-8}\\
\frac{\omega \left(t^2-1\right) (4d_0+2 g_0) \left((2 \omega +1) t^2+1\right)}{8-8 \omega ^2 t^2}\\
0
\end{pmatrix}\,.
\end{equation}

\noindent We will now fix $g_0=-1$ and $d_0=-1$ and show that for this choice this backreaction is exact. Furthermore, the above choice cancels any odd power of $\gamma$ in the corresponding generating functions but leaves all even powers. In the following we will study the even powers in $\gamma$ and analyze whether or not they are trivial in cohomology.

With this choice for the redefinition and expanding in $\beta$ and $\gamma$ up to order $\beta^n$ and $\gamma^m$ one gets the following result:
\begin{equation}
\vec J_{n,m}^{(1)}=\frac18\begin{pmatrix}
\frac{\omega}{t^2\omega^2-1}[A^{(1)}_{n,m}(t)+B^{(1)}_{n,m}(t)t\omega]\\
\frac{2\omega}{t^2\omega^2-1}[A^{(2)}_{n,m}(t)+B^{(2)}_{n,m}(t)t\omega]\\
\frac{2\omega}{ \omega  t+1}[\left((1-t)^m+(t-1)^m\right) (1-t)^{n+1}]\\
\frac{\omega}{t^2\omega^2-1}[A^{(4)}_{n,m}(t)+B^{(4)}_{n,m}(t)t\omega]\\
0\\
\frac{2\omega}{\omega  t+1}[(t+1) \left((1-t)^m+(t-1)^m\right) (1-t)^{n+1}]
\end{pmatrix}\,.
\end{equation}
In the expression above we define
{\footnotesize
\begin{align}
A_{n,m}^{(1)}&=-\left((-u)^m+u^m \right) v^n+\left((-v)^m+v^m \right)u^n \notag \\
B_{n,m}^{(1)}&=-t u^n \left(v^m+(-v)^m\right)-t \left(u^m+(-u)^m\right) v^n \notag \\
A_{n,m}^{(2)}&= u^n \left(t \left( v^m+(-u)^m+(-v)^m\right)+v^m-(-u)^m+(-v)^m\right)-u^{m+1} v^n+(-u)^{m+1} v^n-u^{m+n+1} \notag \\
B_{n,m}^{(2)}&=-t \left(u^n \left(t \left(v^m+(-u)^m+(-v)^m\right)+ v^m-(-u)^m+(-v)^m\right)+u^{m+1} v^n+u (-u)^m v^n-u^{m+n+1}\right) \notag \\
A_{n,m}^{(4)}&=v u^n \left(v^{m+1}+t \left(2 u^m+2 (-u)^m+(-v)^m\right)-2 u^m-2 (-u)^m+(-v)^m\right)-u^2 \left(u^m+(-u)^m\right) v^n \notag \\
B_{n,m}^{(4)}&=(-t) v \, u^n \left( v^{m+1}+t \left(2 u^m+2 (-u)^m+(-v)^m\right)-2 u^m-2 (-u)^m+(-v)^m\right)-t u^2 \left(u^m+(-u)^m\right) v^n \notag 
\end{align}
}\noindent
where again the integration over $t$ is implicit and we defined $u=1-t$ and $v=1+t$.
\noindent To show its exactness it is first important to use the Fierz identity freedom to bring the above expressions to a canonical form. We then distinguish various cases:

\paragraph{For $\boldsymbol{n>0}$ and $\boldsymbol{m>0}$:} Labeling by $\adD J^{(0)}_{m,n}$ the exact term associated with the term $\beta^n\gamma^m$ we get
{\footnotesize \allowdisplaybreaks
\begin{align}
J^{(0)}_{m,n}&=\frac{\omega^2}{8 \left(1-\omega^2\right)}
\int_0^1 \d t\Big[
2 \omega  n \left(\frac{(1-t )^{m+2} (1+t)^n}{(m+1) (\omega  t -1)}+\frac{(t +1)^{m+2} (1-t )^n}{(m+1) (\omega  t +1)}\right)\nonumber\\
&-\frac{4 \omega  \left(1-t ^2\right) (m+n+1) (1-t )^{m+n}}{(m+1) (\omega  t +1)}\nonumber\\
&+\frac{t  (2 m n+m+3 n+2) (1-t )^m (1+t)^n}{(m+1) (n+1)}+\frac{t  (2 m n+m+3 n+2) (1+t)^m (1-t )^n}{(m+1) (n+1)}\nonumber\\
&+\frac{(1-t )^m (1+t)^n (2 (m+1) (n+1)-\omega  (m+n+2))}{\omega  (m+1) (n+1)}\nonumber\\
&+\frac{(1+t)^m (1-t )^n (\omega  (m+n+2)-2 (m+1) (n+1))}{\omega  (m+1) (n+1)}+\frac{2 (m-n) (1-t )^{m+n+1}}{(m+1) (n+1)}\label{shadowredefinition}
\Big]\,,
\end{align}} \noindent
where $m\in 2\mathbb{N}$ is even, otherwise there is no need for an improvement. 

\paragraph{For $\boldsymbol{n>0}$ and $\boldsymbol{m=0}$:} The exact term is given by
{\footnotesize \allowdisplaybreaks
\begin{align}
J^{(0)}_{m,n}&=\int_0^1 \d t\frac{-\frac{\left(1+\frac{1}{\omega }\right)^n (-1+\omega ) (-2+n 
(-1+\omega )+2 t \omega )}{-1+t \omega }+\frac{\left(\frac{-1+\omega 
}{\omega
}\right)^n \left(6+3 n+2 (n+2 t) \omega -(2+n) \omega ^2\right)}{1+t \omega }}{4 \left(-1+t^2\right) \omega} \,,
\end{align}}\noindent 
where again $m\in 2\mathbb{N}$ is even as otherwise there is no need for an improvement. 

\paragraph{For $\boldsymbol{m>0}$ and $\boldsymbol{n=0}$:} Again we can construct an exact term which is given by
{\footnotesize \allowdisplaybreaks
\begin{align}
J^{(0)}_{m,n}&=\int_0^1 \, \d t \frac{\left(\frac{-1+\omega }{\omega }\right)^{1+m}}{2 (1+t) \left(t+\frac{1}{\omega }\right)} \,,
\end{align}}\noindent 
where again $m\in 2\mathbb{N}$.

\noindent Note that we have used slightly different conventions for the sign of $n$ and $m$ here as compared to Appendix~\ref{app:cohomologies} in which they were defined as $\gamma^{-m}$ and $\beta^{-n}$.

\subsubsection{Gauge Sector}
\label{app:decouplingTwistedGaugeSector}
Let us recall that the equations of motion for the twisted gauge fields to second order read
\begin{equation}
(\tadD \somega^{(2)}) \psi=\tilde{\mathcal{V}}(\Omega,\pomega,\pC) \,.
\end{equation}
As explained in Subsection~\ref{sec:twistedSecondOrder} the source term will in general depend on $g_0$, $d_0$, $g_1$ and $d_1$. An explicit calculation shows that the source term is given by
\begin{align}
\tilde{\mathcal{V}}(\Omega,\pomega,\pC)=\phi h^{\ga \ga} \int \d\xi\, \d\eta\, \left\{ K_{\ga \ga}\, \pC(\xi,\phi|x)\, \pomega(\eta,-\phi|x) \, \psi  + \bar{K}_{\ga \ga}\, \pomega(\xi,\phi|x) \, \pC(\eta,\phi|x) \, \psi \right\} \,.
\end{align}
The kernels are given by
{\footnotesize \allowdisplaybreaks
\begin{align}
K_{\ga \ga} =\int_0^1 \d t \, & \left\{ \frac{1}{4} (t^2-1) \, (y_\ga-\eta_\ga-\xi_\ga) (y_\ga-\eta_\ga-\xi_\ga) \, e^{i (y-\eta)(t \xi + \eta)} \notag \right. \\
&\left. + \frac{1}{4} \, (t^2-1) \, g_0  \, (y_\ga y_\ga + \eta_\ga \eta_\ga + \xi_\ga \xi_\ga - 2 y_\ga \eta_\ga) \, \cos(t (y-\eta) \xi)e^{i y \eta} \notag \right. \\
&\left. - \frac{1}{2} \, (t^2-1) \, d_0  \, (y_\ga \xi_\ga + \xi_\ga \eta_\ga) \, \cos(t (y-\eta) \xi)e^{i y \eta} \notag \right. \\
& \left. + \frac12 \eta_\ga (y_\ga -\xi_\ga+(2t-1)\eta_\ga) \, e^{i [(1-t)y-t\xi]\eta}  \right. \\ 
& \left. -\frac14 (t^2-1) (y_\ga - \xi_\ga + \eta_\ga)(y_\ga - \xi_\ga + \eta_\ga) e^{i (t y + \eta)(t y + \xi)} \right. \notag \\ 
& \left. -\frac14 (t^2-1) \, g_1 \, (y_\ga y_\ga + \xi_\ga \xi_\ga + \eta_\ga \eta_\ga - 2 \xi \eta) \, \cos(t y (\xi - \eta)) e^{i \eta \xi} \right. \notag \\
& \left. -\frac12 (t^2-1) \, d_1 \, (y_\ga \eta_\ga - y_\ga \xi_\ga) \, \cos(t y (\xi - \eta)) e^{i \eta \xi} \right\} \,, \notag
\end{align}} \noindent
and also by
{\footnotesize \allowdisplaybreaks
\begin{align}
\bar{K}_{\ga \ga} =\int_0^1 \d t \,  & \left\{ \frac{-1}{4} (t^2-1) (y_\ga+\xi_\ga-\eta_\ga) (y_\ga+\xi_\ga-\eta_\ga) \,  e^{i (y+\xi)(t \eta - y)} \notag  \right.\\
& \left. -\frac{1}{4} (t^2-1) \, g_0 \, (y_\ga y_\ga+\xi_\ga \xi_\ga + 2 y_\ga \xi_\ga +\eta_\ga \eta_\ga)  \,  \cos(t(y+\xi)\eta) e^{i  y \xi} \notag  \right.\\
& \left. -\frac{1}{2} (t^2-1) \, d_0 \, (y_\ga \eta_\ga+\xi_\ga \eta_\ga)  \,  \cos(t(y+\xi)\eta) e^{i y \xi} \notag  \right.\\
& \left. + \frac12 \xi_\ga (-y_\ga+\eta_\ga+(2t-1)\xi_\ga) \, e^{i [(1-t)y-t\eta]\xi}  \right. \\ 
& \left. +\frac14 (t^2-1) (y_\ga - \xi_\ga - \eta_\ga)(y_\ga - \xi_\ga - \eta_\ga) e^{i (t y - \eta)(t y + \xi)} \right. \notag \\
& \left. +\frac14  \,(t^2-1) \, g_1 \, (y_\ga y_\ga + \xi_\ga \xi_\ga + \eta_\ga \eta_\ga + 2 \xi_\ga \eta_\ga) \cos(t y(\xi+\eta)) e^{i \xi \eta} \right. \notag \\
& \left. -\frac12  \,(t^2-1) \, d_1 \, (y_\ga \eta_\ga + y_\ga \xi_\ga) \cos(t y(\xi+\eta)) e^{i \xi \eta} \right\} \notag \,.
\end{align}}\noindent
We will show that this term is exact only for the choice $g_0=d_0=g_1=d_1$ in the following. We will discuss this for the kernel $\bar{K}_{\ga \ga}$ here, but we also checked that the kernel $K_{\ga \ga}$ can be removed. For this purpose it is advantageous to decompose the kernel $K_{\ga \ga}$ as follows
\begin{align}
\bar{K}_{\ga \ga}=\oint &\left( J_1 \, \xi_\ga \xi_\ga + J_2 \, \frac{(\eta_\ga + y_\ga) \xi_\ga}{(s+\tau)}  + J_3 \, \frac{(\eta_\ga-y_\ga)\xi_\ga}{s-\tau} + J_4 \, \frac{(\eta_\ga + y_\ga) (\eta_\ga+y_\ga)}{(s+\tau)(s+\tau)} \right. \notag \\
&\left. + J_5 \,  \frac{(\eta_\ga - y_\ga) (\eta_\ga-y_\ga)}{(s-\tau)(s-\tau)} + J_6 \, \frac{(\eta_\ga+y_\ga)(\eta_\ga-y_\ga)}{(s+\tau)(s-\tau)}  \right)   \exp(i \tau \xi \eta + i s y \xi + i r y \eta) \,.
\end{align}
The contour integration is with respect to the variables $s+\tau$, $s-\tau$ and $r$. As in Appendix~\ref{app:Basis} the $J_i$ are formal series in these three variables
\begin{equation}
J_i=\sum_{n,m} J^{m,n}_i(\tau,r,s)=\sum_{n,m}(s-\tau)^m(s+\tau)^n k_i(r)\,,
\end{equation} 
where we defined $J^{m,n}_i(\tau,r,s)=(s-\tau)^m(s+\tau)^n k_i(r)$.

This decomposition is similar to the one of Appendix~\ref{app:Basis} but we had to slightly modify it as we are now considering the twisted-adjoint covariant derivative $\tadD$ acting on functionals linear in both $\pC$ and $\pomega$ as opposed to the adjoint covariant derivative $\adD$ acting on functionals quadratic in $\pC$ in Appendix~\ref{app:Basis}. 

As discussed in Appendix~\ref{app:Basis} the twisted-adjoint covariant derivative only mixes those $J^{m,n}_i$ which have the same values for $m$ and $n$. By adding an exact term, which we parametrize by $k^{(0)}$, and using the freedom of Fierz identities expressed by three arbitrary functions $\chi_i(r)$ the coefficients $k_i$ change as follows:
\begin{align}
\delta \vec{k}=
\begin{pmatrix}
 k^{(0)} + \chi'_1,\\ 
  k^{(0)} (-1 + s) - n \chi_1 - \chi_2, \\ 
  k^{(0)} (1 + s) - m \chi_1 + \chi'_3, \\
  (-1 + n) \chi_2, \\
  (1-m ) \chi_3, \\ 
  k^{(0)} (-1 + s^2) + m \chi_2 - n \chi_3
\end{pmatrix}
\,, \label{eq:changeUnderFierzAndExact}
\end{align} 
where we denoted $\chi'_i=\partial_r \chi_i$. We distinguish various cases for $n$ and $m$.

\paragraph{For $\boldsymbol{m=1}$ or $\boldsymbol{n=1}$:}By \eqref{eq:changeUnderFierzAndExact} for the choice of $n=1$ or $m=1$ the components $k_4$ and $k_5$ respectively can not be changed by Fierz identities and adding an exact form. These components therefore have to vanish up to polynomials. Upon defining $\ga=r^{-1}$ this leads to
\begin{align}
(n,m)=(1,-1) && \rightarrow && (d_1+g_0-g_1-d_0)  \vec{v}_1 &\sim 0 \notag \,, \\
(n,m)=(1,-2) && \rightarrow && (g_0+g_1-d_0-d_1)  (1-t^2 \ga) \, \vec{v}_1 &\sim 0 \notag \,, \\
(n,m)=(-1,1) && \rightarrow && (g_0-g_1-d_1+d_0)   \vec{v}_2 &\sim 0 \notag \,, \\
(n,m)=(-2,1) && \rightarrow && (d_1-d_0+g_1-g_0)  (1+t^2 \ga) \, \vec{v}_2 &\sim 0 \notag\,,
\end{align}
where $\vec{v}_1=(0,0,0,v,0,0)^T$ and $\vec{v}_2=(0,0,0,0,v,0)^T$ with $v=(-1 + t^2) \frac{\alpha}{8 t^2 \ga^2 -8}$ and integration over $t$ from 0 to 1 is implicit. These equations then imply that
\begin{equation}
g_0=g_1=d_0=d_1 \label{eq:thecatch}
\end{equation}
and it can be easily checked that all other sectors $(1,m)$ and $(n,1)$ for $m,n \leq 0$ also vanish up to polynomials for the choice \eqref{eq:thecatch}.

\paragraph{For $\boldsymbol{n<0}$ and $\boldsymbol{m=0}$:} In this case we can remove the corresponding components of the kernel $\bar{K}_{\ga \ga}$ by adding an exact form with
{\footnotesize
\begin{equation}
k^{(0)} = \int^1_0 \d t \frac{(-1 + t) (1 + t)^{-n} (r + t) (-n + 2 r - (2 - n) t) + (r - 
        t) (2 (r + t) + (1 - t)^{-n} (1 + t) (-n + 
           2 r + (2 - n) t)}{2 i (1 - r^2) (r - t) (r + t)} \,, \notag
\end{equation} 
}where we have restricted ourselves to the choice $g_0=g_1=d_0=d_1=-1$ for the sake of obtaining an expression of reasonable size. However we also made sure that this holds for the general case \eqref{eq:thecatch}.

\paragraph{For $\boldsymbol{m<0}$ and $\boldsymbol{n=0}$:} In this case the source term identically vanishes for the choice $g_0=g_1=d_0=d_1=-1$. We also checked that it is exact for values different from $-1$.

\paragraph{For $\boldsymbol{n<0}$ and $\boldsymbol{m<0}$:} An exact form similar as in the case $n<0$, $m=0$ can be constructed. We will not give its explicit form here as it is quite involved.

By a completely analogous procedure we also checked that the kernel $K_{\ga \ga}$ is exact with respect to the twisted-adjoint covariant derivative $\tadD$. Therefore we have shown that for the choice \eqref{eq:thecatch} we can fully remove the source term $\tilde{\mathcal{V}}(\Omega,\pomega,\pC)$ in \eqref{eq:somega2AfterRedef} by a pseudo-local field redefinition.

\section{Basis}
In this appendix we will summarize a few aspects of two different basis for the various backreactions studied in this paper. We will first focus on the index form in Subsection~\ref{app:IndexForm} and then study the integral form in Subsection~\ref{app:Basis}.

\subsection{Index Basis}
\label{app:IndexForm}
\setcounter{equation}{0}
This appendix is devoted to describing the various details of the index form for the backreactions discussed in Section~\ref{app:technicaldetailsBackreaction}. This form can be obtained by Taylor expanding a backreaction,
\begin{align}
J(y)=\sum_k \frac{1}{k!} J_{\ga(k)} y^\ga...y^\ga\,.
\end{align}
In the following we will discuss in Appendix~\ref{app:indexFormFromFourierSpace} how one can efficiently obtain the index form $J_{\ga(k)}$ from the Fourier-space expression of Section~\ref{app:indexFormFromFourierSpace}. Then we will discuss in Section~\ref{app:conservationIndexForm} how one can derive a conservation identity in this formalism, which we generalize in Section~\ref{app:indexFormDiffDegree0and1} to other form-degrees. As we will see the index representation is only fixed up to Fierz identities which we will discuss in Section~\ref{app:fierz}. In Section~\ref{app:FronsdalCurrents} we will then focus on how to solve the torsion constraint, as discussed in Section~\ref{sec:PresentationBackreactionFronsdal}.

\subsubsection{Obtaining the Index Basis from Fourier Space}
\label{app:indexFormFromFourierSpace}
The most general two-form structure for the physical backreaction reads
\begin{align}\label{mostgentensor}
J_{\ga(2s=m+n)}&= \alpha_1^{n,m,l}H_{\ga\ga} C_{\ga(n-1)\nu(l)}C\fud{\nu(l)}{\ga(m-1)}+
\alpha_2^{n,m,l}H\fdu{\ga}{\gb} C_{\ga(n-1)\gb\nu(l)}C\fud{\nu(l)}{\ga(m)}\notag\\
&+ \alpha_3^{n,m,l}H\fdu{\ga}{\gb} C_{\ga(n)\nu(l)}C\fud{\nu(l)}{\ga(m-1)\gb}+
\alpha_4^{n,m,l}H^{\gb\gb}C_{\ga(n)\gb\gb\nu(l)}C\fud{\nu(l)}{\ga(m)}\\
&+\alpha_5^{n,m,l}H^{\gb\gb}C_{\ga(n)\nu(l)}C\fud{\nu(l)}{\ga(m)\gb\gb}+
\alpha_6^{n,m,l}H^{\gb\gb}C_{\ga(n)\gb\nu(l)}C\fud{\nu(l)}{\ga(m)\gb}\,,\notag
\end{align}
where $\alpha_2$, $\alpha_3$ and  $\alpha_4$, $\alpha_5$ are not really independent unless the fields have additional Yang--Mills indices. There is an additional ambiguity due to Fierz identities, which we will discuss below. 

We can extract these coefficients from expressions in Fourier space which have the following most general form (omitting all integrals):
\begin{align}
H^{\gc\gc} \xi_{\gc(A'')}\eta_{\gc(B'')}y_{\gc(2-A''-B'')} (y\xi)^{A'}(y\eta)^{B'}(\eta\xi)^{C'} \exp{i(ay\xi+by\eta+c\eta\xi)} P(t,q) \, \pC(\xi,\phi) \pC(\eta,-\phi) \,,
\end{align}
where $a,b,c$ are possibly functions of $t,q$, constants or zero. Then the coefficient of
\begin{align}
H\fud{\gb(A''+B'')}{\ga(2-A''-B'')} \pC_{\gb(A'')\ga(n)\nu(l)} \pC\fud{\nu(l)}{\gb(B'')\ga(m)}\,,
\end{align}
is found to be 
\begin{multline}
f^{n,m,l}(A',B',C'|A'',B'')=\frac{(-)^{A''+B''}(-i)^{A''+B''+A'+B'+l+C'}(m+n-A''-B''+2)!}{(n-A')!(m-B')!(l-C')!} \label{eq:coeffFunc} \\\times\int \d t\,\d q\, a^{n-A'}b^{m-B'}c^{l-C'} P(t,q)\,,
\end{multline}
which is related in a simple way to $\alpha_i$  by
\begin{align*}
\alpha_1^{n,m,l}&=f^{n-1,m-1,l}(A',B',C'|0,0)\,, &
\alpha_2^{n,m,l}&=f^{n-1,m,l}(A',B',C'|1,0)\,, \\
\alpha_3^{n,m,l}&=f^{n,m-1,l}(A',B',C'|0,1)\,, &
\alpha_4^{n,m,l}&=f^{n,m,l}(A',B',C'|2,0)\,, \\
\alpha_5^{n,m,l}&=f^{n,m,l}(A',B',C'|0,2)\,, &
\alpha_6^{n,m,l}&=f^{n,m,l}(A',B',C'|1,1)\,. 
\end{align*}
Note that for vanishing parameter $a$ in \eqref{eq:coeffFunc} the  corresponding term has to be replaced by $\delta_{n,A'}$ and analogously for $b$ and $c$. Therefore the coefficients $f^{n,m,l}_{N,M,L}$ of \eqref{eq:fnml} are given by
\begin{equation}
f^{n,m,l}_{N,M} = (-1)^{N+M}(-i)^{N+M+n+m+2l}(m+n-N-M+2)! \,.
\end{equation}

\subsubsection{Conservation in Index Form}
\label{app:conservationIndexForm}
Since there is only one three-form structure
\begin{align}
H \pC_{\ga(n)\nu(l)} \pC\fud{\nu(l)}{\ga(m)}\,,
\end{align}
imposing the conservation leads to a single identity among six $\alpha_i^{n,m,l}$ :

{\footnotesize{
\begin{align}
&\alpha_{4}^{n,m,l}\frac{-\sigma(n+l+2)(n+l+3)}{3}
-\alpha_{4}^{n-1,m+1,l-1}\frac{(m+1)}{3}
+\alpha_{4}^{n-2,m+2,l}\frac{\sigma (m+1)(m+2)}{3}
+\alpha_{4}^{n,m,l-2}\frac{-\sigma}{3}+\notag\\
&\alpha_{5}^{n,m,l}\frac{\sigma(m+l+2)(m+l+3)}{3}
-\alpha_{5}^{n+1,m-1,l-1}\frac{-(n+1)}{3}
+\alpha_{5}^{n+2,m-2,l}\frac{-\sigma (n+1)(n+2)}{3}
+\alpha_{5}^{n,m,l-2}\frac{\sigma}{3}+\notag\\
&\alpha_{6}^{n+1,m-1,l}\frac{-\sigma(n+1)(n+l+3)}{3}
+\alpha_{6}^{n-1,m+1,l}\frac{\sigma(m+1)(m+l+3)}{3}
-\alpha_{6}^{n,m,l-1}\frac{(m-n)}{6}\label{conserv}+\\
&\alpha_{1}^{n+1,m-1,l+2}\frac{-\sigma (l+2)(l+1)}{3}
+\alpha_{1}^{n-1,m+1,l+2}\frac{\sigma(l+2)(l+1)}{3}
+\alpha_{1}^{n-1,m+1,l}\frac{\sigma}{3}
+\alpha_{1}^{n+1,m-1,l}\frac{-\sigma}{3}\notag+\\
&\alpha_{2}^{n+1,m-1,l+1}\frac{-\sigma(l+1)(n+l+3)}{3}
-\alpha_{2}^{n,m,l}\frac{(m+n+2)}{6}
+\alpha_{2}^{n+1,m-1,l-1}\frac{-\sigma}{3}
+\alpha_{2}^{n-1,m+1,l+1}\frac{-\sigma (l+1)(m+1)}{3}\notag+\\
&\alpha_{3}^{n-1,m+1,l+1}\frac{-\sigma(l+1)(m+l+3)}{3}
-\alpha_{3}^{n,m,l}\frac{(m+n+2)}{6}
+\alpha_{3}^{n-1,m+1,l-1}\frac{-\sigma}{3}
+\alpha_{3}^{n+1,m-1,l+1}\frac{-\sigma (l+1)(n+1)}{3}\notag=0\,,
\end{align}}}\noindent 
where $\sigma=i/2$ is the coefficient in the equations of motion
\be
\nabla \pC^{\ga(k)}=-2\sigma \phi h^{\ga\ga}\pC^{\ga(k-2)}+\sigma \phi h_{\gb\gb}\pC^{\ga(k)\gb\gb}\,.
\ee
\subsubsection{Index Form of Differential at Degree $0$ and $1$}
\label{app:indexFormDiffDegree0and1}
The most general one-form structure reads
\begin{align}\label{mostgenexact}
\formK_{\ga(2s=m+n)} &=\beta_1^{n,m,l}h_{\ga\ga} C_{\ga(n-1)\nu(l)}C\fud{\nu(l)}{\ga(m-1)}+
\beta_2^{n,m,l}h\fdu{\ga}{\gb} C_{\ga(n-1)\gb\nu(l)}C\fud{\nu(l)}{\ga(m)}\notag\\
&+\beta_3^{n,m,l}h\fdu{\ga}{\gb} C_{\ga(n)\nu(l)}C\fud{\nu(l)}{\ga(m-1)\gb}+
\beta_4^{n,m,l}h^{\gb\gb}C_{\ga(n)\gb\gb\nu(l)}C\fud{\nu(l)}{\ga(m)}\\
&+\beta_5^{n,m,l}h^{\gb\gb}C_{\ga(n)\nu(l)}C\fud{\nu(l)}{\ga(m)\gb\gb}+
\beta_6^{n,m,l}h^{\gb\gb}C_{\ga(n)\gb\nu(l)}C\fud{\nu(l)}{\ga(m)\gb}\,,\notag
\end{align}
which leads to the following transformations of $\alpha_i$ that parametrize $\adD \formK$ ($\phi$ is omitted):

{\allowdisplaybreaks
{\footnotesize{
\begin{align}\label{differential}
\delta \alpha_2^{n,m,l}&=\beta_4^{n-1,m+1,l-1}\frac{(m+1)}{2}+\beta_4^{n,m,l}\frac{-2\sigma(n+l+3)n}{2}+\beta_4^{n-2,m+2,l}\frac{2\sigma(m+2)(m+1)}{2}\,,\notag\\
\delta \alpha_4^{n,m,l}&=\beta_4^{n-1,m+1,l+1}\frac{2\sigma(m+1)(l+1)}{2}+\beta_4^{n,m,l}\frac{-n}{2}\,,\notag\\
\delta \alpha_6^{n,m,l}&=\beta_4^{n-1,m+1,l}\frac{-(m+1)}{2}+\beta_4^{n,m,l+1}\frac{2\sigma(n+l+4)(l+1)}{2}+\sigma\beta^{n,m,l-1}_4\,,\notag\\
\delta \alpha_3^{n,m,l}&=\beta_5^{n+2,m-2,l}\frac{-2\sigma(n+1)(n+2)}{2}+\beta_5^{n,m,l}\frac{2\sigma m(m+l+3)}{2}+\beta_5^{n+1,m-1,l-1}\frac{-(n+1)}{2}\,,\notag\\
\delta \alpha_5^{n,m,l}&=\beta_5^{n+1,m-1,l+1}\frac{2\sigma(n+1)(l+1)}{2}+\beta_5^{n,m,l}\frac{-m}{2}\,,\notag\\
\delta \alpha_6^{n,m,l}&=\sigma\beta_5^{n,m,l-1}+\beta_5^{n,m,l+1}\frac{2\sigma (m+l+4)(l+1) }{2}+\beta_5^{n+1,m-1,l}\frac{-(n+1)}{2}\,,\notag\\
\delta \alpha_1^{n,m,l}&=\beta_1^{n+1,m-1,l+1}\frac{-2\sigma n(l+1)}{2}+\beta_1^{n-1,m+1,l+1}\frac{-2\sigma m (l+1)}{2}+\beta_1^{n,m,l}\frac{(n+m+2)}{2}\,,\notag\\
\delta \alpha_2^{n,m,l}&=-\beta_1^{n-1,m+1,l+2}\frac{2\sigma (l+1)(l+2)}{2}-\beta_1^{n-1,m+1,l}\sigma\,,\notag\\
\delta \alpha_3^{n,m,l}&=\beta_1^{n+1,m-1,l+2}\frac{2\sigma (l+1)(l+2)}{2}+\beta_1^{n+1,m-1,l}\sigma\,,\notag\\
\delta \alpha_1^{n,m,l}&=\beta_2^{n+1,m-1,l}\frac{-\sigma n(n+l+3)}{2}+\beta_2^{n-1,m+1,l}\frac{\sigma m(m+1)}{2}+\beta_2^{n,m,l-1}\frac{m}{4}\,,\notag\\
\delta \alpha_2^{n,m,l}&=\beta_2^{n+1,m-1,l+1}\frac{-\sigma n(l+1)}{2}+\beta_2^{n,m,l} (1+\frac{m}4)\,,\notag\\
\delta \alpha_3^{n,m,l}&=\beta_2^{n+1,m-1,l+1}\frac{\sigma (n+l+4)(l+1)}{2}+\beta_2^{n+1,m-1,l-1}\frac{\sigma}{2}+\beta_2^{n,m,l}\frac{-m}{4}\,,\notag\\
\delta \alpha_4^{n,m,l}&=\beta_2^{n,m,l}\frac{-\sigma}{2}+ \beta_2^{n,m,l+1}\frac{-\sigma(l+2)(l+1)}{2}\\
\delta \alpha_6^{n,m,l}&=\beta_2^{n+1,m-1,l+2}\frac{\sigma (l+2)(l+1)}{2} + \beta_2^{n+1,m-1,l}\frac{\sigma}{2}\,,\notag\\
\delta \alpha_1^{n,m,l}&=\beta_3^{n-1,m+1,l}\frac{\sigma m(m+l+3)}{2}+\beta_3^{n-1,m+1,l}\frac{-\sigma n(n+1)}{2}+\beta_3^{n,m,l-1}\frac{-n}{4}\,,\notag\\
\delta \alpha_2^{n,m,l}&=\beta_3^{n-1,m+1,l+1}\frac{\sigma (m+l+4)(l+1)}{2}+\beta_3^{n-1,m+1,l-1}\frac{\sigma}{2}+\beta_3^{n,m,l}\frac{-n}{4}\,,\notag\\
\delta \alpha_3^{n,m,l}&=\beta_3^{n-1,m+1,l+1}\frac{-\sigma m(l+1)}{2}+\beta_3^{n,m,l} (1+\frac{n}4)\,,\notag\\
\delta \alpha_5^{n,m,l}&=\beta_3^{n,m,l}\frac{\sigma}{2}+ \beta_3^{n,m,l+1}\frac{\sigma(l+2)(l+1)}{2}\,,\notag\\
\delta \alpha_6^{n,m,l}&=\beta_3^{n-1,m+1,l+2}\frac{-\sigma (l+2)(l+1)}{2} - \beta_3^{n-1,m+1,l}\frac{\sigma}{2}\,,\notag\\
\delta \alpha_2^{n,m,l}&=\beta_6^{n-1,m+1,l}\frac{\sigma (m+1)(m+l+4)}{2}+\beta_6^{n+1,m-1,l}\frac{-\sigma n(n+1)}{2}+\beta_6^{n,m,l-1}\frac{-n}4\,,\notag\\
\delta \alpha_3^{n,m,l}&=\beta_6^{n+1,m-1,l}\frac{-\sigma (n+1)(n+l+4)}{2}+\beta_6^{n-1,m+1,l}\frac{\sigma m(m+1)}{2}+\beta_6^{n,m,l-1}\frac{m}4\,,\notag\\
\delta \alpha_4^{n,m,l}&=\beta_6^{n,m,l+1}\frac{\sigma (l+1)(m+l+4)}{2}+\beta_6^{n,m,l-1}\frac{\sigma}{2}+\beta_6^{n+1,m-1,l}\frac{-(n+1)}{4}\,,\notag\\
\delta \alpha_5^{n,m,l}&=\beta_6^{n,m,l+1}\frac{\sigma (l+1)(n+l+4)}{2}+\beta_6^{n,m,l-1}\frac{\sigma}{2}+\beta_6^{n-1,m+1,l}\frac{-(m+1)}{4}\,,\notag\\
\delta \alpha_6^{n,m,l}&=\beta_6^{n+1,m-1,l+1}\frac{\sigma(n+1)(l+1)}{2}+\beta_6^{n-1,m+1,l+1}\frac{\sigma (m+1)(l+1)}{2}+\beta_6^{n,m,l}\frac{-(n+m)}{4}\,.\notag
\end{align}}}} \noindent 
As a consequence of $\adD\adD\equiv 0$ such $\alpha$'s obey the conservation identity \eqref{conserv}. Applying $\adD$ to the most general zero-form
\begin{align}
\gamma^{n,m,l} \pC_{\ga(n)\nu(l)} \pC\fud{\nu(l)}{\ga(m)}\,, \label{mostgeneraldoubleexact}
\end{align}
gives a variation of $\beta^{n,m,l}_i$ that does not affect $\alpha_i^{n,m,l}$:
\begin{align}
\delta \beta_1^{n,m,l}&=\gamma^{n+1,m-1,l}(-\sigma n(n+1))+\gamma^{n-1,m+,l}(2\sigma m (m+1))\,, \notag\\
\delta \beta_2^{n,m,l}&=\gamma^{n,m,l}(-n)+\gamma^{n-1,m+1,l+1}(2\sigma (l+1)(m+1))\,,\notag\\
\delta \beta_3^{n,m,l}&=\gamma^{n,m,l}(-m)+\gamma^{n-1,m+1,l+1}(2\sigma (l+1)(n+1))\,,\label{differentialC}\\
\delta \beta_4^{n,m,l}&=\gamma^{n,m,l+2}(+\sigma (l+2)(l+1)) +\gamma^{n,m,l} \sigma\,,\notag\\
\delta \beta_5^{n,m,l}&=\gamma^{n,m,l+2}(-\sigma (l+2)(l+1)) -\gamma^{n,m,l} \sigma\,,\notag\\
\delta \beta_6^{n,m,l}&=0\,.\notag
\end{align}

\subsubsection{Fierz Identities}\label{app:fierz}
Not all of the six $\alpha$'s are independent due to the Fierz identities, which also play an important role in simplifying four-fermion terms in the backreaction of Appendix~\ref{app:simplifiedBackreaction}. There are three independent Fierz identities that can be obtained from
\begin{align}
H_{x\mu} \pC_{\ga(n-1)\nu(l+1)} \pC\fud{\nu(l+1)}{\ga(m)}+
H\fdu{x}{\gc} \pC_{\mu\ga(n-1)\nu(l)} \pC\fud{\nu(l)}{\ga(m)\gc}-
H\fdu{x}{\gc} \pC_{\gc\ga(n-1)\nu(l)} \pC\fud{\nu(l)}{\ga(m)\mu}\equiv0\,,
\end{align}
by contracting or symmetrizing $x$ with some of the $\pC$'s and symmetrizing $\mu$ with $\ga$'s:
\besubeqs
\begin{align}
H_{\ga\ga} \pC_{\ga(n-1)\nu(l+1)} \pC\fud{\nu(l+1)}{\ga(m)}+
H\fdu{\ga}{\gc} \pC_{\ga(n)\nu(l)} \pC\fud{\nu(l)}{\ga(m)\gc}-
H\fdu{\ga}{\gc} \pC_{\gc\ga(n-1)\nu(l)} \pC\fud{\nu(l)}{\ga(m+1)}\equiv0\,,\\
H\fud{\gc}{\ga} \pC_{\gc\ga(n-1)\nu(l+1)} \pC\fud{\nu(l+1)}{\ga(m)}+
H^{\gc\gc} \pC_{\gc\ga(n)\nu(l)} \pC\fud{\nu(l)}{\ga(m)\gc}-
H^{\gc\gc} \pC_{\gc\gc\ga(n-1)\nu(l)} \pC\fud{\nu(l)}{\ga(m+1)}\equiv0\,,\\
H\fud{\gc}{\ga} \pC_{\ga(n-1)\nu(l+1)} \pC\fud{\nu(l+1)}{\ga(m)\gc}+
H^{\gc\gc} \pC_{\ga(n)\nu(l)} \pC\fud{\nu(l)}{\ga(m)\gc\gc}-
H^{\gc\gc} \pC_{\gc\ga(n-1)\nu(l)} \pC\fud{\nu(l)}{\ga(m+1)\gc}\equiv0\,.
\end{align}
\esubeqs
This leads to the following transformations of the coefficients that do no affect the expression but only its presentation in terms of $\alpha$'s:
\begin{align}
\delta \alpha^{n,m,l}_2&=\epsilon^{n-1,m,l-1}_1\,,& 
\delta \alpha^{n,m,l}_4&=-\epsilon^{n,m-1,l}_1\,, &
\delta \alpha^{n,m,l}_6&=\epsilon^{n-1,m,l}_1\,,\notag\\
\delta \alpha^{n,m,l}_3&=\epsilon^{n,m-1,l-1}_2\,, &
\delta \alpha^{n,m,l}_5&=\epsilon^{n-1,m,l}_2\,,\label{FierzAlpha}&
\delta \alpha^{n,m,l}_6&=-\epsilon^{n,m-1,l}_2\,,\\
\delta \alpha^{n,m,l}_1&=\epsilon^{n-1,m-1,l-1}_3\,,&
\delta \alpha^{n,m,l}_2&=-\epsilon^{n-1,m-1,l}_3\,,&
\delta \alpha^{n,m,l}_3&=\epsilon^{n-1,m-1,l}_3\,.\notag
\end{align}
Here all $\epsilon$'s are understood to be vanishing for the case of at least one negative index. The conservation identity is invariant under Fierz transformations. Analogous formulas can be derived for Fourier-space representation of two-forms, but it is somewhat difficult to effectively use Fierz identities due to the appearance of fake poles like $y_\ga=\frac{\eta_\ga(y\xi)-\xi_\ga(y \eta)}{(\eta\xi)}$.

There is a natural way of fixing all Fierz transformations. Any one- or two-form $J(y)$ can be decomposed into three zero-forms as
\begin{align}\label{Symmetrizedform}
\formJ(y)= H^{\gb\gb}\pl_\gb\pl_\gb \compA(y)+H\fdu{\ga}{\gb}y^\ga \pl_\gb \compB(y) + H_{\ga\ga}y^\ga y^\ga\compC(y)\,,
\end{align}
which in components corresponds to \eqref{gentwoformdec}.
One can solve for the Fierz transformations that map six $\alpha$'s into just three sets of coefficients as
\begin{align}
\compA(y)&= \sum_{n,m,l} a^{n,m,l} \frac{1}{(n+m+2)!}\pC_{\ga(n+1)\nu(l)}\pC\fud{\nu(l)}{\ga(m+1)} y^{\ga(n+m+2)}\,,\\
\compB(y)&= \sum_{n,m,l} b^{n,m,l} \frac{1}{(n+m)!}\pC_{\ga(n)\nu(l)}\pC\fud{\nu(l)}{\ga(m)} y^{\ga(n+m)}\,,\\
\compC(y)&= \sum_{n,m,l} c^{n,m,l} \frac{1}{(n+m-2)!}\pC_{\ga(n-1)\nu(l)}\pC\fud{\nu(l)}{\ga(m-1)} y^{\ga(n+m-2)}\,,
\end{align}
which requires us to impose the following relations among $\alpha_{2,3,4,5,6}$ by applying Fierz identities:
\begin{align}
&\frac{\alpha_4^{n-1,m+1,l}}{n(n+1)}=\frac{\alpha_5^{n+1,m-1,l}}{m(m+1)}=\frac{\alpha_6^{n,m,l}}{2(m+1)(n+1)}\,, && \frac{\alpha_2^{n,m,l}}{n}=\frac{\alpha_3^{n,m,l}}{m}\,,
\end{align}
that can be solved unambiguously for $\epsilon^{n,m,l}_{1,2,3}$. In fact, there are three invariants $I_4,I_5,I_6$ of the Fierz transformations and the $a,b,c$ coefficients are linear combinations thereof,

{\footnotesize{
\begin{align*}
a^{n,m,l}&=I_{6}^{n,m,l}\,, \\
b^{n,m,l}(m+n)&=I^{n,m,l}_5+I^{n,m,l}_4+\frac{m-n}{m+n+2}I_6^{n,m,l-1} \,,\\
c^{n,m,l}&=\frac{-m}{m+n}\left(\frac{n}{m+n+1}I_6^{n,m,l-2}+\frac{n}{m}I^{n,m,l-1}_5-I_4^{n,m,l-1}\right) \,, 
\end{align*}}}
where we have defined
{\footnotesize{
\begin{align*}
I_{6}&=\alpha^{n,m,l}_6+\alpha^{n+1,m-1,l}_5+\alpha^{n-1,m+1,l}_4\,,\\
I_{5}&=\alpha^{n,m,l}_3-\alpha^{n+1,m-1,l-1}_5-\alpha^{n,m,l+1}_1\,,\\
I_{4}&=\alpha^{n,m,l}_2+\alpha^{n-1,m+1,l-1}_4+\alpha^{n,m,l+1}_1 \,.
\end{align*}}}
In terms of the $a,b,c$ coefficients the conservation identity is given by
{\footnotesize{
\begin{align}
\label{eq:conservationABCBasis}
&\frac{\sigma }{3 (1+m+n) (2+m+n)}((1+m) (2+m)a^{-1+n,1+m,-2+l}-(1+n) (2+n) a^{1+n,-1+m,-2+l})+\\
&\frac{(2+l+m+n) (3+l+m+n) \sigma  }{3 (1+m+n) (2+m+n)}((1+m) (2+m) a^{-1+n,1+m,l}-(1+n)\notag (2+n)a^{1+n,-1+m,l})\\
&-\frac{1}{3} (1+n) \sigma  b^{1+n,-1+m,-1+l}-\frac{1}{3} (1+l) (2+l+m+n) \sigma ((1+m)  b^{-1+n,1+m,1+l}+ (1+n) b^{1+n,-1+m,1+l})\notag\\
&-\frac{1}{6} (m+n) (2+m+n) b^{n,m,l}+\frac{1}{3} (1+l) (2+l) \sigma  (c^{-1+n,1+m,2+l}-c^{1+n,-1+m,2+l})+\notag\\
&\frac{\sigma  c^{-1+n,1+m,l}}{3}-\frac{\sigma  c^{1+n,-1+m,l}}{3}-\frac{1}{3} (1+m) \sigma  b^{-1+n,1+m,-1+l}=0\,.\notag
\end{align}}}

\noindent By only considering the terms proportional to $\sigma$ in the above expression the action of $\nabla$ on \eqref{Symmetrizedform} can be obtained.

\subsubsection{Fronsdal Currents}\label{app:FronsdalCurrents}
In solving for the Fronsdal current from the backreaction one needs to evaluate $\formj=(I-\nabla Q^{-1})\formJ$ as we discussed in Section~\ref{sec:PresentationBackreactionFronsdal}. Assuming that the backreaction is given in the form \eqref{Symmetrizedform}, the middle component of $\formj$ vanishes as discussed in Section~\ref{sec:PresentationBackreactionFronsdal} and hence
\begin{align}
\formj(y)= H^{\gb\gb}\pl_\gb\pl_\gb \compj(y)+ H_{\ga\ga}y^\ga y^\ga \compj'(y)\,,
\end{align}
where
\begin{align}
\compj(y)&= \sum_{n,m,l} a^{n,m,l}_F \frac{1}{(n+m+2)!}\pC_{\ga(n+1)\nu(l)}\pC\fud{\nu(l)}{\ga(m+1)} y^{\ga(n+m+2)}\,,\\
\compj'(y)&= \sum_{n,m,l} c^{n,m,l}_F \frac{1}{(n+m-2)!}\pC_{\ga(n-1)\nu(l)}\pC\fud{\nu(l)}{\ga(m-1)} y^{\ga(n+m-2)}\,.
\end{align}
The coefficients $a^{n,m,l}_F$ and $c^{n,m,l}_F$ for the Fronsdal current $\formj$ can be expressed in terms of those for $\formJ$ given in the basis \eqref{Symmetrizedform}:

{\footnotesize \allowdisplaybreaks
\begin{align}
a^{n,m,l}_{F}&=a^{n,m,l}+\frac{2\sigma}{(n+m+2)(m+n)}(-(2+m)( a^{-1+n,1+m,-1+l}+(1+l) (4+l+m+n) a^{-1+n,1+m,1+l})+\notag\\&\qquad+(2+n) (-a^{1+n,-1+m,-1+l}-(1+l) (4+l+m+n) a^{1+n,-1+m,1+l}))\notag\\
&-  \frac{\sigma}{2} (m+n) (b^{-1+n,1+m,l}-b^{1+n,-1+m,l})+\frac{\sigma}{2} (1+l) (2+l) (m+n) (b^{-1+n,1+m,2+l}-b^{1+n,-1+m,2+l})\,,\notag\\
c^{n,m,l}_F&=c^{n,m,l}+\frac{2\sigma }{(m+n) (2+m+n)}(-m c^{-1+n,1+m,-1+l}-n c^{1+n,-1+m,-1+l}+\label{FronsdalCoefs}\\
&\qquad-(1+l)  (l+m+n)( c^{-1+n,1+m,1+l}- n c^{1+n,-1+m,1+l}))\notag\\
&+\frac{\sigma  (2+m+n) }{2 (m+n) (1+m+n)}(m (1+m) b^{-1+n,1+m,-2+l}-n (1+n) b^{1+n,-1+m,-2+l})+\notag\\
&+\frac{\sigma  (2+m+n) (l+m+n) (1+l+m+n) }{2 (m+n) (1+m+n)}(m (1+m) b^{-1+n,1+m,l}-n (1+n) (2+m+n) b^{1+n,-1+m,l})\,.\notag
\end{align}}

\subsubsection{Local Conserved Tensors} \label{app:localcurrents}
The canonical spin-$s$ conserved tensor has $s$-derivatives and is fixed up to an overall factor. The simplest way to get a generating function for all such tensors is to take  
\begin{align}
T(y|x)=\pC(y,\phi|x)\pC(y,-\phi|x)\,,
\end{align}
which can be checked to obey $\nabla^{\ga\ga}\pl_\ga^y\pl_\ga^y T=0$. The freedom in the relative factors can be taken into account by  
\begin{align}
T^{can}(y|x)=\int dt\, f(t) \pC(yt,\phi|x)\pC(yt,-\phi|x)\,,
\end{align}
where $t$ counts the rank of the conserved tensors. The dual closed two-forms are 
\begin{align}
\formJ^{can}(y|x)=H^{\ga\ga}\pl_\ga \pl_\ga \int dt\, f(t) \pC(yt,\phi|x)\pC(yt,-\phi|x)\,,
\end{align}
and an equivalent form in Fourier space (to be compared with the formulas of Appendix \ref{app:simplifiedBackreaction}) is
\begin{align}
\formJ^{can}(y|x)=H^{\ga\ga} \int dt\, (-t^2f(t))(\xi-\eta)_\ga(\xi-\eta)_\ga e^{ity(\xi-\eta)} \pC(\xi,\phi|x) \pC(\eta,-\phi|x)\,.
\end{align}
The moments $f_n$ of $f(t)$ make relative factors in front of conserved tensors. The index form of the above expression is
\begin{align}
\alpha^{n,m,l}_4&=\frac{(-)^{m}(m+n)!}{m!n!}\delta_{l,0}f_{n+m}\,, &&\alpha_4^{n,m,l}=\alpha_5^{n,m,l}=-\frac12\alpha_6^{n,m,l}\,.
\end{align}
All other $\alpha_i \equiv 0$. The coefficients, of course, obey identity \eqref{conserv} that ensures the conservation (it splits into three independent equations). There is no hook part, $b^{n,m,l}=0$, as discussed around \eqref{eq:vanishingOfBcomp} and the tensor is traceless $c^{n,m,l}=0$ 
\begin{align}
a^{n,m,l}&=-\frac{(-1)^m f_{n+m}(m+n+2)! \delta_{l,0}}{(n+1)!(m+1)!}\,.\label{canonicalCura}
\end{align}
One can use $\formJ^{can}$ in two ways: it is a doublet (with respect to $\phi$) of traceless conserved tensors or one can put $\formJ^{can}$ as a source for $\omega(y,\phi)$. In the latter case one finds a nonzero torsion and solving for the Fronsdal current as explained in Appendix \ref{app:FronsdalCurrents} we have that $c^{n,m,l}_F=0$, i.e. the Fronsdal current is still traceless, and 
\begin{align}\label{Canonicala}
a^{n,m,0}_F&=-\frac{(-)^m f_{m+n} (m+n+2)!}{(m+1)! (n+1)!}\,, & a^{n,m,1}_F&=-\frac{i(-1)^m f_{m+n} (m+n+2)!}{(m+2)(m+1)! (n+1)!}\,,
\end{align}
i.e. it involves $\pC \pC$-terms with no more than one index contracted and hence the expression is local but contains higher derivatives. One can show that the canonical current is exact, i.e. $J=DK$, with $c^{n,m,l}=b^{n,m,l}=0$ and
\begin{align}\label{CanonicalExact}
a^{n,m,l}&=-\frac{2 (-1)^m i^l (4+2 l+m+n) f_{m+n} (m+n+1)! (m+n+2)!}{(m+1)! (n+1)! (3+l+m+n)!}\,.
\end{align}
We thus see that representing it as an exact form requires a pseudo-local expression. However, it contains $l!$ in the denominator, which gives a seemingly good asymptotic behavior. Therefore, the redefinition, which is clearly unphysical appears to be a well-defined expression. It is also possible to represent it as a $\nabla$-exact form with
\begin{align}\label{CanonicalNablaExact}
a^{n,m,l}&=-\frac{(-)^m \left(\frac{i}{2}\right)^l \sqrt{\pi } f_{m+n} (m+n+2)! \Gamma\left[\frac{1}{2} (4+m+n)\right]}{\Gamma\left[1+\frac{l}{2}\right] (m+1)! (n+1)!\Gamma\left[\frac{1}{2} (5+l+m+n)\right]} \,.
\end{align}

\subsubsection{Pseudo-local Conserved Tensors} \label{app:quasilocalcurrents}
\paragraph{Example 1.} A simple example of a pseudo-local conserved tensor shows up in the second-order computations:
\begin{align}
\formJ(y|x)=H^{\ga\ga} \int \d t\, f(t)(\xi-\eta)_\ga(\xi-\eta)_\ga e^{it(y+\eta)(y+\xi)} \pC(\xi,\phi|x)\pC(\eta,-\phi|x)\,,
\end{align}
which is conserved for any $f(t)$, the corresponding coefficients being
\begin{align}
\alpha^{n,m,l}_4&=\frac{(-)^{m}(-i)^{l+2}(m+n+l)!}{m!n!l!}f_{n+m+l}\,, &&\alpha_4^{n,m,l}=\alpha_5^{n,m,l}=-\frac12\alpha_6^{n,m,l}\,.
\end{align} 
It comes from \eqref{PstructureA} and \eqref{PstructureB} terms in the second-order perturbation theory. This gives 
\be \label{PtensorA} a^{n,m,l}= \frac{(-)^m (-i)^lf_{m+n+l}(m+n+2)!}{(m+1)!(n+1)!l!} \,.\ee
Solving for the Fronsdal current as explained in Appendix \ref{app:FronsdalCurrents} we find that $c^{n,m,l}_F=0$, i.e. the current is traceless and 
\begin{align}\label{PtensorB}
a^{n,m,l}_F&=\frac{(-1)^m i^{-l} ((m+n) f_{l+m+n}-  l f_{-1+l+m+n}+(4+l+m+n) f_{1+l+m+n}) (m+n+2)!}{(m+n) l! (m+1)! (n+1)!}\,,
\end{align}
which leads to a pseudo-local expression.

\paragraph{Example 2.} There is another choice of the coefficients corresponding to a conserved backreaction
\begin{align}
\alpha^{n,m,l}_4&=f_{n+m}\,, &&\alpha_4^{n,m,l}=\alpha_5^{n,m,l}=-\frac12\alpha_6^{n,m,l}\,.
\end{align}
These coefficients correspond to a pseudo-local expression and have a considerably worse asymptotic behavior since there are no damping factorials. In the symmetrized form we have
\be c^{n,m,l}=f_{n+m}\,.\label{strangestrA}\ee
The Fronsdal current is found to be a total trace, i.e. $a_F^{n,m,l}=0$, and
\begin{align}\label{strangestrB}
c^{n,m,l}_F&=-\frac{i ((1+2 i)+(1+i) (m+n)+l (1+l+m+n)) f_{m+n}}{2+m+n}\,.
\end{align}

\paragraph{Example 3: Canonical backreaction.} As it was mentioned, if one takes the local conserved tensor $\formj^{can}(y,\phi)$ and uses it as a source for $D\pomega(y,\phi)$ one has to solve for the contorsion tensor. As a result the Fronsdal current has terms with one pair of contracted indices. One can solve the inverse problem: what is $\formJ^{Fr}(y,\phi)$ such that it yields the canonical conserved tensor as a Fronsdal current, i.e. the terms $\pC_{\ga(n)\nu(l)}\pC\fud{\nu(l)}{\ga(m)}$ with $l>0$ are absent in $\formj$, the canonical backreaction. Such $\formJ^{Fr}(y,\phi)$ must be pseudo-local since one pair of contracted indices produced by the contorsion tensor needs to be canceled by $l=1$ term from $\formJ^{Fr}(y,\phi)$, which thereby produces $l=2$ terms and so on. The solution is 
\begin{align}\label{Fronsdalbackreaction}
a^{n,m,l}&=-\frac{(-1)^{l+m} i^l (m+n) (4+2 l+m+n) f_{n+m} (m+n+1)! (2+m+n)!}{(m+1)!(n+1)! (l+m+n+3)!}\,,
\end{align}
and the Fronsdal current is exactly \eqref{canonicalCura}. 

This solution is remarkable in the sense that a pseudo-local expression is necessary in order to get the canonical $s$-derivative conserved tensor on the \rhs of the Fronsdal equations provided that the symmetry $\phi\rightarrow -\phi$ of the equations is not broken, i.e. the same expression $\formJ^{Fr}(y,\phi)$ appears on the right hand side of (HS) torsion and Riemann two-forms. In particular this is true for the $s=2$ case of the Einstein equations. 

The Fronsdal backreaction is exact, i.e. can be represented as $\formJ^{Fr}=\adD \formU^{Fr}$ for some $\formU^{Fr}$. The expression is quite cumbersome and we give its leading behavior only
\begin{align}\label{FronsdabckrExact}
a_U^{n,m,l}&=-\frac{i^l (-1)^{l+m} (m+n+1)! (m+n+2)! G_{m+n,l}}{(m+1)! (n+1)! (l+m+n+3)!} \,, \\
G_{k,l}&=-2k f_k\left(l \log\,l+...+ \frac{k^l}{l!}\right)\,.\notag
\end{align}
Therefore it has again a factorially damped asymptotic behavior.

\subsection{Integral Basis}
\label{app:Basis}

In the following we discuss a basis that we use extensively in our analysis of the cohomologies and cocycles in Prokushkin--Vasiliev Theory, which are discussed in Appendix~\ref{app:cohomologies} and Appendix~\ref{app:twistedBackreactions}. We consider $q$-forms that are either linear or quadratic in physical zero-forms $\pC$ and consist of vielbeins. We will focus on the linear case first.

\subsubsection{Basis Linear in $\boldsymbol{\pC}$}
For the various form-degrees the most general ansatz for objects containing $\pC$ linearly is given by
\besubeqs
\label{eq:MassimoAnsatz}
\begin{align}
J_0=&\oint \d \tau \, J^{(0)}(\tau) \pC(y\tau)\,, \label{eq:MassimoAnsatzZeroForm}\\
J_1=i\phi &\oint \d \tau \, h^{\ga\ga}\left(J_1^{(1)}(\tau)y_\ga y_\ga+J_2^{(1)}(\tau)\tau^{-1}y_\ga \partial^{y}_\ga+J_3^{(1)}(\tau)\tau^{-2}\partial^y_\ga \partial^y_\ga\right) \pC(y\tau)\,,\label{eq:MassimoAnsatzOneForm}\\
J_2=&\oint \d \tau \, H^{\ga\ga}\left(J_1^{(2)}(\tau)y_\ga y_\ga+J_2^{(2)}(\tau)\tau^{-1}y_\ga \partial^{y}_\ga+J_3^{(2)}(\tau)\tau^{-2}\partial^y_\ga \partial^y_\ga\right) \pC(y\tau)\,,\\
J_3=i\phi &\oint \d \tau \, H J^{(3)}(\tau) \pC(y\tau)\,,
\end{align}
\end{subequations}
where we have encoded the arbitrary relative coefficients of the different tensor structures by a formal series in $\tau^{-1}$ given by
\begin{equation}
\label{eq:Ji}
J_i^{(k)}(\tau)=\sum_{l=1}^\infty j_{i,l}^{(k)}\tau^{-l-1}\,. 
\end{equation}
We normalize the integration measure such that the following equation holds
\begin{equation}
\oint \d \tau \, \tau^{-k}=\delta_{1,k}\,.
\end{equation}
For illustration purposes let us briefly outline how using \eqref{eq:Ji} the zero-form ansatz \eqref{eq:MassimoAnsatzZeroForm} can be rewritten as
\begin{align}
\sum_{l=0}^\infty \oint \d \tau \, j^{(0)}_l \tau^{-l} \left( \sum_{k=0}^\infty \frac{1}{k!} \pC_{\ga(k)} \tau^k y^{\ga(k)} \right) = \sum_{l,k=0}^\infty \frac{1}{k!} j^{(0)}_l \left( \oint \d \tau \,  \tau^{-l+k} \right) \pC_{\ga(k)} y^{\ga(k)} = \sum_{k=0}^\infty \frac{1}{k!}  j^{(0)}_{k} \pC_{\ga(k)} y^{\ga(k)} \,. \nonumber
\end{align}
Similarly one obtains the following tensor structure for form-degree $q$ in \eqref{eq:MassimoAnsatz}:
\begin{align}
q=0&:& &j^{(0)}_{k} \pC_{\ga(k)}\,,\\
q=1&:& &2 j^{(1)}_{1,k} h^{\ga\ga}\pC^{\ga(k-2)}+j^{(1)}_{2,k} h_{\gb\gb} \pC^{\gb\gb\ga(k)}+ j^{(1)}_{3,k} h\fud{\ga}{\gc}\pC^{\ga(k-1)\gc}\,,\\
q=2&:& &2 j^{(2)}_{1,k} H^{\ga\ga}\pC^{\ga(k-2)}+j^{(2)}_{1,k} H_{\gb\gb} \pC^{\gb\gb\ga(k)}+ j^{(2)}_{1,k} H\fud{\ga}{\gc}\pC^{\ga(k-1)\gc}\,,\\
q=3&:& &j^{(3)}_{k} H \pC_{\ga(k)}\,.
\end{align}
Where we have have only listed the coefficients of the various powers of $y_\ga$-oscillators dropping an overall factorial $\frac{1}{k!}$.

\subsubsection{Basis Quadratic in $\boldsymbol{\pC}$}
For the discussion of cohomologies with respect to pseudo-local field redefinitions \eqref{eq:pseudolocalfieldredefs} that we present in Appendix~\ref{app:cohomologies} it is useful to consider expressions in Fourier space using \eqref{eq:fourier}. We will again use the convention that the first and the second zero-form $\pC$ is associated with wave-twistor $\xi_\ga$ and $\eta_\ga$ respectively as discussed in Appendix~\ref{subsec:fourierspace} and therefore
\begin{equation}
\int \d^2\xi \d^2\eta \, J^{(p)}(\xi,\eta,y) \, \pC(\xi,\phi|x) \, \pC(\eta,-\phi|x) \,.
\end{equation}
It is convenient to define  
\begin{equation}
\zeta^\pm_\ga=(\xi\pm\eta)_\ga \,.
\end{equation}
One can then express the most general ansatz for $p$-forms consisting of vielbeins and quadratic in the zero-form $\pC$ in terms of $J^{(p)}$ by
{\allowdisplaybreaks
\begin{subequations}\label{nicebasis}
\begin{align}
J^{(0)}=&\oint J^{(0)}(s+r)^{-1}(s-r)^{-1} \tilde{\mathcal{K}}\,,\\
J^{(1)}=\phi &\oint h^{\ga\ga}\left(J^{(1)}_1y{}_\ga y{}_\ga+J^{(1)}_2 (s-r)^{-1}y_\ga \zeta^+_\ga+J^{(1)}_3 (s+r)^{-1}y_\ga \zeta^-_\ga \right.\label{nicebasis2}\\&\left.\hspace{60pt}+J^{(1)}_4(s-r)^{-2}\zeta^+_\ga \zeta^+_\ga+J^{(1)}_5(s+r)^{-2}\zeta^-_\ga \zeta^-_\ga+J^{(1)}_6 (s+r)^{-1}(s-r)^{-1}\zeta^+_\ga \zeta^-_\ga\right)\tilde{\mathcal{K}}\,,\nonumber\\
J^{(2)}=\frac{1}{4} &\oint H^{\ga\ga}\left(J^{(2)}_1y{}_\ga y{}_\ga+J^{(2)}_2 (s-r)^{-1}y_\ga \zeta^+_\ga+J^{(2)}_3 (s+r)^{-1}y_\ga \zeta^-_\ga \right.\\&\left.\hspace{60pt}+J^{(2)}_4(s-r)^{-2}\zeta^+_\ga \zeta^+_\ga+J^{(2)}_5(s+r)^{-2}\zeta^-_\ga \zeta^-_\ga+J^{(2)}_6 (s+r)^{-1}(s-r)^{-1}\zeta^+_\ga \zeta^-_\ga\right)\tilde{\mathcal{K}}\,,\nonumber\\
J^{(3)}=\frac{\phi}{6} &\oint H J^{(3)}(s+r)^{-1}(s-r)^{-1}\tilde{\mathcal{K}}\,,
\end{align}
\end{subequations}
}\noindent
where the contour integrals are with respect to $\tau$, $X=s+r$ and $Y=s-r$. Furthermore we defined
\begin{equation}
\tilde{\mathcal{K}}=\exp\Big[-\frac{i\tau}{2}\zeta^+\zeta^-+\frac{i(s-r)}{2}y\zeta^-+\frac{i(s+r)}{2}y\zeta^+\Big]\,.
\end{equation}
Again the coefficient functions $J_i^{(k)}$ are formal series in $\tau^{-1}$, $X^{-1}=(s-r)^{-1}$ and $Y^{-1}=(s+r)^{-1}$ given by
\begin{equation}
J_i^{(k)}=\sum j_{i}^{(k)}(l,n,m)\tau^{-l}(s-r)^{-m}(s+r)^{-n}\,.
\end{equation}
Note that inverse powers of $\tau$ lead to contractions between the $\pC$ fields. The choice of considering a basis with respect to $\zeta^{\pm}$ is a very practical one as it turns out to diagonalize the covariant derivative $\adD$. This will be explained in more detail in the following. 

\subsubsection{Derivatives for Linear Basis}
In the following we consider the action of the twisted-adjoint covariant derivative $\tadD$ with respect to functionals linear in $\pC$. This will be of great importance in Appendix~\ref{app:cohomologies} in which we will analyze the cohomology of this differential. Using the equations of motion for $\pC$ given by
\begin{equation}
\left[\nabla+\frac{i}{2}\,\phi\,h^{\ga\ga}(y_\ga y_\ga-\partial^y_\ga\partial^y_\ga)\right] \pC(y,\phi)=0\,,
\end{equation}
one obtains the following action of $\tadD$ on the expressions in \eqref{eq:MassimoAnsatz}:
\begin{subequations}
\begin{align}
\tadD J_0&=\frac{i\phi}{2}\oint \d \tau \, (1-\tau^2)J^{(0)}(\tau)h^{\ga\ga}\left( y_\ga y_\ga+\tau^{-2}\partial^y_\ga\partial^y_\ga\right) \pC(y\tau)\,, \label{eq:tadDZeroFormMassimo}\\
\tadD J_1&=\frac{1}{2}\oint \d \tau \,  H^{\ga\ga}\left(\tilde J_1^{(2)}(\tau)y_\ga y_\ga+\tilde J_2^{(2)}(\tau)\tau^{-1}y_\ga \partial^{y}_\ga+\tilde J_3^{(2)}(\tau)\tau^{-2}\partial^y_\ga \partial^y_\ga\right) \pC(y\tau)\,, \label{eq:tadDOneFormMassimo}\\
\tadD J_2&=-\frac{i\phi}{2}\oint \d \tau \, H \tilde J^{(3)}(\tau) \pC(y\tau)\,,
\end{align}
\end{subequations}
with the coefficient functions given by
\begin{align}
&\tilde J_1^{(2)}(\tau)=+\frac{1}2\left[2\tau-(\tau^2-1)\partial_\tau\right]J_2^{(1)}(\tau)\,,\nonumber\\
&\tilde J_2^{(2)}(\tau)=\left[2\tau-(\tau^2-1)\partial_\tau\right](J_3^{(1)}(\tau)-J_1^{(1)}(\tau))\,,\nonumber\\
&\tilde J_3^{(2)}(\tau)=-\frac{1}{2}\left[2\tau-(\tau^2-1)\partial_\tau\right]J_2^{(1)}(\tau)\,,\nonumber\\
&\tilde J^{(3)}(\tau)=\frac{1}{3}\left[(\tau^2-1)\partial_\tau^2-2\tau\partial_\tau+2\right]( J_3^{(2)}(\tau)+J_1^{(2)}(\tau))\nonumber\,.
\end{align}
It is convenient to introduce an equivalence relation for formal series in $\tau^{-1}$ denoted by $g$ and $f$,
\begin{equation}
f(\tau)\sim g(\tau) \; \; \text{iff} \; \; f(\tau)-g(\tau)=P(\tau) \label{eq:equivalenceRel}  \,,
\end{equation}
with $P(\tau)$ being an arbitrary polynomial. The latter equivalence relation is useful since then one has
\begin{equation}
\oint \d \tau \, f(\tau) = \oint \d \tau \, g(\tau)\quad \Longleftrightarrow \quad f(\tau)\sim g(\tau)\,.
\end{equation}
We will use this equivalence relation extensively in Appendix~\ref{app:cohomologies}.

\subsubsection{Derivatives for Quadratic Basis}
In this subsection we will analyze the action of the adjoint covariant derivative $\adD$ in the integral basis and we will keep the freedom in Fierz-transformations as this will be useful for our analysis. To this end we will consider the following choice of the coefficient functions in \eqref{nicebasis}:
\begin{equation}
\label{mandndef}
J^{(q)}_i=J^{(q)}_{m,n}(\tau,r,s)=(s-r)^m(s+r)^n k^{(q)}_i(\tau) \,.
\end{equation}
These coefficient functions therefore contain a fixed number of $(s-r)$ and $(s+r)$ factors but an arbitrary power of $\tau$.
One can determine the action of $\adD$ on the various $p$-forms of \eqref{nicebasis}. After some manipulations and integrations by parts one arrives at the following representation:
{\allowdisplaybreaks\footnotesize
\begin{subequations}
\begin{align}
\adD k^{(0)}&=\frac{i}{2}\begin{pmatrix}\label{D0F}
-k^{(0)}\\
-(1-\tau)k^{(0)}\\
-(1+\tau)k^{(0)}\\
0\\
0\\
-(1-\tau^2)k^{(0)}
\end{pmatrix}\,,\\
\adD \begin{pmatrix}\label{D1F}
k_1^{(1)}\\
k_2^{(1)}\\
k_3^{(1)}\\
k_4^{(1)}\\
k_5^{(1)}\\
k_6^{(1)}
\end{pmatrix}&=\frac12 \begin{pmatrix}
(m \tau+m-n \tau+n-2)k^{(1)}_1-(1+\tau) \partial_\tau k^{(1)}_2-n k^{(1)}_2+(1-\tau) \partial_\tau k^{(1)}_3-m k^{(1)}_3\\
m(1-\tau^2)k_1^{(1)}-2(1+\tau)k_2^{(1)}+2(n-1)k_4^{(1)}+(m-2)k_6^{(1)}\\
n(1-\tau^2)k_1^{(1)}-2(1-\tau)k_3^{(1)}+2(m-1)k_5^{(1)}+(n-2)k_6^{(1)}\\
(m-1)(1-\tau^2)k_2^{(1)}+[m-n-2+(m+n-2)\tau]k_4^{(1)}-(1-\tau^2)\partial_\tau k_4^{(1)}-(m-1)(1-\tau)k_6^{(1)}\\
(n-1)(1-\tau^2)k_3^{(1)}+[m-n+2+(m+n-2)\tau]k_5^{(1)}+(1-\tau^2)\partial_\tau k_5^{(1)}-(n-1)(1+\tau)k_6^{(1)}\\
-2n(1+\tau)k_4^{(1)}-2m(1-\tau)k_5^{(1)}
\end{pmatrix}\,,\\
\adD\begin{pmatrix}\label{D2F}
k_1^{(2)}\\
k_2^{(2)}\\
k_3^{(2)}\\
k_4^{(2)}\\
k_5^{(2)}\\
k_6^{(2)}
\end{pmatrix}=&\,\,\,
\substack{
\displaystyle\frac{i}{2}\big(-4mn(1-\tau^2)k_1^{(2)}\hspace*{\fill} \\
\displaystyle+n[(m+n-2+(m-n)\tau+(1-\tau^2)\partial_\tau ]k_2^{(2)}+m[(m+n-2+(m-n)\tau-(1-\tau^2)\partial_\tau ]k_3^{(2)}\hspace*{\fill}\\
\displaystyle-2n[(n-1)+(1+\tau)\partial_\tau ]k_4^{(1)}-2m[(m-1)-(1-\tau)\partial_\tau]k_5^{(1)}\hspace*{\fill}\\
\displaystyle-[2(n-1)(m-1)+(m-n+(m+n-2)\tau)\partial_\tau-(1-\tau^2)\partial_\tau^2]k_6^{(2)}\big)\,.\hspace*{\fill}}
\end{align}
\end{subequations}} \noindent
It is important to stress here that in the basis \eqref{nicebasis} the covariant derivative $\adD$ does not mix tensor structures corresponding to different $m$ and $n$ values. Put differently, in this basis $\adD$ is diagonal with respect to $m$ and $n$ and not only with respect to spin, which is given by $2s=-(m+n)$. This property is most useful in identifying independently-conserved sectors of the backreaction. 

It can be shown that the above representation of $\adD$ squares to zero and is compatible with the following representation of the Fierz identities:
{\footnotesize\begin{equation}
\label{eq:fierzMassimo}
\delta\begin{pmatrix}
k_1^{(i)}\\
k_2^{(i)}\\
k_3^{(i)}\\
k_4^{(i)}\\
k_5^{(i)}\\
k_6^{(i)}
\end{pmatrix}=\begin{pmatrix}
\partial_\tau\chi_1^{(i)}\\
m\chi_1^{(i)}-\partial_\tau\chi_3^{(i)}\\
-n\chi_1^{(i)}+\partial_\tau\chi_2^{(i)}\\
-(m-1)\chi_3^{(i)}\\
-(n-1)\chi_2^{(i)}\\
m\chi_2^{(i)}+n\chi_3^{(i)}
\end{pmatrix}\,,
\end{equation}} \noindent
where the $\chi^i_{(k)}$ are arbitrary functions of $\tau$. Using these relations we will study the various cohomologies quadratic in $\pC$ in Appendix~\ref{app:cohomologies}.

\subsubsection{Solving the Torsion Constraint}
Below we give the formulas allowing to map the backreaction to Fronsdal currents, as discussed in Section~\ref{sec:PresentationBackreactionFronsdal}, using the $\adD$-diagonal basis introduced above. The action of $Q^{-1}$ in this basis is diagonal with respect to different contractions of the vielbein but it mixes various components within each diagonal subsector. It is given by 
{\footnotesize
\begin{multline}
\vec{k'}^{(1)} = Q^{-1}\vec{k}^{(2)}\\=\frac1{m+n-2}\begin{pmatrix}
2 k_1(\tau)\\
-(n-2) k_2(\tau)-m k_3(\tau)\\
-n k_2(\tau)-(m-2) k_3(\tau)\\
\frac{1}{(m+n)}\,[2 m (m-1)k_5(\tau)-2 m (n-1) k_4(\tau)+(m-1)(n-m-2)k_6(\tau)]\\
\frac{1}{(m+n)}[2n (n-1) k_4(\tau)-2n (m-1) n k_5(\tau)+(n-1) (m-n-2)k_6(\tau)]\\
\frac{1}{(m+n)}[2 n(n-m-2) k_4(\tau) +2m(m-n-2) k_5(\tau)+(m-n-2)(n-m-2)k_6(\tau) ]\\
\end{pmatrix}\,.
\end{multline}}\noindent To evaluate $(I-\nabla Q^{-1})$ we also need the representation for $\nabla$ in this basis:
\begin{equation}
\vec{k'}^{(2)} = \nabla\vec{k}^{(1)}=\begin{pmatrix}
-i \left(\tau k_1(\tau) (m-n)-t \left(k_2'(\tau)+k_3'(\tau)\right)-n k_2(\tau)-m k_3(\tau)\right)\\
i \left(m \left(\tau^2-1\right) k_1(\tau)+2 t k_2(\tau)+2 (n-1) k_4(\tau)+(m-2) k_6(\tau)\right)\\
i \left(n \left(\tau^2-1\right) k_1(\tau)-2 t k_3(\tau)+2 (m-1) k_5(\tau)+(n-2) k_6(\tau)\right)\\
i \left((m-1) \left(\tau^2-1\right) k_2(\tau)-(\tau^2-1) k_4'(\tau)-tk_4(\tau) (m+n-2)-(m-1)t k_6(\tau))\right)\\
i \left((n-1) \left(\tau^2-1\right) k_3(\tau)+\left(\tau^2-1\right) k_5'(\tau)+t k_5(\tau) (m+n-2)+(n-1) t k_6(\tau)\right)\\
2 i t (n k_4(\tau)-m k_5(\tau))
\end{pmatrix}\,.
\end{equation}
In the following we give some examples for Fronsdal currents and study their relation with the Prokushkin--Vasiliev currents.

\subsubsection{Canonical Currents}\label{app:canonicalCurrentsIntegral}
In the following we will give more details on canonical currents.

\paragraph{Canonical Vasiliev's currents:} 
In the following we study in more detail the canonical current sector of Vasiliev's backreaction that sources the $\adD\pomega^{(2)}=\cdots$ equation. In our basis this is associated with:
\begin{equation}
\vec{k}^{(2)}=\begin{pmatrix}
0\\
0\\
0\\
k_4^{(2)}(\tau)\\
0\\
0
\end{pmatrix}\,,
\end{equation}
for $(m,n)=(1,1-2s)$ and with:
\begin{equation}
\vec{k}^{(2)}=\begin{pmatrix}
0\\
0\\
0\\
0\\
k_5^{(2)}(\tau)\\
0
\end{pmatrix}\,,
\end{equation}
for $(m,n)=(1-2s,1)$. These two components can be combined into bosonic and fermionic canonical currents possibly including higher-derivative encoded by higher powers in $\tau^{-1}$ which are all individually conserved. The analysis of the local cohomology\footnote{By local cohomology we mean that we restrict the space of functionals to be polynomial in the derivatives.} suggests that the other components of the backreaction should be interpreted as improvements, being in one to one correspondence with improvements in the metric-like language. This observation is of key importance to study the very complicated Vasiliev backreaction. Indeed, upon solving the torsion constraint one can show that only this sector gives rise to canonical currents together with a possible higher-derivative $\Box$ tail.

In the following we are first going to study more in details the improvement that removes the canonical current.
It is first convenient to make the choice\footnote{The choice $(m,n)=(1-2s,1)$ is equivalent.} $(m,n)=(1,1-2s)$ with the vector $k_i^{(2)}$ given by

{\footnotesize\begin{equation}
\begin{pmatrix}
k_1^{(2)}\\
k_2^{(2)}\\
k_3^{(2)}\\
k_4^{(2)}\\
k_5^{(2)}\\
k_6^{(2)}
\end{pmatrix}=\begin{pmatrix}
0\\
0\\
0\\
C \tau^{-1}\\
0\\
0
\end{pmatrix}\,,
\end{equation}}
setting to zero any higher-derivative tail for the moment.

Conservation is trivial due to the choice $m=1$ and, as well, no Fierz identity can be used to change the constant $C$ because $m=1$. 

In order to show that this term is exact one is then left with a single differential equation to be solved taking into account equivalence up to polynomials in $\tau$. The differential equation then reads:
\begin{equation}
\left[(1-\tau^2)\partial_\tau-2s(1-\tau)+2\right]k_4^{(1)}(\tau)\sim C \tau^{-1}\,.
\end{equation}
The above equation can therefore be conveniently rewritten as:
\begin{equation}
(1+\tau)^{2s } (1-\tau)^{2} \frac{\partial }{\partial \tau}\Big[(1+\tau)^{-2s+1} (1-\tau)^{-1}k_4(\tau)\Big]=C \tau^{-1}+p(\tau)\,,
\end{equation}
or changing variables in terms of $\omega=\tau^{-1}$, as:
\begin{equation}
-\omega^{-2s}(1+\omega)^{2s } (1-\omega)^{2} \frac{\partial }{\partial \omega}\Big[\omega^{2s}(1+\omega)^{-2s+1} (1-\omega)^{-1}k_4(\omega)\Big]=C \omega+p(\omega^{-1})\,.
\end{equation}
In this form one can integrate the above as:
\begin{equation}
k_4(\omega)=-\frac{(1-\omega)(1+\omega)^{2s-1} }{\omega^{2s}}\int^\omega \frac{x^{2s}}{(1+x)^{2s} (1-x)^{2}}\Big[C x+p(x^{-1})\Big]\,.
\end{equation}
Due to the particular form of the solution one can reduce the polynomial function ambiguity that would produce non-polynomial effects on the solution to only three free parameters: $p(\tau)=\alpha (1+\tau)+\beta(1-\tau)+\gamma \tau^{2s+1}$ associated to the possibility of generating single poles in the integrand, all other polynomial being related to these up to a polynomial shift in $k_4(\tau)$. At this point we can drop $\gamma$ since it would give a solution that is not meromorphic in $\omega\sim 0$.

Requiring for instance\footnote{We can also avoid to fix either $\alpha$ or $\beta$. In this case the difference of the corresponding solutions for two different values of the parameters encode non-trivial cohomologies at form degree-1 and hence parametrize ambiguities in defining the corresponding redefinition.} $\alpha=-\frac{C}{2}$ and $\beta=-\tfrac{3C}{2}$ one can recast the solution in terms of the following series:
\begin{multline}
k_4^{(1)}(\omega)\sim C(1-\omega)(1+\omega)^{2s-1}{}_2F_1(2s,2s,2s+1,-\omega)\\=C(1-\omega)(1+\omega)^{2s-1}\sum_{l=0}^\infty(-1)^l\frac{(2s+l-1)!}{l!(2s+1)!(2s+l)} \omega^{l}\,.
\end{multline}

\paragraph{Fronsdal Currents from Vasiliev Currents:}

In the canonical current sector one observes also nice simplifications when solving the torsion constraint. It is indeed easy to see, restricting the attention for simplicity to the case $(m,n)=(1,1-2s)$, that the corresponding Fronsdal current, i.e. the source to the Fronsdal tensor after having solved the torsion constraint, can be obtained from the Vasiliev current that sources $\adD\pomega^{(2)}$ as:
\begin{equation}
\formj^{\text{Fr.}}=-\frac{1}{2(s-1)}\frac{(1-\omega)^{2s} (1+\omega)^{2}}{\omega^{2s}} \frac{\partial }{\partial \omega}\Big[\omega^{2s}(1-\omega)^{-2s+1} (1+\omega)^{-1}k^{(2)}_4(\omega)\Big]\,. 
\end{equation}
The problem of finding which Vasiliev current would give rise to the standard canonical current as source to the Fronsdal tensor upon solving the torsion constraint, becomes then similar to the problem of solving for improvements and we actually already have the solution displaying a one parameter ambiguity. We can indeed integrate the above equation as:
\begin{equation}
k^{(2)}_4(\omega)=-2(s-1)\frac{(1+\omega)(1-\omega)^{2s-1} }{\omega^{2s}}\int^\omega \frac{x^{2s}}{(1-x)^{2s } (1+x)^{2}}\Big(C x+\alpha\Big)\,, \label{eq:atMassimosRequest}
\end{equation}
with $\alpha$ arbitrary. The above covers for a given choice of $\alpha$ the case studied in index form in \eqref{Fronsdalbackreaction}. Notice however that changing $\alpha$ we observe two very different asymptotic behavior of the corresponding coefficients as $l\rightarrow \infty$. The generic asymptotics is $\frac{1}{l^2 l!}$ but for a given choice of $\alpha$ we get the asymptotic behavior $\frac{1}{l^{2s}l!}$. Anyway the above is pseudo-local and it should be what we should match from Vasiliev's backreaction if it would give rise to canonical Fronsdal currents without higher-derivative tail.

\paragraph{The canonical current sector of Vasiliev's backreaction general structure:}

The general structure of the canonical current sector extracted from the Vasiliev backreaction is remarkably simple for any spin. Its structure involves 3 types of terms that combined together sum up to the function $k_4^{(2)}$:
\begin{equation}
k_4^{(2)}\sim \frac{1}{\omega^{2s+1}}\Big[p_1^{2s+1}(\omega)\log(1+\omega)+p_2^{2s+1}(\omega)\log(1-\omega)+p_3^{2s+1}(\omega)\text{Li}_2(\omega)+p_4^{2s+1}(\omega)\text{Li}_2(-\omega)\Big]\,,
\end{equation}
where $p_i^{(2s+1)}(\omega)$ are polynomials of degree at most $2s+1$ and encode the spin-dependence of the result. The structure of the Vasiliev backreaction is remarkably simple and similar to the structure of the Vasiliev current that gives the canonical Fronsdal current. The difference is given by the dilog contribution and the degree of the polynomial coefficients that is one power higher, as opposite to the simple polynomial coefficient $(1+\omega)(1-\omega)^{2s-1}$ that we found in the previous paragraph. The above is true both before and after the twisted-sector decoupling. Notice however that after the redefinition that decouples the twisted sector the polynomial multiplying the Dilog function becomes of lower degree.


\section{More on Lorentz Invariance in The Schwinger--Fock Gauge}
\label{app:llt}

In this appendix we provide some details related to Section~\ref{sec:LorentzCovariance}, where the issue of preserving Lorentz invariance for Prokushkin--Vasiliev Theory in the Schwinger--Fock gauge \rf{SFgauge} is discussed. 

\subsection{Preserving the Schwinger--Fock Gauge}

For the naive Lorentz generators $L^\textsc{yz}$ one finds
\begin{equation}
\delta_\Lambda \MS_\alpha = [L^\textsc{yz},\MS_\alpha]_\star = 2\MS_\beta \Lambda^\beta_{\;\alpha} + \frac{\delta \MS_\alpha}{\delta\B}[L^\textsc{y},\B]_\star\,,
\end{equation}
where the variation $\delta$ on the above right-hand side is a simple functional variation (with respect to $\B$ in this case). The structure of the right-hand side is as follows: in the commutator with $\MS_\alpha$, $L^\textsc{yz}$ can either act on the oscillator $z_\alpha$ that $\MS_\alpha$ is proportional to, or it can act on the rest of the expression \rf{pertBbis} for $\MS_\alpha$. Now, the rest thereof is a Lorentz-invariant quantity (no free oscillator indices), and is fully determined in terms of $\MB$, which is why we can write the second term in the above right-hand side in such a way.

Evidently, the above transformations do not preserve the Schwinger--Fock gauge $z^\alpha \MS_\alpha = 0$. The corrected Lorentz generators $L^\textsc{s}$ of \rf{correctedlgen} do preserve the gauge, and one easily checks it to be true by using, on top of the definition \rf{correctedlgen}, the relations \rf{threedVasilievSS}. The result of taking the variation of $\MS_\alpha$ with respect to the correct generators is
\begin{equation}
\delta_\Lambda \MS_\alpha = \frac{\delta \MS_\alpha}{\delta\B}[L^\textsc{y},\B]_\star\,,
\end{equation}
which is proportional to $z_\alpha$ and hence preserves the Schwinger--Fock gauge $z^\alpha \MS_\alpha = 0$. The above equation is also compatible with the fact that $\MS_\ga$ is an auxiliary field.

\subsection{Recovering the Lorentz Algebra and Covariant Rotation of Fields}

First, let us explain how the commutation relations \rf{Lorentznonclosure} for the Lorentz generators in the Schwinger--Fock gauge are obtained. This is straightforward: looking at the definition \rf{correctedlgen} and recalling \rf{threedVasilievBS}, \rf{threedVasilievSS} one finds \rf{Lorentznonclosure}, where it should be made clear that the dependence of $L^\textsc{s}_{\alpha\alpha}$ on $\B$ is via $\MS_\alpha$. \\

Let us now explicitate how a true Lorentz algebra is recovered when looking at the commutator of local Lorentz transformations on the various master fields. For this we will assume the covariant laws of rotation explicitated in \rf{covariantrotations}. Let us first look at the scalar master field $\MB$. Using \rf{threedVasilievBS} one finds
\begin{equation}
\label{appBcovariantredef}
\delta_{\Lambda_1} \MB = \frac{\delta\MB}{\delta\B}[L_1^\textsc{y},\B]_\star\,,
\end{equation}
where $L_{1}^\textsc{y} \equiv \tfrac{1}{2}\Lambda_{1}^{\alpha\alpha}L^\textsc{y}_{\alpha\alpha}$. From this, applying a second transformation $\delta_{\Lambda_2} \bullet$ to $\MB + \delta_{\Lambda_1} \MB$ and antisymmetrizing with respect to the exchange of $\Lambda_1$ and $\Lambda_2$ one concludes that
\begin{equation}
(\delta_{\Lambda_1}\circ\delta_{\Lambda_2} - \delta_{\Lambda_2}\circ\delta_{\Lambda_1})\MB \equiv [\delta_{\Lambda_1},\delta_{\Lambda_2}]_\star\MB = \frac{\delta\MB}{\delta\B}[[L_1^\textsc{y},L_2^\textsc{y}],\B]_\star\,,
\end{equation}
where one has to use the Jacobi identity. The Lorentz algebra is thus restored on $\MB$ since the generators $L^\textsc{yz}$ truly close to the Lorentz algebra, without extra terms as in \rf{Lorentznonclosure}. 

Let us then look at the auxiliary master field $\MS_\alpha$. We know from Section~\ref{sec:LorentzCovariance} that the variation $\delta_\Lambda \MS_\alpha$ with respect to $L^\textsc{s} \equiv \tfrac{1}{2}\Lambda^{\alpha\alpha}L^\textsc{s}_{\alpha\alpha}$ reads
\begin{equation}
\label{appMScovariantrotation}
\delta_\Lambda \MS_\alpha = \frac{\delta\MS_\alpha}{\delta\B} [L^\textsc{y},\B]_\star\,,
\end{equation}
since this is precisely what allows one to claim that the corrected Lorentz generators $L^\textsc{s}$ preserve the Schwinger--Fock gauge (see previous subsection). Then proceeding as in the above case of the scalar master field $\MB$ one finds 
\begin{equation}
(\delta_{\Lambda_1}\circ\delta_{\Lambda_2} - \delta_{\Lambda_2}\circ\delta_{\Lambda_1})\MS_\alpha \equiv [\delta_{\Lambda_1},\delta_{\Lambda_2}]_\star\MS_\alpha = \frac{\delta\MS_\alpha}{\delta \B} [[L_1^\textsc{y},L_2^\textsc{y}],\B]_\star\,,
\end{equation}
and the Lorentz algebra thus closes on $\MS_\alpha$ too. 

For the one-form master field $\MW$ we recall first the splitting \rf{gaugeconnectionredef} of $\MW$ into its spin-connection and the rest of it:
\begin{equation}
\label{appgaugeconnectionredef}
\MW \equiv \frac{1}{2}\omega^{\alpha\beta} L^{s}_{\alpha\beta} + \mathsf{W}\,,
\end{equation}
where $\mathsf{W}$ does not depend on $\omega^{\alpha\beta}$. Then we note again its law of transformation under a local Lorentz transformation, which is given in \rf{newgaugevariation}. It reads
\begin{equation}
\delta (\mathsf{W}+\frac{1}{2}\omega^{\ga\ga}L^\textsc{s}_{\ga\ga}(\B))=\frac12\left(\d\Lambda^{\ga\ga} - \omega\fud{\ga}{\nu}\Lambda^{\nu \ga}\right)L^\textsc{s}_{\ga\ga}-[\mathsf{W},\frac12\Lambda^{\ga\ga}L^\textsc{yz}_{\ga\ga}]_\star\,. \label{appnewgaugevariation}
\end{equation}
The proof then follows that of the other master fields. 

As we observe, once the covariance rotation of the fields is proven, the closure of the Lorentz algebra on the fields follows automatically by simply making use of the Jacobi identity. The most non-trivial piece of work is thus that of obtaining the transformation laws, and in particular the above form \rf{appnewgaugevariation}. 

\subsection{Identifying the Proper Spin-Connection}

It is convenient to first investigate the zeroth-order implications of the above identification \rf{appgaugeconnectionredef} of the correct spin-connection in the Schwinger--Fock gauge. At zeroth order, as we already pointed out, $ L^\textsc{s} = L^\textsc{y}$ and hence 
\begin{equation}
\bar{\MW} = \frac{1}{2}\varpi^{\alpha\beta} L^\textsc{y}_{\alpha\beta} + \bar{\mathsf{W}} = \frac{1}{2}(\varpi^{\alpha\beta} + h^{\alpha\beta})L^\textsc{y}_{\alpha\beta} \,,
\end{equation}
so that the redefinition boils down to renaming the part of the background connection containing the dreibein as $\mathsf{W}$. This is again in harmony with the fact that, at order zero,  $ L^\textsc{s} = L^\textsc{y}$ means that the background gauge connection $\bar{\MW}$ is already in the form \rf{gaugeconnectionredef}. At first order we have the following relation:
\begin{equation}
\MW^{(1)} = \frac{1}{2}\omega^{(1)\alpha\beta} L^\textsc{y}_{\alpha\beta} + \frac{1}{2}\varpi^{\alpha\beta} L^\textsc{s,(1)}_{\alpha\beta} + \mathsf{W}{}^{(1)}\,,
\end{equation}
and things become more complicated. As one can see, from the standpoint of the naive identification of the spin-connection the equation \rf{appgaugeconnectionredef} amounts to a redefinition thereof, which is nevertheless the identical field redefinition at order zero in perturbation theory. Things are also simpler at order~$1$ since $L^\textsc{s,(1)}_{\alpha\beta}(z=0)=0$. 

\subsection{Possible Sources of Lorentz Non-Manifest Covariance}

A possible source of Lorentz non-covariance is when $\bar{\mathsf{D}}^{\textsc{yz}}$ on the left hand side of \eqref{newWeq}, \eqref{newBeq} acts on terms proportional to $z_\alpha$. But as the spin-connection in \eqref{covDerivativeLLTCov} is now contracted with $L^{\textsc{yz}}_{\ga\ga}$ terms of this form will lead to a contribution of the type
\begin{align}
\left.(d+\frac12\varpi^{\ga\ga}[L^{\textsc{yz}}_{\ga\ga},\bullet])z^\nu f_\nu(y,z)\right|_{z=0}=\left.\frac12\varpi^{\ga\ga}(y_\ga\pl^y_\ga+z_\ga\pl^z_\ga)z^\nu f_\nu(y,z)\right|_{z=0}=0\,.
\end{align}
The $z_\alpha$-dependent terms \eqref{zdepenMasterField} will contribute through the vielbein part of \eqref{covDerivativeLLTCov} as
\begin{align}
\left[h, \begin{pmatrix}
                 f(y,z) \\
                 g(y,z)\psi \\
               \end{pmatrix}
\right]&=\begin{pmatrix}
           \phi h^{\ga\ga}(y_\ga-i\pl^z_\ga)\pl^y_\ga f(y,z) \\
           -
\frac{i}{2}\phi h^{\ga\ga}\left((y_\ga-i\pl^z_\ga)(y_\ga-i\pl^z_\ga)-\pl^y_\ga\pl^y_\ga \right)g(y,z)\psi \\
         \end{pmatrix}\,,
\end{align}
so that on the $z=0$ surface we find for an arbitrary function $z^\nu f_\nu(y,z)+z^\nu g_\nu(y,z)\psi$ a non-vanishing contribution given by
\begin{align}
\left[h, \begin{pmatrix}
                 z^\nu f_\nu(y,z) \\
                 z^\nu g_\nu(y,z)\psi \\
               \end{pmatrix}
\right]_{z=0}&=\begin{pmatrix}
           -i\phi h^{\ga\ga}\pl^y_\ga f_\ga(y,0) \\
           -\phi h^{\ga\ga}(y_\ga-i\pl^y_\ga)g(y,0)\psi \\
         \end{pmatrix}\,.
\end{align}
Therefore, we have shown that all possible sources of Lorentz non-covariance disappear and one can therefore use the perturbation scheme outlined in Section~\ref{subsec:lorentzcovariantperturbation} to recover manifestly Lorentz covariant results at any order in perturbation theory. 

\section{Cohomologies}
\setcounter{equation}{0}
\label{app:cohomologies}
In the following we will discuss cohomologies of the twisted-adjoint covariant derivative $\tadD$ and the adjoint covariant derivative $\adD$. We will analyze cohomologies with respect to functional classes of both linear or quadratic functionals of the scalar field $\pC$ and furthermore for functional classes linear in the physical gauge connection $\pomega$ and scalar field $\pC$. For this purpose we will use the integral basis introduced in Appendix~\ref{app:Basis}.

\subsection{Cohomology Linear in $\pC$}
In the following we consider in detail the cohomology of $\tadD$ with respect to functionals linear in $\pC$. In particular the cohomology at form-degree 1 parameterizes ambiguities in the redefinitions of $\somega$. The analysis of the cohomology at form-degree 2 shows that it is always possible to remove the linear source to $\somega$. The form-degree 0 cohomology is reviewed for completeness.
\subsubsection{Form-Degree 0}
At form-degree 0 the cohomology is entirely fixed by demanding closure. By \eqref{eq:tadDZeroFormMassimo} a zero-form \eqref{eq:MassimoAnsatzZeroForm} is closed if the following equivalence relation holds:
\begin{equation}
(1-\tau^2)J^{(0)}(\tau)\sim0\,.
\end{equation}
The most general solution to the above relation is given by
\begin{equation}
J^{(0)}(\tau)\sim\frac{\alpha+\beta \tau}{1-\tau^{2}}\,,
\end{equation}
since any higher power of $\tau$ in the numerator would only contribute polynomially. We therefore conclude that there are two elements in cohomology. Plugging this result into \eqref{eq:MassimoAnsatzZeroForm} and rewriting the coefficient function as a geometric series we obtain
\begin{align}
\label{eq:zeroFormCalc}
\oint \d \tau \, \frac{\alpha + \beta \tau}{1 - \tau^{-2}} \tau^{-2} \left( \sum_{k=0}^\infty \frac{1}{k!} \pC_{\ga(k)} \tau^k y^{\ga(k)}\right) &= \sum_{m,k=0}^\infty \oint \d \tau \, \left( \frac{\ga}{\tau^{1+2m+1-k}}+ \frac{\gb}{\tau^{1+2m-k}} \right)\frac{1}{k!} \pC_{\ga(k)} y^{\ga(k)}  \\
&= \alpha \pC_\textsc{f}(y) + \beta \pC_\textsc{b}(y) \notag \,,
\end{align}
where we have dropped an overall sign. We thus see that the two elements in cohomology correspond to the bosonic and fermionic components of $\pC(y)$.

\subsubsection{Form-Degree 1}
By \eqref{eq:tadDOneFormMassimo} a closure of a one-form translates to the following relation
\begin{equation}
\left[2\tau+(1-\tau^2)\partial_\tau\right]f(\tau)\sim0\,,
\end{equation}
where $f$ stands for $J^{(1)}_2$ and $J^{(1)}_1 - J^{(1)}_3$ in \rf{eq:MassimoAnsatzOneForm}.
The above operator sends polynomials of degree $n$ into polynomials of degree $(n+1)$:
\begin{equation}
\tau^n\rightarrow -(n-2)\tau^{n+1}+n\tau^{n-1}\,.
\end{equation}
For $n \neq 2$ we can therefore remove an arbitrary monomial $k \tau^{n+1}$  by shifting $f(\tau) \rightarrow f(\tau) + \frac{1}{2-n} k \tau^n$. Note that shifting $f(\tau)$ in this way is allowed since the countour integral of \rf{eq:MassimoAnsatzOneForm} is blind to such polynomials contributions. This allows us to restrict our attention to the following differential equation:
\begin{equation}
\left[2\tau+(1-\tau^2)\partial_\tau\right]f(\tau)=(1+\tau)^2(1-\tau)^2\partial_\tau\Big[(1+\tau)^{-1}(1-\tau)^{-1}f(\tau)\Big]=\alpha+\beta \tau^3\,.
\end{equation}
It is convenient to perform the change of variables $\omega=\tau^{-1}$ which results in the following differential equation:
\begin{equation}
\left[\frac{2}{\omega}+(1-\omega^2)\partial_\omega\right]f(\omega)=\alpha+\frac{\beta}{\omega^{3}}\,,
\end{equation}
whose solutions can be expressed as
\begin{equation}
f(\omega)=\frac{(1+\omega)(1-\omega)}{\omega^2}\int^\omega \d\omega' \frac{\omega'^2\,\left( \alpha+\frac{\beta}{\omega'^{3}}\right)}{(1-\omega')^2(1+\omega')^2}\,,
\end{equation}
and upon integration are given by
\begin{align}
f_\alpha(\tau)\equiv f(\tau) |_{\beta=0}&\sim-\frac{\alpha(1-\omega^2)}{2\omega^2}\tanh ^{-1}(\omega)+\frac{\alpha}{2\omega}=\sum_{k=1}^\infty\frac{\alpha}{(2k+1)(2k-1)}\omega^{2k-1}\,,\\
f_\beta(\tau)\equiv f(\tau) |_{\alpha=0}&\sim \frac{\beta}{2\omega^2}+\frac{\beta}{2}\left(1-\frac{1}{\omega^2}\right)\log\left(\frac{1}{\omega^2}-1\right)\,.
\end{align}
We drop the second solution $f_\beta$ since it is not analytic\footnote{A $\beta\neq 0$ inevitably gives rise to non-analytic solutions at $\omega=0$ due to a pole of order $3$.} around $\omega=0$ and therefore is not a formal series in $\tau^{-1}$. 
By \eqref{eq:tadDZeroFormMassimo} an exact one-form is given by
\begin{equation}
J_1=i\phi\oint q(\tau)h^{\ga\ga}\left(y_\ga y_\ga+\tau^{-2}\partial^y_\ga \partial^y_\ga\right)\pC (y\tau)\,,
\end{equation}
where $q(\tau)$ is an arbitrary function, and therefore corresponds to the choices 
\begin{equation}
J_2^{(1)}\sim 0\,,\qquad J_1^{(1)}\sim J_3^{(1)}\sim q(\tau)\,.
\end{equation}
Therefore for closed one-forms we can set $J_2^{(1)}$ and $J_3^{(1)}-J_1^{(1)}$ independently to be equal to $f_\ga$. Exact one-forms satisfy $J_2^{(1)} \sim 0$ and $J_3^{(1)}-J_1^{(1)} \sim 0$. As a result any choice $\alpha \neq 0$ corresponds to an element in the cohomology. Therefore the cohomology is two-dimensional and we can obtain a particularly useful representative by choosing 
\begin{equation}
J_2^{(1)}\sim f_{d_0} \,,\qquad J_1^{(1)}- J_3^{(1)}\sim f_{g_0}\,,\qquad J_1^{(1)} + J_3^{(1)}\sim 0 \,.
\end{equation}
A calculation similar to \eqref{eq:zeroFormCalc} shows that this results in the following representative:
\begin{align}\label{DtildeCoho}
\frac14 \phi h^{\ga\ga}\int_0^1 \d t\, g_0(t^2-1) \left(y_\ga y_\ga- t^{-2}\pl_\ga\pl_\ga\right)\pC_\textsc{b}(ty)+ \frac12 \phi h^{\ga\ga}\int_0^1 \d t\, d_0(t^2-1)t^{-1} y^{\vphantom{y}}_\ga \pl^y_\ga \pC_\textsc{f}(ty)\,,
\end{align}
where we have used the identity
\begin{equation}
\frac{2}{(2k+1)(2k-1)}=\int^1_0 \d t \, (1-t^2) t^{2k-2}\,.
\end{equation}
This is exactly the ambiguity $R$, given in \eqref{eq:representativeofH}, of the redefinition $M_1$.

\subsubsection{Form-Degree 2}

At form-degree 2 we can also solve the closure condition translated in terms of the following linear ordinary differential equation:
\begin{equation}
\left[(1-\tau^2)\partial_\tau^2+2\tau\partial_\tau-2\right]f(\tau)\sim0\,.
\end{equation}
Taking into account the most general polynomial coefficients that can affect the solution non-polynomially one arrives to the equation
\begin{equation}
\left[(1-\tau^2)\partial_\tau^2+2\tau\partial_\tau-2\right]f(\tau)=\alpha \tau+\beta \tau^2\,.
\end{equation}
To study the behavior at infinity one can again perform the change of variable $\omega=\tfrac{1}{\tau}$ which results in
\begin{equation}
\left[\omega^2(1-\omega^2)\partial_\omega^2+\omega(3-\omega^2)\partial_\omega+2\right]f(\omega)=-\frac{\alpha}{\omega}-\frac{\beta}{\omega^2}\,.
\end{equation}
The exact elements are parametrized by $f\sim0$ while the cohomology is in correspondence with the solutions of the above ordinary differential equation that are analytic at $\omega=0$ up to polynomials in $\omega^{-1}$. The solutions of the homogeneous equation are not analytic at $\omega=0$. In particular the homogeneous solution is a power series in $\omega^{-1}$ and $\ga$ and $\gb$ produce $\log(\omega)$-singularities at $\omega=0$. We therefore conclude that the cohomology at form-degree $2$ is trivial.

\subsection{Cohomology Quadratic in $\pC$}
In this subsection we will analyze the cohomology of the adjoint covariant derivative $\adD$ with respect to pseudo-local field redefinitions, as defined in \eqref{eq:pseudolocalfieldredefs} (also see comments there below). The analysis of form-degree 1 is needed for the study of source terms to the twisted zero-form $\sC^{(2)}$ and form-degree 2 for the backreaction on the higher-spin gauge fields $\pomega^{(2)}$. Form-degree 0 cohomology parameterizes parameterizes the redefinitions of $\pC^{(2)}$.

\subsubsection{Form-Degree 0}

Note that by \eqref{D0F} the covariant derivative $\adD$ does not produce any contributions to $J^{(1)}_4$ and $J^{(1)}_5$ of \eqref{nicebasis2}. Therefore we can impose $m\leq0$ and $n\leq0$ as greater values for $m$ and $n$ in \rf{mandndef} would only lead to contributions that are projected out by the contour integral in \eqref{nicebasis2}.
\noindent The closure condition is required to hold only up to Fierz identities \eqref{eq:fierzMassimo} and therefore takes the form:
{\footnotesize\begin{equation}
\begin{pmatrix}
-k(\tau)-2 i \chi_1'(\tau)\\
-(1-\tau) k(\tau)-2 i \left(m \chi_1(\tau)-\chi_3'(\tau)\right)\\
-(1+\tau) k(\tau)+2 i \left(n \chi_1(\tau)-\chi_2'(\tau)\right)\\
2 i (m-1) \chi_3(\tau)\\
2 i (n-1) \chi_2(\tau)\\
-\left(1-\tau^2\right) k(\tau)-2 i (m \chi_2(\tau)+n \chi_3(\tau))\end{pmatrix}\sim0\,,\label{Dzero+fierz}
\end{equation}}where we have used the notation $\chi'_i = \partial_\tau \chi_i$. We consider the following three cases separately:
\begin{itemize}
\item $m<0$ and $n<0$,
\item $m=0$ and $n<0$ (and $m\leftrightarrow n$),
\item $m=0$ and $n=0$.
\end{itemize}

\paragraph{For $\boldsymbol{m<0}$ and $\boldsymbol{n<0}$:} By summing and subtracting the second and third equations after multiplying them by $n$ and $m$ respectively we can eliminate $\chi_1$ in the second equation:

{\footnotesize\begin{equation}
\label{eq:massimoZeroFormFierzed}
\begin{pmatrix}
-k(\tau)-2 i \chi_1'(\tau)\\
-k(\tau) [\tau (m-n)+m+n]-2 i \left(m \chi_2'(\tau)-n \chi_3'(\tau)\right)\\
k(\tau) [\tau (m+n)+m-n]-2 i \left(2 m n \chi_1(\tau)-m \chi_2'(\tau)-n \chi_3'(\tau)\right)\\
2 i (m-1) \chi_3(\tau)\\
2 i (n-1) \chi_2(\tau)\\
-\left(1-\tau^2\right) k(\tau)-2 i (m \chi_2(\tau)+n \chi_3(\tau))\end{pmatrix}\sim0\,.
\end{equation}}\noindent From the fourth and fifth equation we can conclude that $\chi_2\sim\chi_3\sim0$. The last equation then implies
\begin{equation}
k(\tau)\sim\frac{\alpha+\beta \tau}{1-\tau^2}\,,\label{ksollast}
\end{equation}
as we have learned from the study of the cohomology of $\tadD$ linear in $\pC$ at form-degree~$0$. 
The coefficient $\beta$ must be set to zero since it cannot be removed by $\chi_1'$ in the first equation, because $\chi_1'$ cannot contain a $\tau^{-1}$ pole. This choice of $\beta$ is however incompatible with the second equation, showing that there is no cohomology.

\paragraph{For $\boldsymbol{m=0}$ and $\boldsymbol{n<0}$:} For $m=0$ only $J^{(1)}_2$, $J^{(1)}_4$ and $J^{(1)}_6$ in \eqref{nicebasis2} are not projected out by the contour integrals over $(s-r)$. Therefore we only have to consider the second, fourth and sixth component of \eqref{eq:massimoZeroFormFierzed}. The fourth equation implies $\chi_3 \sim 0$ and the sixth equation has again a solution for $k(\tau)$ of the type \eqref{ksollast}. The parameter $\beta$ has to be tuned to solve the second equation in \eqref{eq:massimoZeroFormFierzed}:
\begin{equation}
n\, k(\tau) \,(1- \tau)\sim 0\,,
\end{equation}
which implies $\alpha=\beta$:
\begin{equation}
k(\tau)\sim\frac{\alpha}{1-\tau}=-\frac{\alpha\omega}{1-\omega}\,,\qquad \omega=\frac{1}{\tau}\,.\label{ksollastA}
\end{equation}
Analogous solutions are obtained for $n=0$ and $m<0$ if one replaces $\tau\rightarrow -\tau$. We therefore find a cohomology for each pair $(n,0)$ and $(0,m)$. This is equivalent to having one bosonic and one fermionic cohomology upon combining $(n,0)$ and $(0,m)$ appropriately.

\paragraph{For $\boldsymbol{m=0}$ and $\boldsymbol{n=0}$:}In this case only $J^{(1)}_6$ in \eqref{nicebasis2} is not projected out by the contour integrals over $(s-r)$ and $(s+r)$. Therefore only the sixth component of \eqref{eq:massimoZeroFormFierzed} has to be considered and the corresponding solution is given by
\begin{equation}
k(\tau)\sim\frac{\alpha+\beta \tau}{1-\tau^2}=-\frac{\alpha\,\omega^2+\beta\, \omega}{1-\omega^2}\,,\qquad \omega=\frac{1}{\tau}\,.\label{ksollast2}
\end{equation}
This implies that we have again one bosonic ($\alpha=0$) and one fermionic ($\beta=0$) cohomology.
The cohomology $(0,0)$ can be seen to correspond to $\text{Tr}(\pC\star \pC)$.

\subsubsection{Form-Degree 1}

By using Fierz identities \eqref{Dzero+fierz} and adding exact forms we remove various components of 

{\footnotesize\begin{equation}
\label{eq:vector}
\vec{k}^{(1)}=\begin{pmatrix}
k_1^{(1)}\\
k_2^{(2)}\\
k_3^{(3)}\\
k_4^{(4)}\\
k_5^{(5)}\\
k_6^{(6)}
\end{pmatrix}\,.
\end{equation}}One can distinguish the following cases for $m$ and $n$ in \rf{mandndef}:
\begin{itemize}
\item $n<0$ and $m<0$. 
One can use an exact form \eqref{D0F} to remove the last component in \eqref{eq:vector} and the term proportional to $\tau^{-1}$ in the first component. Then one can apply a Fierz identity \eqref{eq:fierzMassimo} to remove the fourth and fifth entry and the remaining terms in the first component. This leaves us with 
{\footnotesize\begin{equation}
\vec{k}^{(1)}=\begin{pmatrix}
0\\
f(\tau)\\
g(\tau)\\
0\\
0\\
0\end{pmatrix}\,.
\end{equation}\label{gaugefixed1form}}As we will discuss in the following we find one bosonic and one fermionic cohomology for each spin in this case. 

\item $m=0$ and $n<0$. In this case only $k^{(1)}_2$, $k^{(1)}_4$ and $k^{(1)}_6$ will not be projected out by the contour integral in \eqref{nicebasis}.
By choosing $\chi^{(1)}_3$ appropriately in \eqref{eq:fierzMassimo} we can eliminate the fourth component. We can then remove the sixth by adding an exact form. This results in
{\footnotesize\begin{equation}
\vec{k}^{(1)}=\begin{pmatrix}
0\\
f(\tau)\\
0\\
0\\
0\\
0\end{pmatrix}\,,
\end{equation}}\noindent and the case $n=0$, $m<0$ can be obtained from this one by swapping the 2nd and 3rd components and performing $\tau\rightarrow-\tau$. Combining both cases we find again one bosonic and one fermionic cohomology for each spin.

\item $m=1$ and $n<0$. In this case only $k^{(1)}_4$ is kept by the contour integral. This component is unaffected by an arbitrary Fierz identity and by adding an exact form. Therefore the 1-form representative can be chosen as:

{\footnotesize\begin{equation}
\vec{k}^{(1)}=\begin{pmatrix}
0\\
0\\
0\\
f(\tau)\\
0\\
0\end{pmatrix}\,,
\end{equation}}\noindent and again the case $n=1$, $m<0$ can be obtained from this one by swapping the 4th and 5th components and performing $\tau\rightarrow-\tau$. Combining both cases we find one bosonic and one fermionic cohomology for each spin, as in the previous case. 

\item $m\geq0$ and $n\geq0$. In this case only the sixth component is not projected out by the contour integrals. Therefore all 1-forms are trivially exact.

\end{itemize}

\noindent Let us expand further on these cases:

\paragraph{$\boldsymbol{n<0}$ and $\boldsymbol{m<0}$:} Non-trivial cohomologies are in correspondence with the solution of the following condition:

{\footnotesize\begin{equation}
\adD \vec{k}^{(1)}=-i\begin{pmatrix}
-(\tau +1) f'(\tau )-n f(\tau )-(\tau -1) g'(\tau )-m g(\tau )+2 \chi_1'(\tau)\\
-2 \left((\tau +1) f(\tau )-m \chi_1(\tau )+\chi_3'(\tau )\right)\\
2 \left((\tau -1) g(\tau )-n \chi_1(\tau )+\chi_2'(\tau )\right)\\
-(m-1) \left(\left(\tau ^2-1\right) f(\tau )+2 \chi_3(\tau )\right)\\
-(n-1) \left(\left(\tau ^2-1\right) g(\tau )+2 \chi_2(\tau )\right)\\
2 (m \chi_2(\tau )+n \chi_3(\tau ))\end{pmatrix}\sim0\,.
\end{equation}}\noindent One can easily solve the fourth and fifth equations as
\begin{align}
f(\tau)&\sim\frac{\alpha+\beta \tau+2\chi_3(\tau)}{1-\tau^2}\,,&g(\tau)&\sim\frac{\gamma+\delta \tau+2\chi_2(\tau)}{1-\tau^2}\,.
\end{align}
Summing and substracting the second equation multiplied by $n$ and the third equation multiplied by $m$ we obtain:
\begin{align}
-n (\tau+1) f(\tau)+m (\tau-1) g(\tau)+m \chi_2'(\tau)-n \chi_3'(\tau)&\sim0\,,\\
n (\tau+1) f(\tau)+m (\tau-1) g(\tau)-2 m n \chi_1(\tau)+m \chi_2'(\tau)+n \chi_3'(\tau)&\sim0\,,
\end{align}
where in the second equation we can drop $m \chi_2'(\tau)+n \chi_3'(\tau)$ due to the last component which requires it to be a polynomial.
Substituting the solution for $f(\tau)$ and $g(\tau)$ one arrives to:
\begin{align}
\frac{2 \tau Y(\tau)}{\tau^2-1}-  \left(\frac{n (\alpha +\beta  \tau)}{(m-1) (\tau-1)}-\frac{m (\gamma +\delta  \tau)}{(n-1) (\tau+1)}\right)+\frac{2 X(\tau)}{\tau^2-1}+Y'(\tau)&\sim0\,,\\
\frac{n (\alpha +\beta  \tau)}{(m-1) (\tau-1)}+\frac{m(\gamma +\delta \tau)}{(n-1) (\tau+1)}-2 i m n \chi_1(\tau)-\frac{2 i t X(\tau)}{\tau^2-1}+\frac{2 i Y(\tau)}{\tau^2-1}&\sim0\,,
\end{align}
where we have defined $X(\tau)=m\chi_2(\tau)+n\chi_3(\tau)$ and $Y(\tau)=m\chi_2(\tau)-n\chi_3(\tau)$. While the last equation can be solved to fix $\chi_1(t)$ completely up to polynomials, the first equation can be rewritten as
\begin{equation}
\frac{i n (\alpha +\beta  \tau)}{(m-1) (\tau-1)}-\frac{i m (\gamma +\delta  \tau)}{(n-1) (\tau+1)}-\frac{2 \tau Y(\tau)}{\tau^2-1}+Y'(\tau)\sim0\,.
\end{equation}
after noticing that without loss of generality one can set $X(\tau)=0$. Taking into account the form of the homogeneous part of the equation, any polynomial of the type $(1-\tau^2)p(\tau)$ on the right-hand side can be reabsorbed by a polynomial shift in $Y$. Hence, the most general equation we need to solve is actually
\begin{equation}
Y'(\tau)+\frac{2\tau}{1-\tau^2}Y(\tau)= -\frac{i n (\alpha +\beta  \tau)}{(m-1) (\tau-1)}+\frac{i m (\gamma +\delta  \tau)}{(n-1) (\tau+1)}+C+D \tau\,.
\end{equation}
Its solution can be easily found and, after changing variables to $\omega=\tau^{-1}$, reads:
\begin{align}
Y(\omega)&=\frac{1-\omega ^2}{4 (m-1) (n-1) \omega ^2}\Big[\log (1-\omega ) \Big(2 C (m-1) (n-1)+2 D (m-1) (n-1)\\
+&i (m(m-1) (\gamma +\delta )+n(n-1)  (\alpha -\beta ))\Big)\nonumber\\
+&\log (\omega +1) \Big(-2 C (m-1)(n-1)+2 D (m-1) (n-1)\nonumber\\
-&i (m(m-1) (\gamma +\delta )+n(n-1) (\alpha -\beta ))\Big)-4 D (m-1) (n-1) \log (\omega )\nonumber\\
-&\frac{2 i \left(m(m-1) (\omega -1) (\gamma -\delta )+n(n-1) (\omega +1) (\alpha +\beta )\right)}{\omega ^2-1}\Big]\,.\nonumber
\end{align}
From the above it is clear that any $\adD$ needs to be set to zero while, without loss of generality, it is convenient to make the following choice:
\begin{align}
\alpha&=\frac{m(m-1) (\gamma -\delta )-\beta  n(n-1)}{n(n-1)}\,,& C&=i \left(\frac{\beta  n}{m-1}-\frac{\delta  m}{n-1}\right)\,,
\end{align}
which without affecting the positive powers in $\omega$ up to an overall constant, cancels any pole in $\omega^{-1}$.
Plugging now everything back into the first equation one finally gets
\begin{align}
\frac{(n-1) \left(2 \gamma  (m-1)^2+\beta  (n-1) n (\tau-1)\right)+\delta  (m-1)^2 (n (\tau-1)+2)}{n(m-1) (n-1) \left(\tau^2-1\right)}\sim0\,.
\end{align}
Its solution is given by 
\begin{align}
\beta&=-\frac{\delta  (m-1)^2}{(n-1)^2}\,,& \gamma&=-\frac{\delta  }{n-1}\,.
\end{align}
To summarize we find the following solutions:
\begin{align}
f(\omega)&\sim\sigma\left[\frac{\omega}{1+\omega}-m\tanh^{-1}(\omega)\right]\,,&
g(\omega)&\sim\sigma\left[\frac{-\omega}{1-\omega}-n\tanh^{-1}(-\omega)\right]\,,
\end{align}
where $\sigma$ is an arbitrary overall constant that might depends on $m$ and $n$. This gives rise to one bosonic and one fermionic cohomology for each choice of $m$ and $n$. Recall that the above functions should be interpreted as formal series around $\omega=\tau^{-1}\sim0$.

The above shows how the cohomology in the space of pseudo-local functionals is non-trivial. Considering also the local cohomology, the only possibility to have a local cohomology is to find a local and exact one-form whose improvement is pseudo-local. This amount to solving the condition that the following vector has polynomial components in $\tau^{-1}$: 

{\footnotesize\begin{equation}
\vec{k}^{(1)}=\begin{pmatrix}
-f(\tau)-2 i \chi_1'(\tau)\\
-(1-\tau) f(\tau)-2 i m \chi_1(\tau)\\
-(1+\tau) f(\tau)+2 i n \chi_1(\tau)\\
0\\
0\\
-\left(1-\tau^2\right) f(\tau)\end{pmatrix}\,,
\end{equation}}\noindent when $f(\tau)$ is a non-polynomial function. Gauge fixing the third component using $\chi_1(\tau)$ one can see that no such solution exists so that there is no local cohomology.

\paragraph{$\boldsymbol{m=0}$ and $\boldsymbol{n<0}$:} In this case we can further use exact forms to fix $f(\tau)\sim f_2 \tau^{-2}+O(\tau^{-3})$ so that the term of order $\tau^{-1}$ vanishes.
Non-trivial cohomologies are then in correspondence with the solution of the following equation:

{\footnotesize\begin{equation}
\adD \vec{k}^{(1)}=-i\begin{pmatrix}
0\\
2 i \left((\tau+1) f(\tau)+\chi_3'(\tau)\right)\\
0\\
-i \left(\left(\tau^2-1\right) f(\tau)+2 \chi_3(\tau)\right)\\
0\\
-2 i n \chi_3(\tau)
\end{pmatrix}\sim0\,,
\end{equation}}\noindent where, due to the condition $m=0$, we have set to zero trivially-vanishing pieces.
One can now set to zero $\chi_3$ up to polynomials and arrive to the solution:
\begin{align}
f(\tau)&\sim\frac{\sigma}{1+\tau}=\frac{\sigma\,\omega}{1+\omega}\,.
\end{align}
Therefore, we have again one bosonic and one fermionic cohomology.

In the case of a local cohomology we have to find the non-polynomial functions $f(\tau)$ for which the following is polynomial:

{\footnotesize\begin{equation}
\vec{k}^{(1)}=\begin{pmatrix}
0\\
(\tau-1) f(\tau)\\
0\\
0\\
0\\
\left(\tau^2-1\right) f(\tau)\end{pmatrix}\,.
\end{equation}}\noindent One can then find a solution of the form
\begin{equation}
f(\tau)=\frac{p(\tau^{-1})}{1-\tau}\,,
\end{equation}
for any polynomial $p$. The above infinite solutions are however trivial in local cohomology since
\begin{equation}
p(\tau^{-1})=\tau^{-k}-1=(\tau^{-1}-1)\,\tilde p(\tau^{-1})=\tau^{-1}(1-t)\,\tilde p(\tau^{-1})\,,
\end{equation}
gives a $\vec{k}^{(1)}$ which is locally exact for any $k>0$ and the constant gives rise to a trivial 1-form.
We thus find no cohomology in the space of local functionals.

\paragraph{$\boldsymbol{m=1}$ and $\boldsymbol{n<0}$:} As explained above only the fourth component (or the fifth if we exchange $m$ and $n$) contributes after performing the contour integrals. Non-trivial cohomologies are in correspondence with the solution of the following ODE:
\begin{equation}
(1-\tau^2) f'(\tau)-f(\tau) ((n-1) \tau-n-1)\sim 0\,.
\end{equation}
Looking at the form of the differential operator and at its image on polynomials, it is easy to see that up to polynomials we need to solve:
\begin{equation}
(1-\tau^2) f'(\tau)-f(\tau) ((n-1) \tau-n-1)= \alpha+\beta \tau+\gamma \tau^{-n+2}\,.
\end{equation}
Going to the point at infinity one can drop the $\gamma$ term since it gives rise to poles in $\omega\sim0$ while $\alpha$ and $\beta$ are uniquely fixed up to an overall coefficient, by the requirement that the expansion in $\omega$ is analytic at $\omega=0$ and starts from the linear term. 
One can indeed rewrite the above equation as:
\begin{equation}
(1-\tau)^2(1+\tau)^{-n+1}\partial_\tau\Big[(1-\tau)^{-1}(1+\tau)^{n}f(\tau)\Big]=\alpha+\beta \tau\,,
\end{equation}
whose solution in terms of $\omega$ can be easily integrated as
\begin{equation}
f(\omega)=-\frac{(1-\omega)(1+\omega)^{-n}}{\omega^{-n+1}}\int^\omega \d x\frac{x^{-n+1}(\alpha+\beta x^{-1})}{(1-x)^2(1+x)^{-n+1}}\,.
\end{equation}
The analyticity condition in $\omega$ is automatically satisfied due to the absence of poles at $x=0$.
Performing the integration one can see that modulo polynomials only one constant among $\alpha$ and $\beta$ remains arbitrary, while the solution takes the form
\begin{equation}
f(\omega)\sim\sigma\frac{(1-\omega)(1+\omega)^{-n}}{\omega^{-n+1}}\tanh^{-1}(\omega)\,,
\end{equation}
where $\sigma$ is an arbitrary constant that can depend on $n$. Again we find no local cohomology.

\subsubsection{Form-Degree 2}
At form-degree $2$ we can distinguish three relevant cases:
\begin{itemize}
\item $n<0$ and $m<0$,
\item $n<0$ and $m=0$,
\item $n<0$ and $m=1$.
\end{itemize}
All other cases are either obtainable from the ones above or are trivial.

\paragraph{$\boldsymbol{n<0}$ and $\boldsymbol{m<0}$:}
The preliminary step in order to study the cohomology is to parametrize the most general term that cannot be made exact fixing the freedom in Fierz transformations and exact forms. At this form-degree this freedom amounts to nine arbitrary functions. Three functions $\chi^{(2)}_i$ due to Fierz identities and six functions due to the freedom of adding an exact form $\adD \vec{f}^{(1)}$. The resulting expression is given by

{\scriptsize
\begin{equation}
\vec{k}^{(2)}=\begin{pmatrix}
i \left((\tau+1) f_2'(\tau)+n f_2(\tau)+(\tau-1) f_3'(\tau)+m f_3(\tau)-i k_1(\tau)-2 \chi_1'(\tau)\right)-i f_1(\tau) (m \tau+m-n \tau+n-2)\\
\begin{pmatrix}
i \Big(2 m n \left(\tau^2-1\right) f_1(\tau)+2 n (\tau+1) f_2(\tau)-2 m \tau f_3(\tau)+2 m f_3(\tau)+2 (n-1) n f_4(\tau)+2 m^2 f_5(\tau)-2 m f_5(\tau)+2 m n f_6(\tau)\\
-2 m f_6(\tau)-2 n f_6(\tau)-i n k_2(\tau)-i m k_3(\tau)-2 Y'(\tau)\Big)\end{pmatrix}\\
\begin{pmatrix}
-i \Big(-2 n (\tau+1) f_2(\tau)-2 m (\tau-1) f_3(\tau)+n (-2 (n-1) f_4(\tau)+2 f_6(\tau)+i k_2(\tau)+4 m \chi_1(\tau))\\
+m (2 (m-1) f_5(\tau)-2 f_6(\tau)-i k_3(\tau))-2 X'(\tau)\Big)
\end{pmatrix}\\
\begin{pmatrix}
i \left((m-1) \left(\tau^2-1\right) f_2(\tau)-(\tau-1) \left((\tau+1) f_4'(\tau)+(m-1) f_6(\tau)\right)-f_4(\tau) (\tau (m+n-2)+m-n-2)\right)\\+k_4(\tau)+\frac{i (m-1) (X(\tau)-Y(\tau))}{n}
\end{pmatrix}\\
\begin{pmatrix}
i \left((n-1) \left(\tau^2-1\right) f_3(\tau)+\left(\tau^2-1\right) f_5'(\tau)+f_5(\tau) (\tau (m+n-2)+m-n+2)+(n-1) (\tau+1) f_6(\tau)\right)\\
+k_5(\tau)+\frac{i (n-1) (X(\tau)+Y(\tau))}{m}
\end{pmatrix}\\
i (2 n (\tau+1) f_4(\tau)-2 m (\tau-1) f_5(\tau)-i k_6(\tau)-2 X(\tau))
\end{pmatrix}\,.
\end{equation}}\noindent  Above we have changed variables defining $X(\tau)=m\chi_2(\tau)+n\chi_3(\tau)$ and $Y(\tau)=m\chi_2(\tau)-n\chi_3(\tau)$ while summing and subtracting the first and the second components in $\adD \vec{f}^{(1)}$ after multiplying them with $n$ and $m$ respectively.
At this point it is not hard to see that:
\begin{itemize}
\item The last component can be removed by fixing $X(\tau)$;
\item The third component can be removed by fixing $\chi_1$;
\item The fourth and fifth component can be removed upon choosing $f_2$ and $f_3$ respectively;
\item The first component and the second component can be removed by fixing either $f_6$ or $f_1$ if the conservation condition is enforced.
\end{itemize}
We then conclude that there is no pseudo-local cohomology at form-degree $2$ if $n<0$ and $m<0$. One can also show that the corresponding local cohomology is trivial as well.

\paragraph{$\boldsymbol{n<0}$ and $\boldsymbol{m=0}$:}This case is similar to the previous one except that only the first, third and fifth component of the vector are projected out by the contour integrals over $s+r$ and $s-r$.

\paragraph{$\boldsymbol{n<0}$ and $\boldsymbol{m=1}$:}
In this case only the fourth component contributes and the condition to be an exact form reads:
\begin{equation}
 [(1-\tau)n+1+\tau]f(\tau)+\left(1-\tau^2\right) f'(\tau)\sim k(\tau)\,.
\end{equation}
Notice that the choice $m=1$ makes conservation trivial so that the only condition to solve is whether there exist a solution of the above equation that admits a well-defined expansion around $\omega=\tau^{-1}\sim0$.

We can then study the above question by considering $k(\tau)=\tau^{-k}$ and studying the corresponding solutions:
\begin{equation}
 [(1-\tau)n+1+\tau]f(\tau)+\left(1-\tau^2\right) f'(\tau)= \tau^{-k}+\alpha+\beta \tau\,.
\end{equation}
Re-expressing the equation above in terms of $\omega$ and fixing the ambiguity up to elements belonging to the 1-form cohomology, one recovers the equation:
\begin{equation}
\left(1-\omega ^2\right) f'(\omega )+[1+\omega-n(1-\omega)]\frac{f(\omega )}{\omega}=\omega ^k\,,
\end{equation}
whose solution can be integrated as:
\begin{equation}
f(\omega)=-\frac{(1-\omega)(1+\omega)^{-n}}{\omega^{-n+1}}\int^\omega dx\frac{x^{k-n+1}}{(1-x)^2(1+x)^{-n+1}}
\end{equation}
The solution for $k\geq1$ has the following structure:
\begin{equation}
f(\tau)\sim\frac{(1-\omega)(1+\omega)^{-n}}{\omega^{-n}}\left[A_{n,k}\log(1-\omega)+B_{n,k}\log(1+\omega)\right]\,,
\end{equation}
where for any $n$ and $k$:
\begin{align}
A_{k,n}&=\frac{1}{2\pi i}\oint_{x=1} dx\frac{x^{k-n+1}}{(1-x)^2(1+x)^{-n+1}}\,,\\
B_{k,n}&=\frac{1}{2\pi i}\oint_{x=-1} dx\frac{x^{k-n+1}}{(1-x)^2(1+x)^{-n+1}}\,.
\end{align}
Hence this concludes the proof that the cohomology at form-degree $2$ is trivial in the space of pseudo-local functionals.


\subsection{Cohomologies Linear in $\pC$ and $\pomega$}
The cohomologies $\mathbb{H}^n(\tadD,\pomega \pC)$ and $\mathbb{H}^n(\tadD,\pC \pomega)$ can be trivially obtained from the cohomologies calculated in the previous subsections. This is due to the observation that upon appropriate relabeling of $\xi$, $\eta$ and $y$ the Fourier representation of $\tadD$ acting on functionals linear in $\pC$ and $\pomega$, which are given by \eqref{eq:barredKernels} and \eqref{eq:tadInspector}, reduce to the Fourier representation of the adjoint covariant derivative $\adD$ acting on functionals quadratic in $\pC$ given in \eqref{eq:inspector}. Note however that the form-degree $n$ is shifted by one as $\pomega$ is a one-form.

\section{Asymptotic Behavior}\label{appsec:Prop}

Below we collect in detail the large-$l$ asymptotic behavior of the various expressions that appear in the main text:
{\allowdisplaybreaks \footnotesize \besubeqs 
\begin{align}
\eqref{Canonicala}&: & a^{n,m,l}|_{l\rightarrow \infty}\sim&\, \text{exactly zero}\,,\\
\eqref{CanonicalExact}&: & a^{n,m,l}|_{l\rightarrow \infty}\sim&\frac{-4 l i^l f_{m+n} }{(m+n+l+3)!}\,,\\
\eqref{CanonicalNablaExact}&: & a^{n,m,l}|_{l\rightarrow \infty}\sim& \frac{2^l}{l! l^{\frac{m+n+4}{2}}}\,,\\
\eqref{PtensorA},\eqref{PtensorB}&: & a^{n,m,l}|_{l\rightarrow \infty}\sim& \frac{(-i)^{l}f_{m+n+l}}{l!} &
a_F^{n,m,l}|_{l\rightarrow \infty}\sim& \frac{i^{-l}f_{m+n+l}}{l!}\,,\\
\eqref{strangestrA},\eqref{strangestrB}&: & c^{n,m,l}|_{l\rightarrow \infty}\sim& f_{n+m}  & c^{n,m,l}_F|_{l\rightarrow \infty}\sim&-l^2 f_{n+m}\,,\label{weirdstr}\\
\eqref{Fronsdalbackreaction}&: & a^{n,m,l}|_{l\rightarrow \infty}\sim& \frac{(-i)^lf_{n+m}}{(m+n+l+3)!}\,,\\
\eqref{Lorentzredef} &: & a^{n,m,l}|_{l\rightarrow \infty}\sim&-\frac{i (-i)^l (-1)^m (m+n)!}{l! m! n! (l+m+n+3)^2}\,,\\
\eqref{Simplifiedbackreaction} &: & a^{n,m,l}|_{l\rightarrow \infty}\sim&-\frac{i (-i)^l (-1)^m (m+n)!}{l! l^2 m! n! }\,,\\
\eqref{FronsdabckrExact}&: & a^{n,m,l}|_{l\rightarrow \infty}\sim&\frac{(-i)^l l \log l f_{n+m}}{(l+m+n+3)! }\,,\\
\eqref{susyscalredef} &: & a^{0,0,l}|_{l\rightarrow \infty}\sim&\frac{(i)^l }{l! }\,,\\
\eqref{eq:atMassimosRequest} &: & a^{n,m,l}|_{l\rightarrow \infty}\sim& \frac{1}{l^{2s}l!} \, \\
\eqref{shadowredefinition}& : & a^{n,m,l}|_{l\rightarrow \infty}\sim&\frac{(i)^l }{l! l^q }\,,\qquad q\geq0\,.
\end{align}
\esubeqs}\noindent 
The backreaction obtained from the Prokushkin--Vasiliev theory, \eqref{Simplifiedbackreaction},
is quite complicated but using its large-$l$ asymptotic we see that its $\adD$-exact representation $\formJ^\textsc{PV}=\adD \formU^\textsc{PV}$ has large-$l$ asymptotics which are no worse than $\frac{1}{l!}$. Apart from the artificial example of \eqref{weirdstr}, all large-$l$ asymptotics have the same damping factor $1/l!$.

\section{$D$-Dimensional Theory at $D=3$}\label{sec:ddimtheory}
\label{app:Ddimensionaltheory}
This section is devoted to another three-dimensional higher-spin theory. Namely, we wish to consider the generic $D$-dimensional Vasiliev theory \cite{Vasiliev:2003ev}, which is known not to require the presence of any twisted sector. After some simplifications the $D$-dimensional Vasiliev theory can be reduced to\footnote{We do not give a detailed account of this theory, referring to the original paper  \cite{Vasiliev:2003ev} for definitions, to \cite{Bekaert:2005vh} for a review and to  \cite{Alkalaev:2014nsa} for a brief summary and explanations on how to slightly reduce the field content to the form presented in this appendix.}
\begin{align}
\d \MW+\MW\star \MW&=0\,, & \{\MS_\ga, \MB\star \varkappa\}_\star&=0\,,\\
\d (\MB\star \varkappa)+[\MW, \MB\star \varkappa]_\star&=0 \,, & [\MS_\ga, \MS_\gb]_\star&=-2i\epsilon_{\ga\gb}\left(1+ \MB\star \varkappa \right)\,,\\
\d \MS_\ga+[\MW, \MS_\ga]_\star&=0\,,
\end{align}
supplemented with the so-called kinematical constraints
\begin{align}
[F^0_{\ga\beta}, \MW]_\star &=0\,, & [F^0_{\ga\beta}, \MB]_\star &=0\,, & [F^0_{\ga\beta}, \MS_\gamma]_\star &=\epsilon_{\ga\gamma}\MS_\beta + \epsilon_{\beta\gamma}\MS_\alpha\,,
& [F^0_{\ga\beta}, \varkappa]_\star &=0\,,
\end{align}
where $\varkappa=e^{iyz}$ is the usual Klein operator and $F^0_{\ga\gb}=-\frac{i}4\{y^a_\ga,y_{a\gb}\}+\frac{i}4\{z_\ga,z_{\gb}\}$ are the $\textrm{sp}(2)$ generators of the algebra Howe dual to the AdS algebra $\textrm{so}(D,2)$. The Lorentz and translation generators of the background anti-de Sitter algebra $\textrm{so}(D,2)$ are
\begin{align}\label{generatorsLP}
L^{ab}&=\frac{i}{4}\{y^a_\nu,y^{b\nu}\} \,, &P^a&= \frac{i}{4}\{y^a_\nu,y^{\nu}\}\,,
\end{align}
and by the Howe duality property they commute to the $\textrm{sp}(2)$ generators.

In generic $D$ this system describes interactions of all $s=0,1,2,3,\dots$ fields and there is no need to involve a twisted sector, which we recall is a built-in feature of Prokushkin--Vasiliev Theory studied in this paper. One could, however, choose to add twisted fields and couple them to the $D$-dimensional Vasiliev theory. This is done by enlarging the higher-spin algebra and yields an extended $D$-dimensional theory. The minimalistic option is to add a Klein operator $k$ for $y^a_\ga$, i.e.\footnote{Due to the form of generators \eqref{generatorsLP} there is an ambiguity on how to couple $k$ to the algebra as to ensure $kP^ak=-P^a$. While the simplest option is in the main text, let us give an alternative realization that is in the spirit of the Prokushkin-Vasiliev theory. The alternative relations read:
\begin{equation}\notag
\{y_\ga, k\}=\{ z_\ga, k\}=0\,,  \qquad \{\rho,k\}=0\,,  \qquad [y_\ga,\rho]=[z_\ga,\rho]=0\,,    \qquad[y^a_\ga,k]=[y^a_\ga,\rho]=0\,, \qquad k^2=\rho^2=1\,.
\end{equation}
As in the Prokushkin-Vasiliev theory a truncation needs to be imposed such that $\MW$ and $\MB$ are $\rho$-independent and $\MS_\ga=\rho s_\ga(y^a_\ga,y_\ga,k)$. The vacuum for $\MS_\ga$ is $\rho z_\ga$ and it functions the same way as $z_\ga$ in the original theory thanks to $[\rho z_\ga,k]=0$, $[k,\varkappa]=0$.
}
\begin{equation}
[y_\ga, k]=[z_\ga, k]=0\,,  \qquad\{y^a_\ga,k\}=0\,, \qquad k^2=1\,.
\end{equation}
No modification of the above equations is needed. The vacuum is the canonical one, i.e. $\MB=0$, $\MS_\ga=z_\ga$, with
\begin{align}
\Omega=\frac12 \omega_{a,b} L^{ab}+h_a P^a\,,
\end{align}
where $\omega^{a,b}$ and $h^a$ are the spin-connection and vielbein of AdS space. Now the linearized equations for $\MW=\Omega+\boldsymbol{w}$, $\MB=\flC$ and $\MS_\ga=z_\ga+\boldsymbol{s}_\ga$ reduce to
\begin{align}
D_\Omega\boldsymbol{w}&=0\,, \qquad\pl^z_\ga \boldsymbol{w}=\frac{i}2 D_\Omega \boldsymbol{s}_\ga\,, \qquad D_\Omega (\flC\star \varkappa)=0\,,\\
\pl^z_\ga \flC&=0\,, \qquad \pl^z_\ga \boldsymbol{s}^\ga=\flC\star\varkappa\,.
\end{align}
These equations can be solved as usual. First, $\flC=\flC(y^a,y,k)$ is $z_\alpha$-independent and
\begin{align}
\d \flC+\Omega\star \flC-\flC\star \pi(\Omega)&=0\,,
\end{align}
where $\pi(\Omega)=\varkappa\star\Omega\star\varkappa$ is the automorphism that flips the sign of the translation generator $P^a=\frac{i}2y^a_\nu y^\nu$. The equation
splits into two different equations for the components $\flC=\pC+\sC k$:
\begin{align}
\adD\sC=\d \sC+\Omega\star \sC-\sC\star \Omega&=0\,, \label{ddimKilling}\\
\tadD \pC=\d \pC+\Omega\star \pC-\pC\star \widetilde{\Omega}&=0 \,.
\end{align}
The interpretation is straightforward: $\pC$ obeys the usual equation for higher-spin Weyl tensors and descendants thereof, i.e. it describes the gauge invariant field-strengths of higher-spin fields. Instead, $\sC$ describes an infinite set of totally-symmetric AdS Killing tensors, including the Killing constant.

For completeness we also write the solution for $\boldsymbol{w}$:
\begin{align}
\boldsymbol{w}=\flomega+\frac{i}2 h_b\int (1-t)\d t\,\left(iz_\nu \pl^{b\nu}\pC(y^a,-zt)+z_\nu y^{b\nu} \sC(y^a,-zt)k\right)e^{ityz}\,,
\end{align}
where $\flomega=\flomega(y^a,y,k)=\pomega+\somega k$ also splits into a higher-spin algebra connection $\pomega$ and a twisted $\somega$, which has the same mysterious interpretation as its Prokushkin--Vasiliev cousin. Let us point out that, contrary to Prokushkin-Vasiliev Theory, in $D>3$ there are non-trivial sources on the right-hand sides of the first-order equations of motion:
\begin{align}
\tadD\somega=\d \somega+\Omega\star \somega-\somega\star \widetilde{\Omega}&=-\frac{i}4h_a\wedge h_b\, y^a_\nu y^{b\nu}\sC(y^m,0)\,,\label{ddimshadow}  \\
\adD \pomega=\d \pomega+\Omega\star \pomega-\pomega\star {\Omega}&=+\frac{i}4h_a\wedge h_b\, \pl^a_\nu \pl^{b\nu}\pC(y^m,0)\,,
\end{align}
which for $\pomega$ amount to the nontriviality of higher-spin Weyl tensors in $D>3$. The sources disappear at $D=3$ except for $s=1$, which will be discussed below.

\subsection{Confronting two Three-Dimensional Higher-Spin Theories}

From now on we will consider the case $D=3$. We are thus left with a three-dimensional higher-spin theory which also involves a twisted sector, and it is therefore natural to compare it with the Prokushkin--Vasiliev theory studied in the main text. We anticipate the fact that in Prokushkin--Vasiliev Theory the twisted fields are built in whereas here we add them 'by hand' seems to indicate that the two theories should differ. We find it however enlightening to compare them precisely, which we comment on in the following.

The above theory at $D=3$ also involves Killing tensors, but there is only one Killing constant therein while there are two in the Prokushkin--Vasiliev theory, the doubling being due to the $\phi$-dependence. The field $\pC$ describes one scalar field and a spin-one field which are present, while the Weyl tensors for $s\geq2$ vanish identically. In contrast, in Prokushkin--Vasiliev Theory there is no dynamical spin-one field, which is also clear by noticing that the interactions among higher-spin fields are Chern--Simons-like. We also see that the spectrum of the higher-spin algebras do not match, the difference being that there are two Lorentz scalars in the physical sector of Prokushkin--Vasiliev while there is only one such scalar in the $D$-dimensional theory at $D=3$ because the spin-one Weyl tensor is equivalent to a three-dimensional vector.\footnote{The difference is however not drastic since a vector field is dual to a scalar in dimension $3$.}

Again let us stress that the key difference with Prokushkin--Vasiliev Theory is that the twisted sector of the extended $D$-dimensional theory is not built-in, which we have shown to be not straightforward in Prokushkin--Vasiliev. The truncation is achieved by requiring all fields not to depend on $k$.

When going to the second order in the extended $D$-dimensional theory we find that the structure of the backreaction is different. For example, $\tadD\pC^{(2)}\sim \pC\pC+\sC\sC$ and $D\sC^{(2)}\sim \pC \sC$, while in Prokushkin--Vasiliev we have instead $D\sC^{(2)}\sim \pC \pC+\sC\sC$ and $\tadD\pC^{(2)}\sim \pC\sC$, which illustrates the general statement that the twisted sector can be truncated away in the extended $D$-dimensional theory but not in Prokushkin--Vasiliev Theory.

As mentioned above another difference lies in the field content. The $D$-dimensional theory extrapolated to $D=3$ has degrees of freedom associated with $s=0$ and $s=1$, the corresponding energies, which can be read off from the general formulas $E=D+s-3$, which gives $E=0$ for $s=0$ and $E=1$ for $s=1$. The AdS masses are $m^2=0$ and $m^2=-2$. In dimension $3$ a vector field is dual to a scalar with $m^2=0$, so we have two scalars of the same mass $m^2=0$. This should be compared with the masses of the scalars in the Prokushkin--Vasiliev theory, where $m^2=-1+\lambda^2$.

The precise truncation of Prokushkin--Vasiliev Theory we consider in the paper corresponds to $\lambda=\frac12$ and should be dual to the $W_{\frac12}$-minimal model. The $D$-dimensional theory fits $\lambda=1$ and should be dual to a $2D$ free boson theory as it generically occurs in higher dimensions.\footnote{The oscillator realization, found in \cite{Vasiliev:2003ev}, gives the HS algebra as a subquotient with respect to certain ideal. We note that both $\textrm{hs}(\lambda)$ at $\lambda=1$ and the realization of \cite{Vasiliev:2003ev} for $\lambda=1$ are equivalent and share the property that the generators with $s>0$ form an ideal, which can be seen from the bilinear form \cite{Pope:1989sr,Vasilev:2011xf}. Such decoupling is expected since the $s=0$ component is of conformal weight-$0$ and has the logarithmic mode. Therefore, the formation of the ideal at $\lambda=1$ is in accordance with AdS/CFT. While one might face certain difficulties in trying to factorize in the realization of \cite{Vasiliev:2003ev} at the interacting level, our linear analysis above is unaffected as well as general statements on the mixing of twisted and physical sectors. } Both $\lambda=1$ and $\lambda=\frac12$ seem to be generic from the bulk point of view, which only makes it even more surprising to have such different behaviors of the twisted sectors in both cases. \\

\subsection{Twisted Sector} 

Let us have a closer look at the twisted sector. This discussion applies
both to $D$-dimensional Vasiliev Theory and to Prokushkin--Vasiliev Theory. We first look at the gauge twisted fields $\somega$ and discuss the field content, gauge symmetries and possible gauge invariant equations and then 
show how the twisted zero-forms $\sC$ source $\somega$ via \eqref{ddimshadow}.

\paragraph{Twisted one-forms.} When decomposed into Lorentz tensors \eqref{ddimshadow} splits into an infinite set of equations that involve
\begin{align}
&\somega^{a(s+k),b(s)}\,, \qquad k=0,1,2,\dots\,,
\end{align}
for every $s=0,1,2,\dots$. There exists a standard technique, the $\sigma_-$-cohomology, used to analyze the content of any unfolded equation \cite{Lopatin:1987hz,Bekaert:2005vh}. The procedure consists of taking the part of the differential that lowers the degree of a fiber tensor. In our case the relevant operator is $\tadD$, so that
\begin{equation}
(\sigma_- \somega)^{a(s+k),b(s)}=h_c\wedge\left(\somega^{a(s+k)c,b(s)}+\frac{1}{k+2}\,\somega^{a(s+k)b,b(s-1)c}\right)\,,
\end{equation}
which is dubbed $\sigma_-$ and can be checked to be nilpotent. The $\sigma_-$-cohomology for all kinds of $\textrm{AdS}$-modules was computed in \cite{Bekaert:2005vh, Boulanger:2008kw, Skvortsov:2009nv}. 
Following standard techniques, the metric-like content of $\somega$ is given by $\mathbb{H}^1(\sigma_-)$. It is easy to see that for $s>1$ the only tensor in the kernel of $\sigma_-$ is given by
\begin{align}
\somega^{a(s),b(s)}&=h_c\Phi^{a(s),b(s),c}\,,
\end{align}
where $\Phi^{a(s),b(s),c}$ is not traceless but has only one non-vanishing trace so that the independent metric-like fields described by the twisted sector are $\Phi^s$, $s=1,2,\dots$:
\begin{align}
&\Phi^{a(s),b(s),c}\,, \qquad \Phi\fud{a(s),b(s-1)n,}{n}\,.
\end{align}
In addition to the above elements, the case $s=0$ is degenerate due to the non-trivial kernel of $\sigma_-$ given by $h^a\,\Phi$. Also for $s=1$ there is an additional element given by\footnote{These cocycles of $\sigma_-$ become trivial if the trace constraints on the fiber tensors are relaxed.} $h^{[b}\Phi^{a]}$, where however $\Phi^{a}$ cannot be identified with the trace of $\Phi^{a,b,c}$. 

To summarize, the indpendent components of $\somega$ are given by 
\begin{align}
\somega^{a(s),b(s)}&=h_c\Phi^{a(s),b(s),c}\,,& \somega^{a,b}&=h^{[a}\Phi^{b]}\,,& \somega^a&=h^a\Phi\,.
\end{align}
The above fields, being one-forms, are gauge fields whose metric-like components transform as
\begin{align}
\delta \Phi^{a(s),b(s),c}&=\nabla^c \xi^{a(s),b(s)} +\mbox{permutations}-\mbox{trace}\,,
\end{align}
with a traceless gauge parameter whose physical components belong to $\mathbb{H}^0(\sigma_-)$, and hence in the metric-like formalism has the same index structure as the higher-spin Weyl tensors. Remarkably, the rigid symmetries associated with these gauge fields are the infinite-dimensional Weyl modules themselves, in contrast with the physical higher-spin fields whose rigid symmetries are given by Killing tensors and hence are finite-dimensional at a fixed spin. For $s=0$ one finds in particular
\begin{align}
\delta \Phi=(\square +2(d-2))\xi\,,
\end{align}
and rigid symmetries are given by $(\square +2(d-2))\xi=0$, i.e. correspond to an on-shell scalar field. 

The rest of the Lorentz components of $\somega$ are either pure gauge or expressed as derivatives of $\Phi^s$. Possible gauge-invariant equations
are given by $\mathbb{H}^2(\sigma_-)$. One finds three independent first order operators:
\begin{align}
&E^{a(s),b(s),c,d}\,, \qquad E^{a(s),b(s-1),c}\,, \qquad E^{a(s-1),b(s-1)}\,, \label{possibleshadeq}
\end{align}
corresponding to the irreducible components of a two-form with the index structure of a Weyl tensor:
\begin{equation}
E^{a(s),b(s)}{}_{c,d}h^c\wedge h^d\,.
\end{equation}
Note that only the trace of $\Phi^s$ contributes to the last operator in \eqref{possibleshadeq}, $\nabla_m\Phi\fud{a(s-1)m,b(s-1)n,}{n}$. For the degenerate case $s=0$ there are no equations possible since the cohomology is empty.

\paragraph{Twisted zero-forms.} As it was already said, \eqref{ddimKilling} describes Killing tensors encoded in $\sC$. The Killing equation \eqref{ddimKilling} also splits into an infinite set of equations for a finite number of fields:
\begin{align}
& \sC^{a(s-1),b(k)}\,, \qquad k=0,....,s-1\,.
\end{align}
In particular the first equation of each chain,
\begin{align}
\nabla_\mm \sC^{a(s-1)}+h_{b\mm} \sC^{a(s-1),b}&=0\,,\label{firstKillingeq}
\end{align}
implies, after symmetrizing the indices, the standard Killing equation $\nabla^a \sC^{a(s-1)}=0$. There is also a degenerate case $s=1$ for which we have a Killing constant $\sC$.

\paragraph{Equations for twisted one-forms.} Now we see that \eqref{ddimshadow} sets to zero the first two operators from \eqref{possibleshadeq}, imposing equations thereon. The last operator matches the symmetry of one of the Killing tensor components and yields one more equation:
\begin{align}
\nabla_m\Phi\fud{a(s-1)m,b(s-1)n,}{n}&=\sC^{a(s-1),b(s-1)}\,.
\end{align}
In particular the Killing constant appears as a source for the $s=1$ field $\nabla_m \Phi^{m} =\sC(y=0)$. Nothing dramatic happens for $D=4$ and all the conclusions above are still true.

The Killing equations \eqref{ddimKilling} or \eqref{firstKillingeq} can be easily solved using the ambient space technique, see e.g. \cite{Eastwood:2002su}, the solution for a spin-$s$ tensor being a polynomial in the boundary coordinates with the powers of the Poincare coordinate $z$ ranging from $-(s-1)$ to $(s-1)$.

As in Prokushkin--Vasiliev Theory, at second order one finds on the right-hand side of Fronsdal equations
some currents built out of the first order fields that include $\sC\sC$. Therefore, even if the physical scalars and higher-spin fields are switched off at first order there is a non-trivial source for higher-spin fields at second order due to the Killing tensors.

In Appendix~\ref{app:cohomologies} we observe that the definition of the twisted one-forms $\somega$ is ambiguous due to an option to shift them by physical fields of the form $h\wedge \pC$. Differently put the cohomology $\mathbb{H}^1(\tadD)$, with coefficients $\pC$ in the twisted-adjoint module of the higher-spin algebra, is not empty. Despite the difference between the way the twisted and physical fields couple to each other in Vasiliev Theory and Prokushkin--Vasiliev, it is easy to see that $\mathbb{H}^1(\tadD)$ is non-trivial also in the former.  

\subsection{Invariant Definition of Twisted Sectors}

Let us conclude this appendix with a small remark on the algebraic interpretation of the twisted sector. It turns out that, at least algebraically, the definition and realization of the twisted sector does not require any extra ingredients as compared to those already present in any higher-spin theory. 

Any known higher-spin algebra comes as an associative algebra on which we then define the Lie bracket to be the commutator of the corresponding associative product. Moreover, it comes equipped with an automorphism, $\pi$, that flips the sign of $\textrm{AdS}$-translations, which in the conformal basis can be seen to exchange translations with boosts and flip the sign of the dilatation generator. This automorphism allows one to construct the twisted-adjoint representation, where the action of the higher-spin algebra on itself is twisted by $\pi$, i.e. $\widetilde{\textrm{ad}}_a x=a\star x- x\star \pi(a)$. The twisted-adjoint representation is the one used to describe degrees of freedom, e.g. scalar fields in the Prokushkin--Vasiliev theory. Given an order-two automorphism $\pi$ one can build an extended associative algebra that is $\textrm{hs}\oplus\textrm{hs}$ as a linear space equipped with the following product:
\begin{align}
(a,x)\star (b,y)&=(a\star b+x\star \pi(y), x\star \pi(b)+a \star y)\,, \qquad a,b,x,y\in\textrm{hs}\,.
\end{align}
This is the algebra upon which the higher-spin theory extended with twisted fields is built. The adjoint representation of the extended algebra contains both the usual adjoint and the twisted-adjoint representations of the higher-spin algebra it was built from:
\begin{align}
[(a,x),(b,y)]_\star&=([a, b]_\star+x\star \pi(y)-y\star \pi(x), a\star y - y\star \pi(a) +x\star \pi(b)-b \star x) \,.
\end{align}
Given a Klein operator $k$ that implements the automorphism via $\pi(x)=k\star x\star k$, $k^2=1$, the extended algebra is just the algebra of $a+x\star k$.

The algebraic interpretation of the twisted sector is then related to the fact that the Klein operator realizes the inversion operator $I$: $kK_i k=P_i$, $kP_ik=K_i$ and $kDk=-D$, which are the exact same identities that follow from $K_i=IP_iI$, where $P_i, K_i, D, L_{ij}$ are the generators of the $\textrm{AdS}$ algebra in the conformal basis. For conformal fields we have $I\phi(x)=(x^2)^{-\Delta}\phi(\frac{x}{x^2})$. Therefore, the twisted sector can be interpreted as describing the same field content as usual but viewed from the point at infinity. All that being said it is still not clear what is the physical meaning of the twisted sector in the $\textrm{AdS}$ dual theory. What is clear is that the above inversion which sends lowest-weight representations to highest-weight ones clashes with unitarity.

An interesting question to ask is which of the symmetries does a higher-spin theory extended with a twisted sector realize. The AdS background is given by a flat connection $\Omega$ of the anti-de Sitter algebra whose twisted part is identically zero. Decomposing the global symmetry equation $\delta \Omega=0$ into physical and twisted parts we find
\begin{align}
0&=\d\Mxi+[\Omega,\Mxi]\,, & 0&=\d\widetilde{\Mxi}+[\Omega,\tilde{\Mxi}]\,.
\end{align}
Therefore, the global symmetry algebra is the extended higher-spin algebra $\textrm{hs}\oplus_\pi\textrm{hs}$. The vacuum value of the physical $\MB$ in AdS is zero, which leads to the following additional constraints on the global symmetry algebra of the vacuum unless $\tilde{\MB}=0$:
\begin{align}\label{Btrans}
0&=\tilde{\MB}\star \pi(\tilde{\Mxi})-\tilde{\Mxi}\star \pi(\tilde{\MB})\,, & 0&=\tilde{\MB}\star \pi(\Mxi)-\Mxi \star \tilde{\MB}\,.
\end{align}
Here lies one of the crucial differences with the Prokushkin--Vasiliev theory, in which we have, rather,
\begin{align}
0&=[\tilde{\MB},\tilde{\Mxi}\psi]\,,& 0&=[\tilde{\MB},\Mxi]\,.
\end{align}
In the latter case $\tilde{\MB}$ can be non-zero along the Killing constant without having to restrict the global symmetry algebra.\footnote{In \cite{Prokushkin:1998bq} it was argued that $\tilde{\Mxi}$-transformations mix physical fields with twisted fields, whose solutions are not normalizable, and on these grounds should be eliminated, i.e. one has to impose $\tilde{\Mxi}=0$. This may need to be reconsidered in view of physical fields yielding non-trivial backreactions on the twisted sector.} In the $D$-dimensional case the second equation of \eqref{Btrans} implies that the only $\tilde{\MB}$ that does not restrict $\Mxi$ is $\tilde{\MB}=0$. This leaves us with only one option~---~$\tilde \MB=0$~---~to preserve the full higher-spin algebra in the vacuum even in the situation in which the backreaction on the twisted sector can be trivialized (in which one could in principle treat $\tilde{\MB}$ as a set of coupling constants).

\end{appendix}


\begingroup
\setlength{\emergencystretch}{8em}
\printbibliography
\endgroup

\end{document}